\documentclass[11pt,a4paper,preprintnumbers]{article} 
\pdfoutput=1
\usepackage{jheppub}
\usepackage[T1]{fontenc}
\usepackage{lmodern}
\usepackage{graphicx}	
\usepackage{bm}	
\usepackage{mathtools}
\usepackage{amsfonts}
\usepackage{amsmath}
\usepackage{dsfont}
\newcommand{\up}[1]{\textsuperscript{#1}}		
\newcommand{\tabref}[2][]{table{#1}~\ref{tab:#2}}		
\newcommand{\figref}[2][]{figure{#1}~\ref{fig:#2}}		
\newcommand{\sectref}[2][]{section{#1}~\ref{sect:#2}}	
\newcommand{\appref}[2][x]{appendi{#1}~\ref{app:#2}}	
\renewcommand{\eqref}[2][]{eq{#1}.~(\ref{eq:#2})}		
\newcommand{\citeR}[2][]{ref{#1}.~\cite{#2}}		
\newcommand{\ORCID}[2]{\footnotetext[#1]{\href{https://orcid.org/#2}{ORCID: #2}}} 
\newcommand{\eten}[1]{\ensuremath{\times 10^{#1}}}			
\newcommand{\Tr}[1]{\ensuremath{\text{Tr}\left[ #1 \right]}	}	
\newcommand{\lb}{\ensuremath{\left}}		
\newcommand{\rb}{\ensuremath{\right}}
\newcommand{\nl}{\nonumber \\ & \qquad }
\newcommand{\Umat}{\ensuremath \,\mathcal{U}}
\newcommand{\Lag}{\ensuremath\mathcal{L}}
\newcommand{\suTC}{\ensuremath SU(N)_{\textrm{TC}}}
\newcommand{\suC}{\ensuremath SU(3)_{\textrm{C}}}
\newcommand{\suW}{\ensuremath SU(2)_{\textrm{W}}}
\newcommand{\uY}{\ensuremath U(1)_{\textrm{Y}}}
\newcommand{\suL}{\ensuremath SU(3)_{\textrm{L}}}
\newcommand{\suR}{\ensuremath SU(3)_{\textrm{R}}}
\newcommand{\suLR}{\ensuremath \suL \times \suR}
\newcommand{\piTC}{\ensuremath \pi_{\textsc{tc}}}
\newcommand{\barpiTC}{\bar{\pi}_{\textsc{tc}}}
\newcommand{\omegaTC}{\ensuremath \omega_{\textsc{tc}}}
\newcommand{\kappaTC}{\ensuremath \kappa_{\textsc{tc}}}
\newcommand{\LambdaTC}{\ensuremath \Lambda_{\textsc{tc}}}
\newcommand{\LambdaLow}{\ensuremath \Lambda_{\text{low}}}
\newcommand{\LambdaQCD}{\ensuremath \Lambda}
\newcommand{\sinc}{\ensuremath \mkern3mu\textrm{sinc}\mkern1mu}
\newcommand{\omegaTCvev}{\ensuremath \langle \omega_{\textsc{tc}} \rangle}
\newcommand{\kappaTCvev}{\ensuremath \langle \kappa_{\textsc{tc}} \rangle}
\newcommand{\hvev}{\ensuremath \langle h \rangle}
\newcommand{\phicrit}{\ensuremath \phi_{\text{crit.}}}
\newcommand{\MPlred}{\ensuremath m_{\text{Pl.}}}

\begin{document}

\title{Relaxation of the Composite Higgs Little Hierarchy}

\author[a]{Brian Batell,} 
\emailAdd{batell@pitt.edu}

\author[b,c,d,1]{Michael A.~Fedderke,} 
\ORCID{1}{0000-0002-1319-1622}
\emailAdd{mfedderke@uchicago.edu}

\author[b,c,d]{and Lian-Tao Wang}
\emailAdd{liantaow@uchicago.edu}

\affiliation[a]{Pittsburgh Particle Physics, Astrophysics, and Cosmology Center, Department of Physics and Astronomy, University of Pittsburgh, 3941 O'Hara Street, Pittsburgh, PA 15260, USA}
\affiliation[b]{Department of Physics, The University of Chicago, 5640 S Ellis Ave, Chicago, IL 60637, USA}
\affiliation[c]{Enrico Fermi Institute, The University of Chicago, 5640 S Ellis Ave, Chicago, IL 60637, USA}
\affiliation[d]{Kavli Institute for Cosmological Physics, The University of Chicago, 5640 S Ellis Ave, Chicago, IL 60637, USA}


\arxivnumber{1705.09666}
\preprint{\texttt{PITT-PACC-1706}}

\abstract{%
We describe a composite Higgs scenario in which a cosmological relaxation mechanism naturally gives rise to a hierarchy between the weak scale and the scale of spontaneous global symmetry breaking.
This is achieved through the scanning of sources of explicit global symmetry breaking by a relaxion field during an exponentially long period of inflation in the early universe.
We explore this mechanism in detail in a specific composite Higgs scenario with QCD-like dynamics, based on an ultraviolet $SU(N)_{\textrm{TC}}$ `technicolor' confining gauge theory with three Dirac technifermion flavors. 
We find that we can successfully generate a hierarchy of scales $\xi \equiv \langle h \rangle^2 / F_\pi^2 \gtrsim 1.2 \times 10^{-4}$ (i.e., compositeness scales $F_\pi \sim 20$\,TeV) without tuning.
This evades all current electroweak precision bounds on our (custodial violating) model.
While directly observing the heavy composite states in this model will be challenging, a future electroweak precision measurement program can probe most of the natural parameter space for the model.
We also highlight signatures of more general composite Higgs models in the cosmological relaxation framework, including some implications for flavor and dark matter.
}

\maketitle
\flushbottom 

\section{Introduction}
\label{sect:Introduction}
The cosmological relaxation scenario of Graham, Kaplan, and Rajendran~\cite{Graham:2015cka} provides a novel approach to the hierarchy problem of the Standard Model (SM). 
In this scenario, the vacuum expectation value (vev) of an axion field \cite{Weinberg:1977ma,Wilczek:1977pj}, dubbed the \emph{relaxion}, slowly rolls through a trans-Planckian excursion down a very flat shift-symmetry-breaking potential during an exponentially long period of low-scale inflation, in the process dynamically `scanning' the value of the Higgs squared-mass parameter.
Although the bare Higgs squared-mass parameter can be assumed natural (i.e., positive and of the order of the cutoff), it is eventually scanned through zero, triggering spontaneous electroweak symmetry breaking (EWSB).
The breaking of electroweak symmetry gives rise to a back-reaction%
\footnote{%
	In the simplest realization \cite{Graham:2015cka}, the back-reaction is supplied by the emergence of the periodic QCD vacuum potential \cite{Polyakov:1975rs,Belavin:1975fg,tHooft:1976snw,tHooft:1976rip,Polyakov:1976fu,Jackiw:1976pf,Callan:1976je,Coleman:1985aos,Weinberg:1995qft} following EWSB, with $V_{\textsc{qcd}}  \propto m_\pi^2 f_\pi^2 \propto m_{u,d} \propto v$.
	} %
on the flat relaxion scanning potential which, combined with energy dissipation from Hubble friction, stalls the relaxion rolling, dynamically locking in a small, technically natural value of the Higgs vev and Higgs mass.%
\footnote{%
	Related ideas utilizing Hubble friction and back-reaction from field dynamics are employed in the warm inflation scenario~\cite{Berera:1995ie}.
	} %
While some of the required ingredients may appear rather exotic from an effective field theory perspective, the scenario offers a fresh perspective on the hierarchy problem, in the spirit of the self-organized criticality proposal of \citeR{Giudice:2008bi}, and is worthy of further exploration (see \citeR[s]{Espinosa:2015eda,Hardy:2015laa,Antipin:2015jia,Patil:2015oxa,Jaeckel:2015txa,Gupta:2015uea,Batell:2015fma,Matsedonskyi:2015xta,DiChiara:2015euo,Ibanez:2015fcv,Fonseca:2016eoo,Evans:2016htp,Kobayashi:2016bue,Farakos:2016hly,Hook:2016mqo,Higaki:2016cqb,Choi:2016luu,McAllister:2016vzi,Choi:2016kke,Flacke:2016szy,Lalak:2016mbv,You:2017kah,Evans:2017bjs,Agugliaro:2016clv,Beauchesne:2017ukw} for some recent studies). 
See also \citeR[s]{Abbott:1984qf,Dvali:2003br,Dvali:2004tma,Arkani-Hamed:2016rle,Arvanitaki:2016xds} for other cosmological approaches to naturalness.

The relaxion models presented in \citeR{Graham:2015cka} are based on the QCD axion, or an extended strong dynamics with a non-QCD axion.
These simple models are able to extend the cutoff of the SM to scales that are parametrically larger than the weak scale, but still well below the GUT or Planck scales.
In other words, these simple models are not able to fully address the `big' hierarchy problem, but instead can offer a solution to the `little' hierarchy problem---i.e., a way to understand the absence of new particles at the LHC, as well as deviations from the SM predictions in precision flavor, electroweak, and Higgs measurements.
While it is possible that more complex models (perhaps with additional scanning fields~\cite{Espinosa:2015eda}) can extend the cutoff further into the ultraviolet (UV) and perhaps even all the way to the Planck scale, one can also imagine that the new physics that emerges at the cutoff is of a more conventional type, such as supersymmetry or compositeness, which shields the Higgs against arbitrary short-distance physics.
Supersymmetric completions of the relaxion were investigated in~\citeR[s]{Batell:2015fma,Evans:2016htp}.

In this paper, we consider the relaxion scenario in the context of composite Higgs (CH) models~\cite{Kaplan:1983fs,Kaplan:1983sm,Dugan:1984hq,ArkaniHamed:2002qy,Agashe:2004rs,Giudice:2007fh}%
\footnote{%
	See also \citeR[s]{Terazawa:1976xx,Terazawa:1979pj} for some early work proposing the Higgs as a bound state of constituent fermions.
	} %
(see \citeR{Panico:2015jxa} for a recent review).
In such models, the big hierarchy problem is ameliorated by the assumption that the Higgs is a composite object of heavy `technifermions' bound together by the agency of strong `technicolor' (TC) gauge dynamics~\cite{Weinberg:1975gm,Weinberg:1979bn,Susskind:1978ms}.
Above the confinement scale, the theory is one of free fermion constituents whose masses are technically natural.
As in QCD, dimensional transmutation accounts for the hierarchy between the ultimate cutoff scale (e.g., the GUT or Planck scale) and the confinement scale of the composite theory.
Below the confinement scale, the theory is that of the pseudo-Nambu--Goldstone bosons (pNGBs) \cite{Nambu:1960ui,Nambu:1960xd,Goldstone1961,Nambu:1961tp,Nambu:1961fr,Goldstone:1962ma}---four of which comprise the Higgs doublet---of a spontaneous global symmetry breaking triggered by the strong dynamics when it confines. 
Due to explicit global symmetry breaking, the pNGB Higgs develops a potential, and vacuum misalignment arguments dictate that the Higgs vev in such models is expected to be of the same order as the compositeness scale, whereas phenomenological viability of CH models demands that the Higgs vev, $\langle h \rangle = 246$ GeV, should lie somewhat below the compositeness scale, $F_\pi$.
This is summarized by the well-known requirement
\begin{align}
\xi &\equiv \langle h \rangle^2/F_\pi^2 \ll 1,
\label{eq:xi}
\end{align}
which encapsulates the little hierarchy problem in CH models.

Our aim in this paper is to demonstrate, within the context of an explicit CH model, that a large hierarchy between the weak scale and the global symmetry breaking scale, \eqref{xi}, can be achieved in a technically natural fashion by invoking the cosmological relaxation mechanism.
The essential feature is that as the relaxion evolves in the early universe, it scans the techniquark masses, which provide a source of explicit global symmetry breaking.
Since the Higgs potential is controlled by such explicit symmetry breaking, this manifests in the low energy effective theory as a scanning of the Higgs potential, allowing the relaxation mechanism to be implemented in a manner similar to \citeR{Graham:2015cka}.

UV completions of CH models based on strong technicolor dynamics generally give rise to the cosets $SU(N_F)/SO(N_F)$, $SU(N_F)/Sp(N_F)$, and $[SU(N_F)\times SU(N_F)]/SU(N_F)$, when $N_F$ technifermions are in a real, pseudoreal, or complex representation, respectively, of the technicolor gauge group~\cite{Peskin:1980gc,Preskill:1980mz}.
While the relaxation mechanism can be implemented with any of these cosets, we will construct and investigate a concrete model with QCD-like dynamics, based on an $\suTC$ gauge group with $N_F = 3$ Dirac flavors (an `$L + N$' model).
This leads to the global symmetry breaking pattern $SU(3)\times SU(3) \times U(1) \rightarrow SU(3) \times U(1) \supset SU(2)_{\textsc{W}} \times U(1)_{\textsc{Y}}$.
Indeed, this theory can in many ways be viewed as a scaled up copy of QCD.
Interestingly, this is the smallest in the class of QCD-like cosets, $[SU(N_F)\times SU(N_F)]/SU(N_F)$, which furnishes a Higgs doublet.
However, this coset is not usually considered for CH models since it does not contain the custodial symmetry group, $SU(2) \times SU(2)$~\cite{Sikivie:1980hm}, which protects against large tree-level corrections to the electroweak precision $T$ parameter~\cite{Peskin:1990zt,Peskin:1991sw,Altarelli:1990zd}. 
In our scenario, however, this coset can indeed be viable since the relaxation mechanism will naturally generate the large hierarchy in \eqref{xi}, allowing the $T$ parameter to be adequately suppressed and compatible with precision electroweak measurements.

In our construction, the relaxion is taken to have axion-like couplings to both the  technicolor $\suTC$ gauge group 
and  the QCD $\suC$ gauge group.
An appropriate chiral rotation of the technifermion fields leads to a coupling of the relaxion to the techniquark masses. 
As the techniquark mass terms explicitly break the global symmetry and contribute to the composite Higgs potential, this coupling provides the basis for the scanning mechanism. 
We construct the low-energy Chiral Lagrangian, taking into account the large radiative corrections to the Higgs potential due to the top quark, and show that the potential contains the requisite relaxion--Higgs couplings to effect electroweak symmetry breaking and halt the relaxion evolution once it 
dynamically rolls through some critical value.
The strong-CP problem is addressed as in \citeR{Graham:2015cka} with a slope-drop mechanism: 
we assume the scanning potential for the relaxion arises from a coupling to the inflaton, such that it dominates the rolling during inflation but disappears post-inflation, allowing the effective QCD $\theta$-angle to relax to small values.

By design, the relaxation mechanism pushes the dynamics stabilizing the weak scale to higher scales, making experimental confirmation of the scenario more challenging. 
While there is no guarantee that the framework can be fully tested with near-term experiments, there are certainly some experimental opportunities worth pursuing. 
In the specific model studied here, the spectrum of the pNGB sector in our model consists of a light composite Higgs state, and four additional ultra-heavy composite technimesons, which are either neutral or only charged under the electroweak (EW) gauge group.
It will be challenging to directly probe such a heavy spectrum, even at proposed future hadron colliders such as the SPPC \cite{preCDR} or FCC-hh \cite{Benedikt:2015kqj,Benedikt:2016qcy,Benedikt:2238248}.
More promisingly, future improvements in the measurements of electroweak precision observables (EWPO) such as the $T$ parameter at the ILC \cite{Behnke:2013xla,Baer:2013cma,Adolphsen:2013jya,Adolphsen:2013kya,Behnke:2013lya}, CEPC \cite{preCDR}, or FCC-ee \cite{Gomez-Ceballos:2013zzn} have the potential~\cite{Fan:2014vta} to probe this model over most of the natural parameter space.  
More generally, there is potentially a diverse set of experimental probes for this and other composite Higgs theories within the cosmological relaxion framework,
including tests of flavor and CP violation, electroweak precision measurements, dark matter and axion searches, and collider searches for new states. We will highlight some of these opportunities.

Our work is not the first to consider the cosmological relaxation mechanism of \citeR{Graham:2015cka} in the context of composite Higgs models; however, important details of our model differ significantly from previous work \cite{Antipin:2015jia,Agugliaro:2016clv}. 
In particular, we consider in our work only a single composite Higgs doublet, whereas \citeR[s]{Antipin:2015jia,Agugliaro:2016clv} both analyze Type-I Two Higgs Doublet Models with one elementary and one composite Higgs doublet.
Another major difference is that we require the additional axion-like coupling of the relaxion to QCD to stall its rolling, following \citeR{Graham:2015cka}, whereas in \citeR[s]{Antipin:2015jia,Agugliaro:2016clv}, the relaxion rolling is stalled by virtue of the Higgs-vev-dependent barriers for $\phi$ that appear as a result of the non-QCD strong gauge dynamics.

The rest of this paper is structured as follows: 
we begin in \sectref{cartoon} with a discussion of a simplified `cartoon' picture of the mechanism we wish to explore in this paper, in order to orient the reader before we delve into the detailed construction and analysis of our explicit model; in this section we also review both the post-inflation slope-drop mechanism of \citeR{Graham:2015cka} as it applies to our model to solve the strong-CP problem, and the clockwork mechanism~\cite{Choi:2015fiu,Kaplan:2015fuy} that may potentially generate the requisite super-Planckian axion decay constants for our model.
In \sectref{constituent_model} we begin our explicit model construction, by presenting the constituent UV model that defines the underlying theory for the CH sector; we also specify the effective four-fermion interactions that give rise to the CH Yukawa couplings, specify the relaxion sector, and make some initial manipulations to the model to allow construction of the low-energy Chiral Lagrangian. 
In \sectref{chiral_lag} we explicitly construct the Chiral Lagrangian describing the composite states of the theory defined in \sectref{constituent_model}, and we extract those terms from the Chiral Lagrangian which are required to obtain the spectrum of the theory and understand its vacuum structure.
Section \ref{sect:effective_potential} contains our detailed analysis of the effective potential for the model, along with our analysis of the properties of the broken and symmetric electroweak phases of the theory.
The relaxion potential is discussed in \sectref{relaxion_potential}.
We present a summary of our analytical results and a numerical investigation of the model parameter space in \sectref{summary_numerical}.
A general discussion of some additional phenomenologically interesting considerations applicable to both our model, and more general composite Higgs models with large $F_\pi$, is given in Section \ref{sect:other_considerations}.
We conclude in \sectref{conclusion}.
Appendix \ref{app:SU3exponentiation} gives the closed-form expression for a general exponentiated $SU(3)$ matrix, which is of some utility in our analysis.
Appendix \ref{app:unequal_masses} contains a more general analysis of the EWSB dynamics of our full model, in which we relax one of the simplifying assumptions made in \sectref{effective_potential}.

\section{A Simplified `Cartoon' Model}
\label{sect:cartoon}
In order to facilitate a better understanding of the ideas we will explore in this paper, we present in this section a cartoon picture of the mechanism that we develop in greater detail in the sections to follow.

In order to successfully implement the relaxion mechanism of \citeR{Graham:2015cka} in a composite Higgs model, we need to engineer three essential components: (a) a CH--relaxion coupling, (b) a potential for the relaxion which is sufficiently flat and which causes the field to slow-roll in the correct direction in field space to trigger dynamical EWSB, and (c) a mechanism to create barriers in the relaxion potential that stall its slow-roll once EWSB is triggered.

We will achieve (a) and (c) by assuming that the relaxion $\phi$ is an axion of both the strongly coupled TC gauge group that confines to yield the composite Higgs state, and of QCD (see \sectref{clockwork}):
\begin{align}
\Lag \supset \frac{g_s^2}{16\pi^2} \lb[ \frac{\phi}{f} - \theta_\textsc{qcd} \rb] \Tr{G_{\mu\nu}\widetilde{G}^{\mu\nu}}+ \frac{g_{\textsc{tc}}^2}{16\pi^2} \frac{\phi}{F}  \Tr{G_{\textsc{tc}\, \mu\nu}\widetilde{G}^{\mu\nu}_{\textsc{tc}}},
\end{align}
where $f$ and $F$ are dimensionful parameters with $F \gg f$. 

Appropriate to the level of our cartoon picture in this section, we will discuss only an approximate low-energy pseudo-Nambu--Goldstone boson (pNGB) description of the composite sector of the theory, via the Chiral Lagrangian.
If $\Umat$ is the matrix-valued field of pNGBs which include among them the composite Higgs state $h$, then once the axion-type couplings are rotated into the technifermion mass matrices in the underlying constituent theory, and we include the large, dominant radiative effects of the top quark,%
\footnote{\label{ftnt:gauge_loops}%
	There are also subdominant effects from gauge loops.
	These will not change the qualitative picture we explore in this paper, and we ignore them.
	} %
the following terms will appear in the effective potential for the model: 
\begin{align}
V &\sim - c_m \LambdaTC F_\pi^2 \Tr{ M \Umat e^{i\phi/F} + \text{h.c.} } - \frac{c_ty_t^2 N_c \LambdaTC^2 F_\pi^2 }{16\pi^2} | \Tr{ \Umat \cdot \Delta } |^2 + V_\phi(\phi) + V_{\textsc{qcd}} \label{eq:cartoonV1} \\
&\sim - c_m \LambdaTC F_\pi^2 m \cos\lb( \frac{h}{F_\pi} \rb)\cos\lb( \frac{\phi}{F} \rb) - \frac{c_ty_t^2 N_c \LambdaTC^2 F_\pi^2 }{16\pi^2} \sin^2\lb( \frac{h}{F_\pi} \rb) + V_\phi(\phi) + V_{\textsc{qcd}},\label{eq:cartoonV2}
\end{align}
where $c_t$ and $c_m$ are perturbatively incalculable $\mathcal{O}(1)$ numbers, $F_\pi$ is the compositeness scale associated with spontaneous global flavor symmetry breaking, $\LambdaTC \approx (4\pi/\sqrt{N}) F_\pi$ is the cutoff scale of the composite theory, and we have taken $m$ to be a representative mass of the technifermions.
Furthermore, in \eqref{cartoonV2}, $V_\phi(\phi)$ is an additional relaxion potential which will be discussed in \sectref{cartoonRelaxionV}, and $\Delta$ is the appropriate projection operator that extracts the part of $\Umat$ to which the top quark couples (i.e., the Higgs doublet). 
We also emphasize that \eqref{cartoonV2} is highly schematic---much of the development in the following sections is precisely to deal with the more complicated structures that actually appear when evaluating \eqref{cartoonV1} in a realistic theory.
Nevertheless, this simplified picture captures the essential features of the model.
It also suffices for the present discussion to merely assume that $V_{\textsc{qcd}}$ is a cosine periodic potential:
\begin{align}
V_{\textsc{qcd}} \sim - \LambdaQCD^4 \cos\lb( \frac{\phi}{f} - \theta_\textsc{qcd} \rb), \label{eq:cartoonVQCD}
\end{align}
where $\LambdaQCD^4$ depends \emph{linearly} on the Higgs vev $\hvev$ through its dependence on the quark masses $m_q$: $\LambdaQCD^4 \sim m_\pi^2 f_\pi^2 \propto m_q \propto \hvev$.

\subsection{Electroweak Symmetry Breaking}
\label{sect:cartoonEWSB}
The dynamical picture to bear in mind is that while the relaxion is slow-rolling down its potential during an exponentially long period of low-scale inflation, the other fields respond by assuming their instantaneous equilibrium vacuum expectation values, such that the effective potential is minimized with $\phi$ held fixed.
Therefore, before we return to a discussion of the relaxion rolling, consider first the dynamics of the $h$ field per \eqref{cartoonV2}; we will ignore the dynamics of any other states in the theory---this topic will consume much of our attention in the concrete model we analyze in the following sections.

\begin{figure}[t]
\centering
\includegraphics[width=\textwidth]{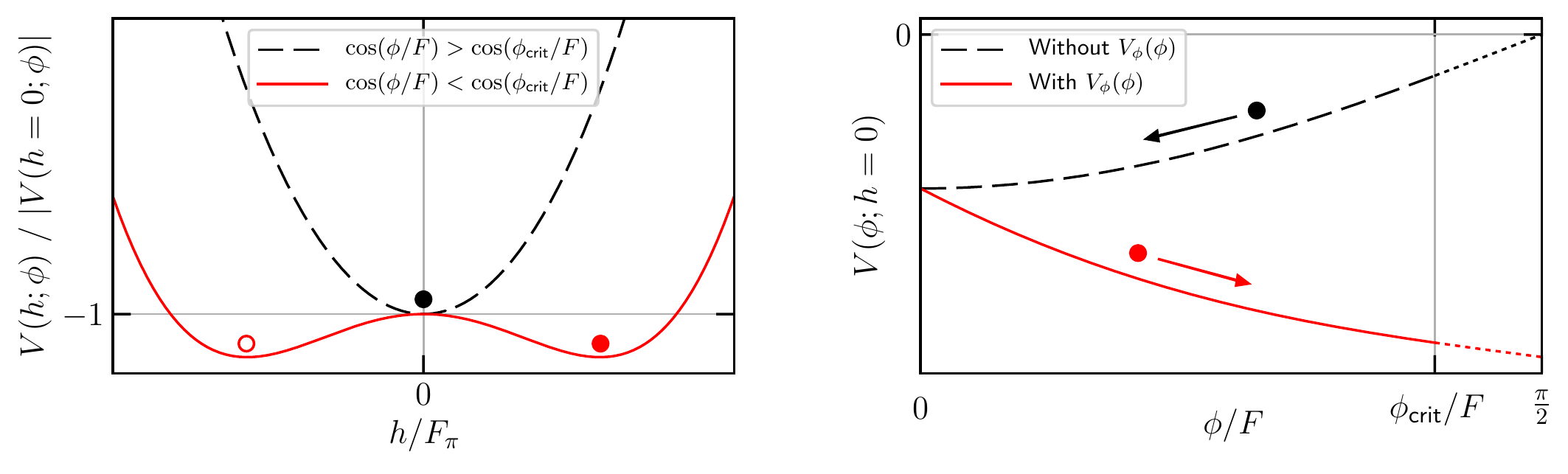}
\caption{\label{fig:cartoon_plots}%
	\textbf{Left panel:} a sketch plot of the potential for the Higgs at fixed $\phi$-field value, $V(h;\phi)$. 
	For $\phi < \phicrit$, the potential $V(h;\phi)$ has a stable minimum at the origin $h=0$, yielding an EW-symmetric vacuum (dashed black line).
	For $\phi > \phicrit$, dynamical EWSB is triggered, and the Higgs field $h$ rolls to a minimum displaced from the origin (solid red line).
	The dots indicate the stable minima; the EWSB minima are degenerate, and we select $h>0$ by convention.\qquad
	\textbf{Right panel:} a sketch plot of the relaxion potential in the EW-symmetric phase,\protect\linebreak $V(\phi;h=0)$.
	The solid (red) line shows the potential with the $V_\phi(\phi)$ term (plotted here for $\gamma \sim 1$; see \sectref{cartoonRelaxionV}), and the dashed (black) line shows the potential without the $V_\phi(\phi)$ term [in both cases, the lines are dotted in the region $\phi \in [\phicrit,\pi/2]$, as the EW-symmetric $(h=0)$ potential is not relevant for the $\phi$-rolling after EWSB had been triggered].
	As indicated by the respective dots and arrows, without the $V_\phi(\phi)$ term, the relaxion field naturally wants to roll from its initial value back toward the origin at $\phi=0$; with the $V_\phi(\phi)$ term, the rolling direction is reversed: $\phi$ will roll toward $\phicrit$, triggering dynamical EWSB.
	}
\end{figure}

If we define $\cos(\phicrit/F)\equiv ( c_t y_t^2 N_c \LambdaTC ) / (8\pi^2 c_m m)>0$ [we assume $c_t$, $c_m>0$], then the minimization condition for the potential \eqref{cartoonV2} in the $h$-direction is
\begin{align}
\partial_h V &\propto \sin\lb( \frac{\hvev}{F_\pi} \rb) \lb[ \frac{\cos(\phi/F)}{\cos(\phicrit/F)} - \cos\lb( \frac{\hvev}{F_\pi} \rb) \rb] =0, \label{eq:dVHiggsCartoon}
\end{align}
where we have ignored correction terms $\sim \LambdaQCD^4 / (\LambdaTC F_\pi^2 \hvev) \ll 1$.
We see that $\hvev=0$ is always a solution to \eqref{dVHiggsCartoon}; whether or not there are other solutions in the region near $\hvev/F_\pi\approx 0$ depends on the relative sizes of $\cos(\phi/F)$ and $\cos(\phicrit/F)$.
For $\cos(\phi/F) > \cos(\phicrit/F)$, the $[\,\cdots]$-bracket in \eqref{dVHiggsCartoon} cannot be set to zero for any value of $h$, so no additional solution(s) exists; however, for $\cos(\phi/F) < \cos(\phicrit/F)$, two additional solutions to \eqref{dVHiggsCartoon} appear symmetrically around $h=0$.
A graphical sketch of the situation is shown in the left panel of \figref{cartoon_plots}, both for $\cos(\phi/F) > \cos(\phicrit/F)$ and vice versa.
Clearly, if $\phi$ rolls to larger values from some initial value satisfying $\phi<\phicrit$,%
\footnote{\label{ftnt:no_tuning}%
		We do not view this as a tuning.
		The initial value of $\phi$ must merely satisfy $\cos(\phi_{\text{init.}}/F) > \cos(\phicrit/F)$ to ensure a stable EW-symmetric vacuum to start with; this occurs for a large fraction of the available parameter space.
		See also \citeR{Graham:2015cka}, wherein an analogous mild assumption about the qualitative size of the initial value of the relaxion field is made.
	} %
as it crosses $\phicrit$ it triggers a dynamical destabilization of the $h=0$ solution leading to a dynamically generated spontaneous EWSB.
Per the mechanism developed in \citeR{Graham:2015cka}, once the $h$ field gets a non-zero vev $\hvev$, the QCD barriers grow in size and rapidly stall the slow-roll of the relaxion field $\phi$ in the vicinity of $\phicrit$, locking in a small, technically natural $\hvev$.

Note that for the discussion in the previous paragraph to work, we must demand that $0<\cos(\phicrit/F)\leq1$, which implies a lower bound on the masses of the fermions charged under the strong dynamics (assuming that $c_t,c_m>0$):
\begin{align}
m \geq N_c  \frac{c_t}{c_m} \frac{y_t^2}{8\pi^2} \LambdaTC. \label{eq:CartoonMrestr}
\end{align}

\subsection{The Higgs Mass}
\label{sect:cartoonHiggsMass}
Some mild residual tuning is required to obtain the correct Higgs mass.
From \eqref[s]{cartoonV2} and (\ref{eq:dVHiggsCartoon}), it is straightforward to derive that%
\footnote{%
	We remind the reader that $\mkern-3mu\sinc(x)  \equiv \sin(x)/x \simeq 1 - x^2/6 + \mathcal{O}(x^4)$ for small $x$. 
	} %
\begin{align}
m_h^2 &= 4 c_t \lb(\frac{N_c}{N}\rb) \lb[ \frac{y_t}{\sqrt{2}} \hvev \sinc \lb( \frac{\hvev}{F_\pi} \rb) \rb]^2 = 4 c_t \lb(\frac{N_c}{N} \rb) m_t^2 .
\end{align}
Therefore, assuming that the relaxion mechanism has already selected the correct value of the Higgs vev $\hvev$ (and by implication the correct top mass $m_t$), the residual tuning can be estimated by comparing the expected $c_t \sim \mathcal{O}(1)$ with the required
\begin{align}
c_t \sim\lb( \frac{N}{N_c}\rb) \lb( \frac{1}{2} \frac{m_h}{m_t} \rb)^2 \sim 0.1\lb(  \frac{N}{3} \rb).
\end{align}
The residual tuning for the Higgs mass is thus at worst on the order of an additional one-in-ten tuning, and may even be much milder if $N\sim 10$. 
\subsection{The Relaxion Potential}
\label{sect:cartoonRelaxionV}
In order to achieve condition (b) and obtain a sufficiently flat relaxion potential \emph{which also drives the relaxion field toward the critical value}, we will need to add in an extra potential term for the relaxion, $V_\phi(\phi)$. 
This is illustrated in the right panel of \figref{cartoon_plots}: without an additional term $V_\phi(\phi)$ in the potential, during the EW-symmetric phase the relaxion would roll in the incorrect direction in field space to give rise to the dynamical EWSB we have just discussed.
Following \citeR{Graham:2015cka}, we thus add a linear potential for the relaxion, to obtain the correct rolling.
For convenience, we will choose to parametrize this term as 
\begin{align}
V_\phi(\phi) = - \gamma \frac{ c_m \LambdaTC F_\pi^2 m }{ F }\, \phi,
\label{eq:VphiParametrization}
\end{align}
where $\gamma$ is an entirely free parameter of as-yet-unknown size controlling the slope of the linear potential.%
\footnote{\label{ftnt:gamma_parametrization}%
		We emphasize that this parametrization is only for convenience in writing expressions like \eqref{cartoonVphi}, and should be considered with care: $\gamma$ may depend on other parameters in the theory in such a way that any na\"ive conclusions drawn from \eqref{VphiParametrization} about the behaviour of $V_{\phi}$ in various parameter limits may be wrong; in particular, we do not intend to imply that $V_\phi(\phi)$ vanishes in the $m\rightarrow0$ or $F \rightarrow \infty$ limits. 
	} %
In the EW-symmetric phase, we would then have
\begin{align}
V(\phi;h=0) &\sim - c_m \LambdaTC F_\pi^2 m \lb[ \cos\lb( \frac{\phi}{F} \rb) + \gamma \frac{\phi}{ F} \rb] \label{eq:cartoonVphi};
\end{align}
assuming slow-roll in the EW-symmetric phase this yields
\begin{align}
\partial_t \phi \propto - \partial_\phi V(\phi;h=0) &=  c_m\frac{\LambdaTC F_\pi^2 m}{F} \lb[  \gamma - \sin\lb( \frac{\phi}{F} \rb) \rb], \label{eq:dVphiCartoon}
\end{align}
which provides a lower bound $\gamma \gtrsim 1$ such that $\partial_t {\phi}>0$ for $\phi\in[0,\phicrit]$. 
With an appropriate $V_\phi$ added in, the rolling direction is now correct, as is illustrated again in the right panel of \figref{cartoon_plots} (which is shown schematically for $\gamma \sim 1$). 

In order for the QCD barriers to be effective in stopping the rolling of the relaxion field at $\phi^* = \phicrit + \delta \phi$ where $0<\delta\phi \ll \phicrit$, the slope of the QCD barriers and the `relaxion rolling slope' must match approximately at the stopping point during inflation.
If we na\"ively%
\footnote{%
	Since the dynamical origin of the additional slope will in general be different from the strong TC dynamics, $\gamma \sim 1$ would appear to require some accidental coincidence.
	} %
were to consider that $\gamma\sim1$ then, up to $\mathcal{O}(1)$ numbers,
\begin{align}
c_m\frac{\LambdaTC F_\pi^2 m}{F} &\sim \frac{\LambdaQCD^4}{f} &\Rightarrow&& F &\sim c_m  \frac{\LambdaTC F_\pi^2 m}{\LambdaQCD^4} \mkern1mu f \qquad\qquad (\gamma \sim 1) .\label{eq:slopeMatchingCartoongamma1}
\end{align}
Taking $\LambdaTC\sim80$\,TeV, $F_\pi \sim \LambdaTC \sqrt{N} / 4\pi \sim20$\,TeV for $N\sim10$, $m \sim 3$\,TeV, $\LambdaQCD \sim \sqrt{ m_\pi f_\pi } \approx 110$\,MeV,%
\footnote{%
	We used the QCD neutral pion mass $m_\pi \approx 135$\,MeV \cite{Olive:2016xmw}, and the QCD pion decay constant $f_\pi \approx 93$\,MeV.
	} %
and $c_m =1$, this implies that $F \sim (7 \times 10^{20})\, f$.
This means that with the usual QCD Peccei--Quinn \cite{Peccei:1977ur,Peccei:1977hh,Kim:1979if,Zhitnitsky:1980tq,Shifman:1979if,Dine:1981rt} symmetry breaking scale $f\sim 10^{11}$\,GeV, this model will require $F \sim 7 \times 10^{31}$\,GeV if we want the compositeness scale on the order of $20$\,TeV; we will comment on the viability of such a large dimensionful scale ($F \gg \MPlred$) in \sectref{clockwork}.

\begin{figure}[t]
\centering
\includegraphics[width=\textwidth]{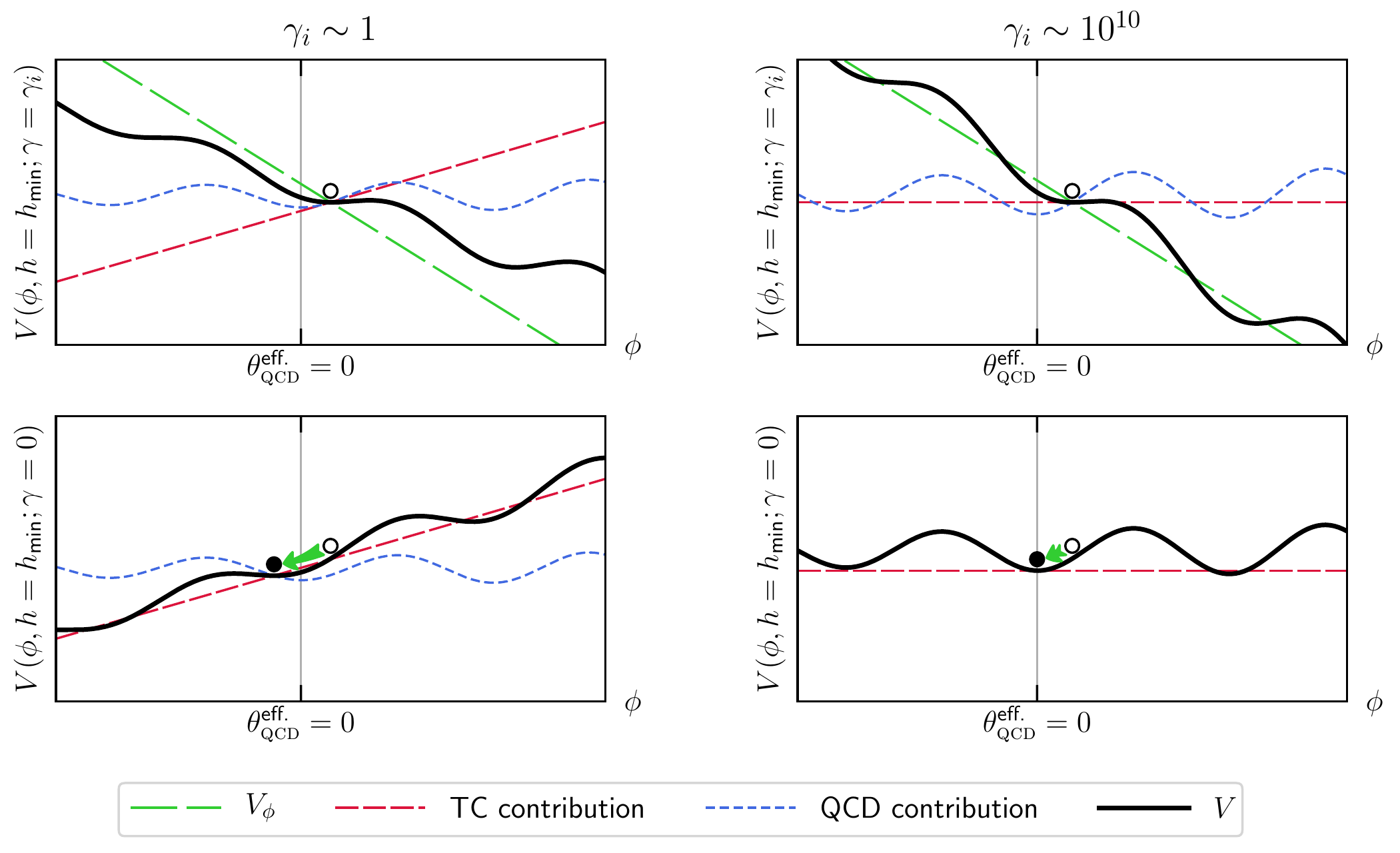}
\caption{\label{fig:cartoon_plots_slope_drop}%
	Sketch plot of the effects of post-inflation slope-drop for the cases of $\gamma_i \sim 1$ (left column) and $\gamma_i \sim 10^{10}$ (right column); the top row shows the potentials during inflation ($\gamma = \gamma_i$), while the bottom row shows the situation after slope drop ($\gamma \rightarrow 0$).
	In all panels, the solid, thick (black) curve shows the total $\phi$ potential with $h$ pinned to its minimum, with the open (black) circle showing where the relaxion stalls rolling before slope-drop (i.e., with $\gamma = \gamma_i$), and the solid (black) circle (bottom row only) showing where the relaxion would settle after slope-drop ($\gamma = 0$).
	The (green) arrows (bottom row only) indicate how the relaxion rolls from its initial stalling position to the position it assumes once the scanning slope drops.
	The long-dashed (green) line (top row only) shows the contribution to the total potential from the `scanning potential' before slope drop (i.e., with $\gamma = \gamma_i$), the medium-dashed (red) line shows the contribution from the strong dynamics, and the short-dashed (blue) line shows the QCD contribution.
	In order to display all the individual potential contributions on the same axes, the zero-point of each individual contribution to the potential has been independently shifted so that all the contributions cross at the point where the relaxion stalls [i.e., the three dashed (colored) lines sum up to the solid (black) line plus a constant offset], such that the important slope information from each contribution at that point is clearly visible.
	The vertical line marked $\theta_{\textsc{qcd}}^{\text{eff.}} = 0$ is the point where the effective QCD $\theta$-angle would vanish in the dip in the periodic QCD potential just prior to the initial relaxion stalling point.
	}
\end{figure}

However, we know from \citeR{Graham:2015cka} that using the QCD barriers to stop the relaxion rolling results in a severe strong-CP problem if a significant non-QCD relaxion slope persists to the present day, because the stopping point for the relaxion is displaced from the minimum of the QCD potential.
In order to alleviate this constraint, we will utilize the mechanism of post-inflation slope-drop proposed in \citeR{Graham:2015cka}: the basic idea of this mechanism is that the slope of the relaxion scanning potential $V_\phi(\phi)$ should originate via a coupling to a field $\sigma$ during inflation, $\gamma = \gamma(\sigma)$, such that $\gamma = \gamma_i$ during inflation, but when the $\sigma$ field rolls to end inflation, it causes the slope of the scanning potential to disappear: $\gamma \rightarrow 0$.
Thus, if we were to na\"ively assume that $\gamma_i\sim1$  during inflation, and that slope-drop mechanism sends $\gamma \rightarrow 0$ at the end of inflation as $\sigma$ rolls, the strong-CP problem would not be alleviated (see \figref{cartoon_plots_slope_drop}).

Suppose then that instead of considering the parameter regime $\gamma_i \sim 1$, we consider $\gamma_i \gg 1$, such that the relaxion rolling slope is entirely dominated by the linear contribution from the coupling to the $\sigma$-field.
Then the estimate \eqref{slopeMatchingCartoongamma1} is modified by an additional factor of $\gamma_i$: 
\begin{align}
\gamma_{i}\mkern1mu c_m\frac{\LambdaTC F_\pi^2 m}{F} &\sim \frac{\LambdaQCD^4}{f} &\Rightarrow&& F &\sim \gamma_{i}\mkern1mu c_m  \frac{\LambdaTC F_\pi^2 m}{\LambdaQCD^4}\mkern1mu f \qquad \qquad (\gamma_i \gg 1).\label{eq:slopeMatchingCartoon}
\end{align}
Now, taking the same parameter estimates as just below \eqref{slopeMatchingCartoongamma1}, this implies that $F \sim \gamma_i \times (7\times 10^{31}) $\,GeV.
For large $\gamma_i$, $F$ is proportionally larger than the estimate at \eqref{slopeMatchingCartoongamma1}, so that the overall slope of the additional potential term, $\partial_\phi V_\phi = \gamma_i ( c_m \LambdaTC F_\pi^2 m / F ) \propto \gamma_i/F$, remains of the correct size to cancel against the QCD barriers and stop the rolling, while the contribution to the relaxion potential from the strong TC dynamics is highly suppressed.
It is this suppression of the contribution from the strong dynamics in this region of parameter space that allows the slope-drop mechanism to work (see \figref{cartoon_plots_slope_drop}).

The estimate for the post-inflation settling point for the relaxion is obtained from \eqref[s]{cartoonV2}, (\ref{eq:cartoonVQCD}) and (\ref{eq:cartoonVphi}) with $\gamma \rightarrow0$:
\begin{align}
\partial_\phi V &= 0 &\Rightarrow && \lb| \frac{c_m \LambdaTC F_\pi^2 m}{F} \cos\lb( \frac{\hvev}{F_\pi} \rb) \sin \lb(\frac{\phi}{F}\rb) \rb| &\sim \lb| \frac{\Lambda^4}{f} \sin\lb( \frac{\phi}{f} - \theta_{\textsc{qcd}} \rb) \rb|.
\end{align}
Post slope-drop, the relaxion will only roll a distance $|\Delta \phi| \sim f$ from its initial stopping point (see \figref{cartoon_plots_slope_drop}), so the changes in $\sin\lb( \phi / F\rb)$ and in $\hvev$ will be negligible. 
Taking $\sin\lb( \phi / F\rb) \sim \mathcal{O}(1)$, using that the scanning mechanism selects $\cos( \hvev / F_\pi ) \sim 1$ (i.e., $\xi \ll 1$) for the purposes of this estimate,
and noting that $\theta_{\textsc{qcd}}^{\text{eff.}} = \lb( \theta_{\textsc{qcd}} -  \phi/f \rb)\!\! \mod 2\pi$, we estimate that
\begin{align}
\frac{c_m \LambdaTC F_\pi^2 m}{F} \sim \frac{\Lambda^4}{f} \lb|\mkern2mu \sin\lb( \theta_{\textsc{qcd}}^{\text{eff.}} \rb) \rb|;
\end{align}
but using \eqref{slopeMatchingCartoon}, we then have
\begin{align}
\lb|\mkern2mu  \sin\lb( \theta_{\textsc{qcd}}^{\text{eff.}} \rb) \rb| &\sim \frac{1}{\gamma_i} & \Rightarrow && \lb|\mkern2mu  \theta_{\textsc{qcd}}^{\text{eff.}} \rb| &\sim \frac{1}{\gamma_i} . \label{eq:thetaQCDestCartoon}
\end{align}
Therefore, we must have $\gamma_i \sim 10^{10}$ to obtain an appropriately small QCD $\theta$-angle, $\lb|\mkern1mu  \theta_{\textsc{qcd}}^{\text{eff.}} \rb|  \sim 10^{-10}$.
We emphasize that the large dimensionless parameter $\gamma_i$ is merely an artifact of our parametrization of $V_\phi(\phi)$; the physical content of the statement that $\gamma_i \gg 1$ is that the slope of the relaxion potential contributed by the strong dynamics should be highly suppressed compared to the slope of $V_\phi(\phi)$; i.e., $F$ must be taken larger than it would be were $\gamma_i \sim 1$ (see \sectref{clockwork}).

\subsection{Self-Consistency}
\label{sect:cartoonSelfConsistency}
In addition to the `slope matching' condition \eqref{slopeMatchingCartoon}, \citeR{Graham:2015cka} presented a number of restrictions on the relaxion mechanism which must be satisfied to achieve self-consistency.
As applied to this cartoon model, these restrictions are as follows:\\[2ex]
\textbf{Vacuum energy domination.} The total change in the relaxion energy density as $\phi$ rolls must be a sub-leading correction to the energy density driving inflation.
The relaxion $\phi$ must roll from an initial position in field space near $\phi=0$ to a value near $\phicrit$ in order to trigger EWSB.
While the exact value for $\phicrit$ depends on parameter choices, we can generically assume that, because $\phi$ enters the Higgs potential as $\cos(\phi/F)$, $\phicrit\mkern-2mu$ will not be orders of magnitude different from $F$.
Therefore,
\begin{align}
V_I &= 3 H_I^2 \MPlred^2 \gg \Delta V(\phi,h) \sim  \gamma_{i}\mkern1mu c_m \LambdaTC F_\pi^2 m& \Rightarrow && H_I &\gtrsim \lb(   \gamma_{i} \frac{c_m\LambdaTC F_\pi^2 m}{3\MPlred^2} \rb)^{\frac{1}{2}}, \label{eq:domination}
\end{align}
where $\MPlred = 1/ \sqrt{8\pi G_{\textsc{n}}} \approx 2.4\eten{18}\,$GeV is the reduced Planck mass, $V_I$ is the energy density driving inflation, and $H_I$ is the Hubble constant during inflation.\\
\textbf{Barrier formation.} The Hubble scale must be low enough that the QCD barriers form:
\begin{align}
H_I \lesssim \LambdaQCD. \label{eq:barriers}
\end{align}
\textbf{Classical beats quantum.} It is necessary to impose a constraint such that classical rolling of the relaxion dominates over quantum fluctuations, so that following inflation each patch of the universe obtains a vev of order the weak scale. In a Hubble time,
\begin{align}
\Delta \phi_{\text{cl.}} &\sim \dot\phi \Delta t_H \approx \frac{1}{3} \frac{ | \partial_\phi V|}{H_I^2} \sim \gamma_i \frac{c_m \LambdaTC F_\pi^2 m}{3F H_I^2}, &
\text{while}&&
\Delta \phi_{\text{quantum}} &\sim H_I
\end{align}
so
\begin{align}
\Delta \phi_{\text{cl.}} &\gtrsim \Delta \phi_{\text{quantum}} & \Rightarrow && H_I \lesssim \lb( \gamma_i \frac{c_m \LambdaTC F_\pi^2 m}{3F} \rb)^{\frac{1}{3}}. \label{eq:clas_vs_quant}
\end{align}
\textbf{Sufficiently many $\bm{e}$-folds.} For $\phi$ to roll a distance on the order of $F$ in field space%
\setlength{\footnotemargin}{3.5mm}%
\footnote{\label{ftnt:no_tuning2}%
		To avoid the assumption of a situation in which the initial value of $\phi$ is tuned to be near $\phicrit$, we consider this conservative condition; were $\phi_{\text{init.}}$ accidentally closer to $\phicrit$, the relaxion would not need to roll so far.
		See also footnote \ref{ftnt:no_tuning}.
		} %
given $N_e$ $e$-folds of inflation requires that
\begin{align}
F \sim \Delta \phi &\approx \dot\phi\Delta t \approx \dot\phi \frac{N_e}{H_I} \approx \frac{1}{3} \frac{ |\partial_\phi V|}{H_I^2} N_e && \Rightarrow  &
N_e &\sim \frac{3H_I^2F^2}{ \gamma_i c_m \LambdaTC F_\pi^2 m} . \label{eq:Efolds}
\end{align}
\\[1ex]
Combining \eqref[s]{slopeMatchingCartoon}, (\ref{eq:domination}), and (\ref{eq:clas_vs_quant}) gives an upper limit on $F$:%
\footnote{%
	Using \eqref{barriers} in place of \eqref{clas_vs_quant} gives a much weaker upper limit for all reasonable values of $f$.
	} %
\begin{align}
F &\leq \lb( \frac{3 \MPlred^6 f}{\LambdaQCD^4} \rb)^{\frac{1}{3}} = \lb( 8 \eten{41}\,\text{GeV} \rb) \lb( \frac{f}{10^{11}\,\text{GeV}} \rb)^{\frac{1}{3}} \label{eq:F_strong_limit}.
\end{align}
Using \eqref{slopeMatchingCartoon}, $\LambdaTC \approx (4\pi/\sqrt{N})F_\pi \approx 4 F_\pi \times\lb(10/N\rb)^{1/2} $, and writing%
\footnote{%
	These estimates are consistent with the parameter choices appearing just below \eqref{slopeMatchingCartoon}.
	} %
\begin{align}
m \approx N_c\frac{c_t}{c_m} \frac{ y_t^2 }{8\pi^2} \LambdaTC \approx \frac{c_t}{25c_m} \LambdaTC
\end{align}
---which saturates \eqref{CartoonMrestr}---transforms \eqref{F_strong_limit} into an upper limit on $\LambdaTC$, the UV cutoff of the composite theory:
\begin{align}
\LambdaTC &\lesssim \lb( \frac{128\pi^4 \sqrt[3]{3}}{c_t y_t^2NN_c} \rb)^{\frac{1}{4}} \gamma_{i}^{-\frac{1}{4}} \lb( \frac{\LambdaQCD^4 \MPlred^3}{f} \rb)^{\frac{1}{6}} \label{eq:LambdaTC_limit}
\end{align}
\begin{align}
&= \lb( 3\eten{7}\,\text{GeV} \rb) \lb( \frac{N}{10} \rb)^{-\frac{1}{4}} \lb( \frac{1}{\gamma_{i}} \rb)^{\frac{1}{4}} \lb( \frac{f}{10^{11}\,\text{GeV}} \rb)^{-\frac{1}{6}} \label{eq:LambdaTC_limit2}\\
&= \lb( 8\eten{4}\,\text{GeV} \rb) \lb( \frac{N}{10} \rb)^{-\frac{1}{4}} \lb( \frac{10^{10}}{\gamma_{i}} \rb)^{\frac{1}{4}} \lb( \frac{f}{10^{11}\,\text{GeV}} \rb)^{-\frac{1}{6}}\label{eq:LambdaTC_limit3} \\
&= \lb( 8\eten{4}\,\text{GeV} \rb) \lb( \frac{N}{10} \rb)^{-\frac{1}{4}} \lb( \frac{  \theta_{\textsc{qcd}}^{\text{eff.}}  }{10^{-10}} \rb)^{\frac{1}{4}} \lb( \frac{f}{10^{11}\,\text{GeV}} \rb)^{-\frac{1}{6}}.\label{eq:LambdaTC_limit4}
\end{align}
We therefore find that the relaxation mechanism can indeed explain in a technically natural fashion a large hierarchy between the UV cutoff of the composite theory and the weak scale; that is, we have provided a realization of the picture advocated from the outset in which compositeness addresses the big hierarchy problem while the relaxion explains the little hierarchy.
We also remark here that the parameter point discussed earlier, $\LambdaTC \sim 80$\,TeV, does not obviously run afoul of any of the self-consistency conditions when the effective QCD $\theta$-angle is appropriately small. 

Additionally, \eqref[s]{slopeMatchingCartoon} and (\ref{eq:clas_vs_quant}) together provide the most stringent available upper bound on the scale of inflation:
\begin{align}
V_I^{\frac{1}{4}} \lesssim \lb(\frac{ \sqrt{3} \LambdaQCD^4}{f} \rb)^{\frac{1}{6}} \MPlred^{\frac{1}{2}} = \lb( 6 \eten{6}\,\text{GeV} \rb) \lb( \frac{f}{10^{11}\,\text{GeV}} \rb)^{-\frac{1}{6}},
\end{align}
while \eqref[s]{slopeMatchingCartoon} and (\ref{eq:domination}) together provide a lower bound on the scale of inflation:
\begin{align}
V_I^{\frac{1}{4}} \gtrsim\LambdaQCD \lb( \frac{F}{f} \rb)^{\frac{1}{4}} &= \lb( 6 \eten{6}\,\text{GeV} \rb) \lb( \frac{f}{10^{11}\,\text{GeV}} \rb)^{-\frac{1}{4}} \lb( \frac{F}{8\eten{41}\,\text{GeV}} \rb)^{\frac{1}{4}}. 
\end{align}
The number of $e$-folds is bounded by \eqref[s]{domination} and (\ref{eq:Efolds}):
\begin{align}
N_e \gtrsim \lb( \frac{F}{\MPlred} \rb)^2 &= \lb( 1 \eten{47} \rb) \lb( \frac{F}{8\eten{41}\,\text{GeV}} \rb)^{2}.
\end{align}
As in the original relaxion model \cite{Graham:2015cka}, these results indicate that an \emph{extremely} long period of low-scale inflation, and an ultra-trans-Planckian relaxion field excursion are required to make the model viable.

Beginning in \sectref{constituent_model}, we devote significant effort to the presentation and detailed analysis of a concrete model that realizes the mechanism which we have just described schematically.

\subsection{Clockwork Mechanism}  
\label{sect:clockwork}
Achieving the hierarchy $F\gg f$ required for the viability of our model requires further model building. 
One possibility is the `clockwork' mechanism of \citeR[s]{Choi:2015fiu,Kaplan:2015fuy}, which we will briefly review here as it applies to our model; 
see \citeR[s]{Kim:2004rp,Choi:2014rja,delaFuente:2014aca} for earlier work in the context of inflation, and \citeR[s]{Giudice:2016yja,Kehagias:2016kzt,Farina:2016tgd,Ahmed:2016viu,Hambye:2016qkf,Craig:2017cda} for further recent theoretical and phenomenological investigations of the clockwork.
The clockwork mechanism postulates the existence of $(M+1)$ complex scalar fields $\varphi_j$ ($j=0,1,\ldots\!,M$) interacting via the Lagrangian
\begin{align}
\mathcal{L} &\supset \sum_{j=0}^M \lb( |\partial_\mu \varphi_j|^2 + \mu^2  | \varphi_j |^2 - \lambda  |\varphi_j|^4 \rb) + \epsilon \sum_{j=0}^{M-1} \lb( \varphi_j^\dagger \varphi^3_{j+1} + \text{h.c.} \rb), \label{eq:clockwork_L}
\end{align}
where $\epsilon \ll \lambda$. 
If $\epsilon = 0$, this Lagrangian would exhibit a global $[U(1)]^{M+1}$ symmetry; all $(M+1)$ of these global symmetries are spontaneously broken such that $\varphi_j \equiv \frac{1}{\sqrt{2}} (f_0 + \rho_j ) \exp[ - i \pi_j / f_0 ]$ with $f_0\equiv \mu^2/\lambda$, giving rise to $(M+1)$ NGBs, $\pi_j$.
When $\epsilon \neq 0$, $M$ of the $U(1)$ symmetries are additionally explicitly broken, which masses-up $M$ of the NGBs. 
The residual unbroken global $U(1)$ has an interesting pattern of charges, with $\varphi_j$ having charge $q_j = 3^{-j}$; the corresponding massless NGB is taken to be the relaxion.
The radial modes have masses $m_{\rho_j} \approx \sqrt{2\lambda} f_0$.

In order to obtain couplings of the relaxion to both QCD and the strong TC dynamics, we assume a KSVZ-type \cite{Shifman:1979if,Kim:1979if} axion model, with vector-like fermions charged under QCD coupled to the field $\varphi_K$ (where $K$ is an integer obeying $0\leq K \ll M$), and fermions charged under the strong TC dynamics coupled to the field $\varphi_M$.
These fermions will obtain masses $m_{K,M} = y_{K,M} f_0 / \sqrt{2}$, where $y_{K,M}$ are the Yukawa couplings of the $\varphi_{K,M}$ fields to the fermions.
Integrating out these fermions in the usual way generates the usual axion couplings of the $\pi_{K}$ and $\pi_{M}$ fields to, respectively, the QCD and TC field strength tensors.

Ref.~\cite{Kaplan:2015fuy} supplied the general procedure for the diagonalization of the tri-diagonal mass matrix for the $\pi_j$ that arises from the explicit breaking terms in \eqref{clockwork_L}; this procedure was elaborated on in detail in \citeR[s]{Giudice:2016yja,Farina:2016tgd}.
Letting the massless NGB be the relaxion $\phi$, and calling the other mass-eigenstate pNGB fields $a_n$ (these fields are sometimes called the `gears' of the clockwork mechanism), it is straightforward to show that the Lagrangian for the (p)NGBs after the vector-like fermions are integrated out can be expressed in terms of the mass-eigenstate fields as
\begin{align}
\mathcal{L} &\supset \frac{1}{2} (\partial_\mu \phi )^2 + \frac{ (g_s)^2}{16\pi^2} \lb[ \frac{\phi}{f} - \theta_{\textsc{qcd}}^0 \rb] \Tr{ G_{\mu\nu}\widetilde{G}^{\mu\nu} } + \frac{ (g_{\textsc{tc}})^2}{16\pi^2} \frac{\phi}{F} \Tr{ {G}_{\textsc{tc}\, \mu\nu}\widetilde{G}^{\mu\nu}_{\textsc{tc}} } \nonumber \\[1ex] &
 + \sum_{n=1}^M \lb[ \begin{array}{l} \displaystyle
 					 \frac{1}{2} (\partial_\mu a_n )^2 - \frac{1}{2}m_n^2 a_n^2 
 					+ \frac{ (g_s)^2}{16\pi^2}  \frac{a_n}{f_K^{(n)}}  \Tr{ G_{\mu\nu}\widetilde{G}^{\mu\nu} }
 					+ \frac{ (g_{\textsc{tc}})^2}{16\pi^2}  \frac{a_n}{f_M^{(n)}} \Tr{ {G}_{\textsc{tc}\, \mu\nu}\widetilde{G}^{\mu\nu}_{\textsc{tc}} } 
					\end{array} \rb], 
\end{align}
where $m_n$ are pNGB masses that obey $ 2\sqrt{\epsilon}  f_0 \leq m_n \leq 4\sqrt{\epsilon} f_0$; $f_j^{(n)}$ are decay constants which are generically roughly of the order $\sqrt{ (M+1)/2 } f_0$; and, crucially,
\begin{align}
f &\equiv  \frac{3}{2\sqrt{2}} \, 3^K f_0 \approx 3^K f_0 
&\text{and}&&
F &\equiv  \frac{3}{2\sqrt{2}} \,   3^M f_0  \approx 3^M f_0.
\end{align}
With moderate $K$ and $M$, this exponential enhancement of the decay constants makes it straightforward to engineer large hierarchies: $F\gg f \gg $ the weak scale.

Two scenarios suggest themselves: (a) $f_0 \sim f$ [i.e., $K=0$ and $M\gg1$], and (b) $f_0\sim$ a few (tens of) TeV $\ll f$ [i.e., $1\ll K\ll M$].
\paragraph{Scenario (a): $\bm{f_0 \sim f}$.}
To obtain $F \sim \gamma_i \times (6.6\times 10^{31}) \,\text{GeV}\sim 6.6\times 10^{41} $\,GeV for $\gamma_i \sim 10^{10}$ with $f\sim10^{11}$\,GeV, we need $M \sim 65$.
The only light field is the relaxion, while the radial modes and gears have masses of order $\sqrt{2\lambda} f$ and $3\sqrt{\epsilon} f$, respectively. 
The radial modes are thus heavy and, provided $\epsilon$ is not exponentially small, so too are the gears.
Additionally, the vector-like fermions charged under QCD and the strong TC dynamics---which were integrated out to give rise to the dimension-5 axion couplings---have masses on the order of $y f/\sqrt{2}$; assuming that $y \sim \mathcal{O}(1)$ (or at least not exponentially small), these too are unobservably heavy.
While the mechanism thus allows for the exponential scale separation $F\gg f \gg $ the weak scale, there are no additional experimental signatures which are accessible at any current or proposed collider, as all the additional new states are extremely massive.

\paragraph{Scenario (b): $\bm{f_0}\sim$ a few (tens of) TeV $\bm{\ll f}$.}
Taking $F \sim 6.6\times 10^{41} $\,GeV, $f\sim10^{11}$\,GeV, and $f_0 \sim 10$\,TeV, we find that we need $K \sim 15$ and $M \sim 79$. 
The radial modes and gears have masses of order $\sqrt{2\lambda} f_0$ and $3\sqrt{\epsilon} f_0$, respectively.
While the radial modes masses are thus around a few (tens of) TeV since $\lambda \sim 1$, the gears could easily have masses below a TeV for reasonably small $\epsilon$.
Moreover, the couplings of the gears and radial modes to the QCD and TC field strength tensors have `decay constants' of order $\sim 3f_0$ and $\sim 6f_0$, respectively.
Additionally, the colored and TC-charged vector-like fermions that were integrated out to give rise to the dimension-5 axion couplings would also have masses on the order of $yf_0/\sqrt{2}$.
Therefore, in this scenario we expect that additional experimental signatures would be accessible at current or future colliders: the gears and radial modes, and the colored fermions from the KSVZ mechanism, could all presumably be produced strongly if they are light enough.

\section{Constituent Model}
\label{sect:constituent_model}
We begin our concrete model construction by specifying in detail the underlying constituent UV model for the technicolor dynamics.
This underlying constituent model can be broken into three distinct components: (a) the CH sector, (b) the terms which give rise to the CH--SM Yukawa couplings, and (c) the relaxion sector.
We discuss each of these components in turn in \sectref[s]{CH-sector}--\ref{sect:relaxion}.

\subsection{Composite Higgs Sector}
\label{sect:CH-sector}
Our CH model is constructed to exhibit the global (TC-flavor) symmetry breaking pattern $\suLR \times U(1)_{\textrm{V}} \rightarrow SU(3)_{\textrm{V}} \times U(1)_{\textrm{V}} \rightarrow SU(2)\times U(1)\times U(1)$, where the first step arises from spontaneous chiral symmetry breaking, and the second arises due to the addition of explicit breaking terms (technifermion masses, Yukawas, and gauge couplings).
A remaining global $SU(2)\times U(1)$ is gauged and identified as the SM electroweak gauge group, $\suW \times \uY$. 

The gauge group which will confine to give rise to the composite Higgs state is assumed to be an $SU(N)$ group under which all SM field content transforms as singlets.%
\footnote{%
		See, e.g., \citeR[s]{Katz:2003sn,Galloway:2010bp,Barnard:2013zea,Ferretti:2013kya,Cacciapaglia:2014uja,Vecchi:2015fma,Ma:2015gra,Sannino:2016sfx,Galloway:2016fuo} for some recent studies of UV-complete composite Higgs models with strong TC dynamics.
	} %
The SM is augmented by three (Dirac) technifermions which transform in the fundamental of this gauged TC group, hereinafter referred to as $\suTC$.
In order to allow for the existence of a bound state of the technifermions with the correct SM quantum numbers to be interpreted as the Higgs, we demand that the three (Dirac) technifermion species consist of an $\suW$ doublet $\mathbb{L}$, and an $\suW$-singlet $\mathbb{N}$; additionally, we demand that $\mathbb{L}$ carry SM $\uY$ hypercharge of $+1/2$, while $\mathbb{N}$ is taken to be neutral under $\uY$.
For the remainder of this paper, we will write all fermion fields as left-handed%
\footnote{%
		That is, transforming under the $(\frac{1}{2},0)$ representation of the Lorentz group \cite{Weinberg:1995tws,Dreiner:2008tw}.%
	} %
two-component Weyl spinors;%
\footnote{%
		We generally follow the notational conventions of \citeR{Dreiner:2008tw}.
		Explicitly, in the Weyl basis, a Dirac fermion $\mathbb{F}$ can be expressed in terms of the two-component Weyl fermions $F$ and $F^c$ as 
		\begin{align*}
		\mathbb{F} = \begin{pmatrix} F_\alpha \\ \lb[ \lb( F^c \rb)^\dagger \rb]^{\dot{\alpha}} \end{pmatrix},
		\end{align*}
		where $\alpha$ is a $(\frac{1}{2},0)$ Lorentz spinor index, and $\dot{\alpha}$ is a $(0,\frac{1}{2})$ Lorentz spinor index.
	} %
our naming conventions for the two-component Weyl fermions, and the matter-field gauge charges, are given in \tabref{gauge_charges}.
Note that given those charges, $LN^c \sim ( \bm{1} , \bm{1} , \bm{2} )_{+1/2}$, where contraction of the TC fundamental and anti-fundamental indices on $L$ and $N^c$, respectively, is understood.
As these are the quantum numbers of the SM Higgs, the requisite composite Higgs state will be formed after confinement.

Introducing the $N$ additional hypercharged $\suW$-doublets $L$ and $L^c$ will modify the running of the couplings $g_1$ and $g_2$.
If $N$ is taken too large, there is a possibility that Landau poles in these couplings may occur below the Planck scale.
A simple check indicates that with $N\lesssim 14$, no such poles should appear in either coupling.

\begin{table}[t]
\centering
\begin{tabular}{l|l|l}
\textbf{Description}			& 		\textbf{Field Name}							&		$\bm{\lb(\suTC ,\,  \suC ,\,  \suW  \rb)_{\uY}}$					\\ \hline
SM quark doublet			&		$Q \equiv \begin{pmatrix} U \\ D \end{pmatrix}$		& 		$( \bm{1} , \bm{3} , \bm{2} )_{+1/6}$			\\
SM down-type quark singlet	&		$D^c$									& 		$( \bm{1} , \bm{\bar{3}} , \bm{1} )_{+1/3}$		\\
SM up-type quark singlet		&		$U^c$									& 		$( \bm{1} , \bm{\bar{3}} , \bm{1} )_{-2/3}$		\\ \hline
Technifermion `L' doublet		&		$L$										&		$( \bm{N} , \bm{1} , \bm{2} )_{+1/2}$			\\
Technifermion `R' doublet		&		$L^c$									&		$( \bm{\bar{N}} , \bm{1} , \bm{\bar{2}} )_{-1/2}$	\\
Technifermion `L' singlet		&		$N$										&		$( \bm{N} , \bm{1} , \bm{1} )_{0}$			\\
Technifermion `R' singlet		&		$N^c$									&		$( \bm{\bar{N}} , \bm{1} , \bm{1} )_{0}$
\end{tabular}
\caption{\label{tab:gauge_charges}%
	Gauge charges and representations for the fermion matter content.
	For the non-Abelian factors we give the representation, and for the hypercharge group we give the charge.
	All fields are assumed to be two-component left-handed [i.e., $(\frac{1}{2},0)$] Weyl fermions. 
	`Singlet' and `doublet' in the descriptions refer to the $\suW$ representations of the matter fields; `L' and `R' in the technifermion descriptions refer to which of the $SU(3)_{\textsc{L,R}}$ TC-flavor groups the fields transform under.
	SM generation indices are suppressed.
	A superscript ${}^c$ is considered to be an integral part of the field name and merely denotes that the field transforms in the anti-fundamental of the relevant gauge groups; i.e., $L$ and $L^c$ each represent two independent, unrelated degrees of freedom ($L^c$ is not a conjugate of $L$).
	We omit the SM lepton fields here as they are not relevant to our discussion.
	}
\end{table}

In order to impose the global $\suLR$ TC-flavor symmetry in our Lagrangian and make it manifest, we arrange $L$ and $N$ into a $(\bm{3},\bm{1})$ TC-flavor multiplet $\chi$, and we arrange $L^c$ and $N^c$ into a $(\bm{1},\bm{\bar{3}})$ TC-flavor multiplet $\chi^c$:
\begin{align}
\chi &\equiv \begin{pmatrix} L \\ N \end{pmatrix} & \text{and}&&
\chi^c &\equiv \begin{pmatrix} L^c \\ N^c \end{pmatrix}.
\end{align}
If $U_{\textsc{L}} \equiv \exp\lb[ i \alpha_{\textsc{L}}^a T^a \rb]$ and $U_{\textsc{R}} \equiv \exp\lb[ i \alpha_{\textsc{R}}^a T^a \rb]$ are, respectively, forward $\suL$ and $\suR$ transformations, we then have $\chi \rightarrow U_{\textsc{L}} \chi$ and $\chi^c \rightarrow \chi^c U_{\textsc{R}}^\dagger$.
Note also that the $\chi$ and $\chi^c$ transform in the fundamental and anti-fundamental, respectively, of the gauged $\suTC$ group, given the representations assigned in \tabref{gauge_charges}.

Thus far, the TC-gauge-coupling and kinetic terms for the $\chi$ and $\chi^c$ fields can be written in manifestly TC-flavor and $\suTC$ invariant fashion as
\begin{align}
\Lag \supset i \chi^\dagger \lb( \bar{\sigma} \cdot D \rb) \chi + i \chi^c \lb( \sigma \cdot D \rb) \lb( \chi^c \rb)^\dagger, \label{eq:lag_kinetic}
\end{align}
where  
\begin{align}
D_\mu \lb\{ \chi , (\chi^c)^\dagger \rb\} &\supset \lb( \partial_\mu - i g_{\textsc{TC}} A_\mu^{\textsc{TC}}  \rb) \lb\{ \chi , (\chi^c)^\dagger \rb\},
\end{align}
where we have suppressed all indices, and have written the matrix-valued TC gauge field $A_\mu^{\textsc{TC}} \equiv (A_\mu^{\textsc{TC}})^{n}\lb( T_{\textsc{TC}}\rb)^n$, where $n=1,\ldots\!,(N^2-1)$, with $T_{\textsc{TC}}$ the fundamental-representation generators for the $\suTC$ group.

Gauging the $\suW$ subgroup of the TC-flavor group is straightforward.
We assume that the matrix representatives of the generators $T^a$ of the $SU(3)_{\textrm{L,R}}$ transformations are given by $T^a = \frac{1}{2} \lambda^a$, where $\lambda^a$ are the usual Gell-Mann matrices \cite{Georgi:1999lap} (the Dynkin index is $\frac{1}{2}$).%
\footnote{\label{ftnt:L_R_index_supression}%
	We have suppressed TC-flavor indices here and have just written $T^a$ as the generators for either simple $SU(3)_{\textsc{L,R}}$ factor in the TC-flavor group. 
	At the risk of being pedantic, we should really write separate generators $T^a_{\textsc{L}}$ and $T^a_{\textsc{R}}$, and be consistent in the usage of each throughout.
	The matrix representatives of these generators are numerically equal as matrices, but as generators they are distinct objects as they carry different types of indices.
	} %
In this convention, $\lambda^{\tilde{a}} = \begin{pmatrix} \sigma^{\tilde{a}} & 0 \\ 0 & 0 \end{pmatrix}$ where (here, and throughout) $\tilde{a} = 1,2,3$, and $\sigma^{\tilde{a}}$ are the usual Pauli matrices.
Therefore, $\suW$ is gauged by simply adding to the covariant derivative $D_\mu$ the term $D_\mu \lb\{\chi,(\chi^c)^\dagger\rb\} \supset -i g_2 W_\mu^{\tilde{a}} T^{\tilde{a}} \lb\{\chi,(\chi^c)^\dagger\rb\} $, and demanding that under a forward $\suW$ gauge transform parametrized by $\alpha^{\tilde{a}}$ we have 
\begin{align}
\lb\{ \chi , (\chi^c)^\dagger  \rb\} &\rightarrow U_V \lb\{ \chi , (\chi^c)^\dagger  \rb\} &\text{and}&&
W_\mu &\rightarrow U_V W_\mu U_V^\dagger + \frac{i}{g_2} U_V \partial_\mu U_V^\dagger,
\end{align}
where
\begin{align}
U_V &\equiv \exp\lb[ i \alpha^{\tilde{a}} T^{\tilde{a}} \rb] &
\text{and}
&& W_\mu &\equiv W_\mu^{\tilde{a}} T^{\tilde{a}}.
\end{align}

Gauging $\uY$ is marginally more complicated. 
Recall that $[T^8,T^{\tilde{a}}]$ = 0 for the $SU(3)$ generators as defined above; this may lead one to conclude that $T^8$ is the $\uY$ generator because it commutes with the $\suW$ generators.
However, the matrix representative of $T^8$ is
\begin{align}
T^8 = \frac{1}{2} \lambda^8 &= \dfrac{1}{2\sqrt{3}} \begin{pmatrix} 1 & \phantom{-}0 & \phantom{-}0 \\ 0 & \phantom{-}1 & \phantom{-}0 \\ 0 & \phantom{-}0 & -2 \end{pmatrix};
\end{align}
since the 3-3 component of this representative is non-vanishing, $T^8$ will act non-trivially on the $N$ component of $\chi$.
But this cannot then be the SM $\uY$ hypercharge generator since $N$ is uncharged under $U(1)_{\textsc{Y}}$.
This problem is easily remedied by noting that the true global symmetry of \eqref{lag_kinetic} is $U(3)_{\textsc{L}} \times U(3)_{\textsc{R}} = \suLR \times U(1)_{\textsc{V}} \times U(1)_{\textsc{A}}$ at the classical level; while it is well known that the $U(1)_{\textsc{A}}$ is anomalous \cite{Polyakov:1975rs,Belavin:1975fg,tHooft:1976snw,tHooft:1976rip,Polyakov:1976fu,Jackiw:1976pf,Callan:1976je,Coleman:1985aos,Weinberg:1995qft}, the $U(1)_{\textrm{V}}$ symmetry remains good at the quantum level.
This additional $U(1)_{\textrm{V}}$ has a generator equal to the identity on TC-flavor space: $T^{\textrm{X}} \equiv \mathds{1}_{3}$, which obviously commutes with all the $T^a$. 
We can thus form the hypercharge generator
\newcommand{\QX}{\ensuremath Q_{\textrm{X}}}%
\begin{align}
Y &\equiv \frac{1}{\sqrt{3}} T^8  + \mkern1mu \QX T^{\textrm{X}} = \begin{pmatrix} ( \QX + \frac{1}{6} )\mkern1mu \mathds{1}_{ 2 \times 2 } & 0 \\[1ex] 0 & \,\QX - \frac{1}{3} \end{pmatrix},
\end{align}
where $\QX$ is chosen appropriately to the field on which $Y$ acts.
For the $\chi$ and $(\chi^c)^\dagger$ fields, we need $\QX [\chi]=\QX\! \lb[ (\chi^c)^\dagger \rb]= \frac{1}{3}$ to obtain the hypercharge assignments for $L^{(c)}$ and $N^{(c)}$ in \tabref{gauge_charges}.
We then gauge $\uY$ by adding to the covariant derivative $D$ the term $D_\mu \lb\{\chi,(\chi^c)^\dagger\rb\} \supset -i g_1 B_\mu Y \lb\{\chi,(\chi^c)^\dagger\rb\}$, and demanding that under a forward $\uY$ transformation parametrized by $\alpha$, we have
\begin{align}
\lb\{ \chi , (\chi^c)^\dagger  \rb\} &\rightarrow \exp\lb[ i \alpha Y \rb] \lb\{ \chi , (\chi^c)^\dagger  \rb\} &
\text{and}&&
B_\mu &\rightarrow B_\mu + \frac{1}{g_1} \partial_\mu \alpha. \label{eq:gaugeY}
\end{align}

We can also include mass terms for the $\chi$ and $\chi^c$ fields by including the explicit $\suLR$-breaking terms
\begin{align}
\Lag \supset - \chi^c M \chi + \text{h.c.} =  - \Tr{ M \chi  \chi^c }+ \text{h.c.}, \label{eq:mass_terms}
\end{align}
where $\Tr{\,\cdots\,}$ is over the TC-flavor indices, and where $M$ is a mass matrix in TC-flavor space which in the mass-eigenbasis must respect the residual global $SU(2)\times U(1)$ symmetry in order not to explicitly break the electroweak gauging:
\begin{align}
M = \begin{pmatrix} m_L \mathds{1}_{2\times 2} & 0 \\ 0 & m_N \end{pmatrix}.
\end{align}
Note that while $M$ explicitly breaks $\suLR \rightarrow SU(2)_{\textrm{V}} \times U(1)_{\textrm{V}}$,%
\footnote{%
	At least if $m_L \neq m_N$; if $m_L = m_N$, it only breaks $\suLR \rightarrow SU(3)_{\textrm{V}}$.
	} %
if $M$ is given the usual spurionic transformation $M \rightarrow U_R M U_L^\dagger$, then \eqref{mass_terms} is a spurionic TC-flavor invariant. 

To summarize, the underlying constituent theory for the CH sector is specified by the Lagrangian terms
\begin{align}
\Lag &\supset i \bar\chi^\dagger \lb( \bar{\sigma} \cdot D \rb) \chi + i \chi^c \lb( \sigma \cdot D \rb) \lb( \chi^c \rb)^\dagger -  \Tr{ M \chi  \chi^c }+ \text{h.c.},\label{eq:lag_kinetic2}
\end{align}
where
\begin{align}
D_\mu \lb\{ \chi , (\chi^c)^\dagger  \rb\}&= \lb(  \partial_\mu - i g_{\textsc{TC}} A_\mu^{\textsc{TC}}  -i g_2 W_\mu^{\tilde{a}} T^{\tilde{a}} -i g_1 B_\mu Y \rb) \lb\{ \chi , (\chi^c)^\dagger  \rb\}.\label{eq:covarDeriv}
\end{align}
We return to the low-energy description of this sector by means of the Chiral Lagrangian in \sectref{chiral_lag}.

\subsection{Contact Interactions that lead to Yukawas}
\label{sect:contactForYukawa}
In order for the low-energy Chiral Lagrangian describing the CH model to exhibit the correct Higgs Yukawa couplings to the SM quarks, we must write couplings of the technifermions to the SM quarks in the constituent theory.
In this work we simply write down the required four-fermion operators coupling the SM fermions to the techniquark bilinear condensate, remaining agnostic about their underlying UV origin.
Various mechanisms exist to generate such couplings, such as extended technicolor~\cite{Eichten:1979ah,Dimopoulos:1979es}, partial compositeness~\cite{Kaplan:1991dc}, or bosonic technicolor~\cite{Simmons:1988fu,Samuel:1990dq,Dine:1990jd}; exploring their detailed consequences in this context goes beyond the scope of this work.

In order to write these couplings in a fashion which will allow spurionic TC-flavor symmetries to be made manifest, we arrange the SM field content into incomplete TC-flavor multiplets.
There are of course multiple ways to do this. 
We assign $U^c$ to an incomplete $(\bm{1},\bm{3})$ of TC-flavor which we denote $U^c_{\bm{3}_R}$, and we assign $D^c$ to an incomplete $(\bm{\bar{3}},\bm{1})$ of TC-flavor which we denote $D^c_{\bm{\bar{3}}_L}$: 
\begin{align}
U^c_{\bm{3}_R} &\equiv \begin{pmatrix} 0\\0\\U^c \end{pmatrix} &\text{and}&& D^c_{\bm{\bar{3}}_L} &\equiv \begin{pmatrix} 0\\0\\D^c \end{pmatrix}. \label{eq:UcDcDefn}
\end{align}
The spurionic $\suLR$ transformations of these incomplete multiplets are $U^c_{\bm{3}_R} \rightarrow U_R U^c_{\bm{3}_R}$ and $(D^c_{\bm{\bar{3}}_L})^\dagger \rightarrow U_L (D^c_{\bm{\bar{3}}_L})^\dagger$.
In order to obtain the correct hypercharge assignments for $U^c$ and $D^c$ under the gauging prescription developed above, we set $\QX\!\lb[U^c_{\bm{3}_R}\rb] = - \frac{1}{3}$ and $\QX\!\lb[(D^c_{\bm{\bar{3}}_L})^\dagger\rb] = 0$.

In close analogy to the requirement in the SM to use the two opposite-hypercharge $\suW$-fundamental fields $H$ and $\widetilde{H}$ (following the notation of \citeR{Schwartz:2014qft}) to write the SM Yukawas, we will need to embed the quark doublets in incomplete multiplets in two different ways.
If $Q$ is in the fundamental $\bm{2}$ of $\suW$, then the field $\widehat{Q} \equiv i \sigma^2 Q$ (in components, $\widehat{Q}^i \equiv \epsilon^{ij} Q_j$ where $\epsilon$ is the anti-symmetric invariant symbol of $SU(2)$ with $\epsilon^{12} = +1$) is in the conjugate%
\footnote{\label{ftnt:pseudoreal}%
	 Since the spinorial $\bm{2}$ of $SU(2)$ is a pseudo-real representation \cite{Georgi:1999lap}, it is unitarily equivalent to the conjugate $\bm{\bar{2}}$ representation.
	 There is thus no distinction between $\bm{2}$ and $\bm{\bar{2}}$; nevertheless, we maintain the bar to match up with the notation for the $\bm{3}$ and $\bm{\bar{3}}$ representations of $SU(3)$ into which the fermions are embedded.
	} %
anti-fundamental $\bm{\bar{2}}$ of $\suW$.
We will embed $Q$ in an incomplete $(\bm{1},\bm{3})$ of TC-flavor which we denote $Q_{\bm{3}_R}$, and embed $\widehat{Q}$ in an incomplete $(\bm{\bar{3}},\bm{1})$ of TC-flavor which we denote $Q_{\bm{\bar{3}}_L}$:
\begin{align}
Q_{\bm{3}_R} &\equiv \begin{pmatrix} Q \\ 0 \end{pmatrix} &\text{and}&& Q_{\bm{\bar{3}}_L} &\equiv \begin{pmatrix} \widehat{Q} \\ 0 \end{pmatrix}.
\end{align}
The spurionic $\suLR$ transformations of these incomplete multiplets are thus $Q_{\bm{3}_R} \rightarrow U_R Q_{\bm{3}_R}$ and $(Q_{\bm{\bar{3}}_L})^\dagger \rightarrow U_L ( Q_{\bm{\bar{3}}_L})^\dagger$; note also that the action of the gauged $\suW$ subgroup of TC-flavor automatically gives the correct $\suW$ action on the quark doublets given these embeddings since we have placed the fundamental $\bm{2}$ of $\suW$ in the fundamental $\bm{3}$ of $\suR$, and the anti-fundamental\up{\ref{ftnt:pseudoreal}} $\bm{\bar{2}}$ of $\suW$ in the anti-fundamental $\bm{\bar{3}}$ of $\suL$.
In order to obtain the correct hypercharge assignment for $Q$ under the gauging prescription developed above, we set $\QX\!\lb[Q_{\bm{3}_R}\rb] = 0$ and $\QX\!\lb[ (Q_{\bm{\bar{3}}_L})^\dagger \rb] = -\frac{1}{3}$.

Armed with these embeddings of the SM quarks, and noting that $\chi\chi^c \sim (\bm{3},\bm{\bar{3}})$ under $\suLR$, there are only two independent four-fermion spurionic TC-flavor invariants with zero net $Q_{\textrm{X}}$ charge which can be formed which couple $\chi\chi^c$ to SM quarks:
\begin{align}
&D^c_{\bm{\bar{3}}_L} \lb( \chi\chi^c \rb) Q_{\bm{3}_R} &\text{and}&& Q_{\bm{\bar{3}}_L} \lb( \chi\chi^c \rb) U^c_{\bm{3}_R} ,
\end{align}
along with their Hermitian conjugates.
By construction, the spurionic TC-flavor invariance (and net zero $Q_{\textrm{X}}$ charge) of these terms implies actual invariance under the gauged electroweak subgroup of the TC-flavor group.

We thus add the following contact terms to the Lagrangian%
\footnote{%
	To be explicit, the full index structure here is 
    \begin{align}
    \Lag \supset &+ \frac{{(Y_{\textsc{d}})_{p}}^{q}}{\Lambda_y^2} \lb[D^c_{\bm{\bar{3}}_L} \rb]^{\alpha,\bar{j},a,p} \lb[ \chi \rb]_{\bar{j},m}^\beta \lb[ \chi^c \rb]_\beta^{\hat{i},m} \lb[ Q_{\bm{3}_R} \rb]_{\alpha,\hat{i},a,q} + \text{h.c.} \nonumber \\
    &+ \frac{{(Y_{\textsc{u}})_{p}}^{q}}{\Lambda_y^2} \lb[ Q_{\bm{\bar{3}}_L} \rb]_{a,q}^{\alpha,\bar{j}} \lb[ \chi \rb]^{\beta}_{\bar{j},m} \lb[ \chi^c \rb]^{\hat{i},m}_{\beta} \lb[ U^c_{\bm{3}_R} \rb]^{a,p}_{\alpha,\hat{i}} + \text{h.c.} , \nonumber
    \end{align}
    where $\alpha, \beta=1,2$ are $(\frac{1}{2},0)$ spinorial Lorentz indices, $a=1,2,3$ is an $\suC$ color index, $p,q=1,2,3$ are SM generation indices, $\bar{j}=1,2,3$ is an $\suL$ index, $\hat{j}=1,2,3$ is an $\suR$ index, and $m=1,\ldots\!,N$ is an $\suTC$ index.
    Per the conventions of \citeR{Dreiner:2008tw}, for all indices other than the spinorial Lorentz indices, a lowered position denotes a fundamental index and a raised position denotes an anti-fundamental index.
    Repeated indices are obviously summed.
	} %
\begin{align}
\Lag &\supset + \frac{{(Y_{\textsc{d}})_{p}}^{q}}{\Lambda_y^2} \lb[D^c_{\bm{\bar{3}}_L} \rb]^p \lb( \chi\chi^c \rb) \lb[ Q_{\bm{3}_R} \rb]_q + \text{h.c.}
+ \frac{{(Y_{\textsc{u}})_{p}}^{q}}{\Lambda_y^2} \lb[ Q_{\bm{\bar{3}}_L} \rb]_{q} \lb( \chi\chi^c \rb) \lb[ U^c_{\bm{3}_R} \rb]^{p} + \text{h.c.} ,
\label{eq:contact_terms_1}
\end{align}
where $\Lambda_y$ is a dimensionful scale that will be fixed later to obtain canonically normalized Yukawas in the Chiral Lagrangian, and $Y_{\textsc{u},\textsc{d}}$ are Yukawa matrices in SM-generation space, which we have indexed explicitly: $p,q=1,2,3$.

While \eqref{contact_terms_1} is useful in that it displays manifest spurionic TC-flavor invariance, an alternative form not expressed in terms of incomplete TC-flavor multiplets is more useful for computational use.
To write this alternative form, we define projection operators
\begin{align}
{\big[ P_k \big]_{\hat{i}}}^{\bar{j}} &\equiv \delta_{\hat{i}}^3 {\delta^{\bar{j}}}_k &\text{and}&& {\big[ \tilde{P}^k \big]_{\hat{i}}}^{\bar{j}} \equiv \delta_{\hat{i}}^k \delta^{\bar{j}}_3,
\end{align}
where $\delta$ is the Kronecker-$\delta$ symbol and where the raised $\bar{j}=1,2,3$ is an $\suR$ anti-fundamental index, the lowered $\hat{i}=1,2,3$ is an $\suL$ fundamental index, and the lowered (raised) $k=1,2$ is an (anti-)fundamental $\suW$ index.
Then \eqref{contact_terms_1} can be alternatively written as
\begin{align}
\Lag &\supset + \frac{1}{\Lambda_y^2} \lb(D^c Y_{\textsc{d}} Q_j \rb) \Tr{ \tilde{P}^j \lb( \chi\chi^c \rb) } + \text{h.c.}
- \frac{1}{\Lambda_y^2} \lb( U^c Y_{\textsc{u}} Q_i \rb) \epsilon^{ij} \Tr{ P_j \lb( \chi\chi^c \rb) } + \text{h.c.},
\label{eq:contact_terms_2}
\end{align}
where $\Tr{\,\cdots\,}$ is a trace over $\suLR$ indices, $\epsilon$ is as before the antisymmetric invariant symbol of $SU(2)$, and we have suppressed the implied SM-generation indices in this form.
Although the SM quark fields are no longer in (incomplete) multiplets of TC-flavor, we can maintain a spurionic invariance of this expression under $\suLR$ by assigning the spurionic transformation rule $P \rightarrow U_R P U_L^\dagger$, where $P\in \{ P_j , \tilde{P}^j \}$.

In the low-energy Chiral Lagrangian, the terms in \eqref{contact_terms_2} will give rise to, \emph{inter alia}, the requisite quark Yukawa terms.

\subsection{Relaxion Sector}
\label{sect:relaxion}
The relaxion sector of the theory consists of a real pseudoscalar field $\phi$, the relaxion \cite{Graham:2015cka}, which we take to have dimension-5 effective axion couplings both to QCD and to the $\suTC$ gauge group (see \sectref{clockwork}); additionally, we allow for the existence of an additional potential $V_\phi(\phi)$ for the relaxion field:
\begin{align}
\Lag \supset \frac{1}{2} (\partial_\mu \phi)^2 + \frac{ (g_s)^2}{16\pi^2} \lb[ \frac{\phi}{f} - \theta_{\textsc{qcd}}^0 \rb] \Tr{ G_{\mu\nu}\widetilde{G}^{\mu\nu} } + \frac{ (g_{\textsc{tc}})^2}{16\pi^2} \frac{\phi}{F} \Tr{ {G}_{\textsc{tc}\, \mu\nu}\widetilde{G}^{\mu\nu}_{\textsc{tc}} } - V_\phi(\phi),\label{eq:L_relaxion}
\end{align}
where $G$ and $G_{\textsc{tc}}$ are, respectively, the matrix-valued gauge field strength tensors for the $\suC$ and $\suTC$ gauge groups, $\widetilde{G}_{\textsc{(tc)}\, \mu\nu} \equiv \frac{1}{2} \epsilon_{\mu\nu\alpha\beta} G_{\textsc{(tc)}}^{\alpha\beta}$ are the respective dual field strength tensors, and $g_s$ and $g_{\textsc{tc}}$ are the respective gauge couplings.
In \eqref{L_relaxion}, $\theta_{\textsc{qcd}}^0$ is a bare QCD $\theta$-term; we do not write an analogous independent $\theta^0_{\textsc{tc}}$ angle, as it could be absorbed into an unobservable shift of $\phi$ and $\theta_{\textsc{qcd}}^0$.
Additionally, $f$ is the usual QCD Peccei--Quinn (PQ) symmetry breaking scale, and $F$ is a PQ symmetry breaking scale which, as we mentioned in \sectref{cartoon}, needs to be taken exponentially larger than $f$: $F\gg f$.
Indeed, as we discussed in \sectref{cartoon} and as we will find in more detail in \sectref{summary_numerical}, $F$ will be required to be many orders of magnitude larger than the Planck scale; see \sectref{clockwork} for further discussion. 

The additional potential $V_\phi(\phi)$ is required to obtain the correct dynamical rolling of the relaxion field and will be discussed in more detail in \sectref{relaxion_potential}.

\subsection{Chiral Rotations}
\label{sect:chiral_rotation}
Before constructing the Chiral Lagrangian, we perform a $U(1)_{\textrm{A}}$ chiral rotation of the $\chi,\chi^c$ fields to rotate the $\suTC$--relaxion coupling into the mass matrix:
\begin{align}
\chi 		&\rightarrow e^{i \phi / 6F } \chi \\
\chi^c 	&\rightarrow e^{i \phi / 6F } \chi^c .
\end{align}
Following the method of Fujikawa \cite{Fujikawa:1979ay,Fujikawa:1980eg} to include the anomalous \cite{Bell:1969ts,Adler:1969gk} transformation of the measure of the functional integral under this transformation, we find that this rotation results in the following form for the Lagrangian:
\begin{align}
\Lag &= {\cal L}_{\textsc{sm},H=0} +  \frac{1}{2} (\partial_\mu \phi)^2 - V_\phi(\phi) 
		+ \frac{(g_s)^2}{16\pi^2} \lb[ \frac{\phi}{f} - \theta_{\textsc{qcd}}^0 \rb] \Tr{ G^{\mu\nu} \widetilde{G}_{\mu\nu} } \nl
		+ i \chi^\dagger \bar{\sigma}^\mu  D_\mu \chi 
		+ i \chi^c  \sigma^\mu D_\mu (\chi^c)^\dagger   
		- \Tr{ M ( \chi \chi^c) } e^{i \phi/3F} + \text{h.c.} \nl
		+ \frac{ 1 }{\Lambda_y^2} \lb( D^c Y_{\textsc{d}} Q_j \rb)  \Tr{ P^j  (\chi \chi^c)  } e^{i \phi/3F} + \text{h.c.} \nl
		- \frac{ 1 }{\Lambda_y^2} \lb( U^c Y_{\textsc{u}} Q_i \rb) \epsilon^{ij}  \Tr{ P_j  ( \chi  \chi^c ) } e^{i \phi/3F}  + \text{h.c.} \nl
		- \frac{\partial_\mu\phi }{6F} (J_A^0)^\mu 
		- \frac{\phi}{F} \lb[ \dfrac{(g_2)^2}{16\pi^2}  \frac{N}{3}  \Tr{ W_{\mu\nu} {\widetilde{W}}^{\mu\nu}}
										+ \dfrac{(g_1)^2}{16\pi^2} \frac{N}{3} \frac{1}{2}  B_{\mu\nu} {\widetilde{B}}^{\mu\nu} 
									 \rb],
\end{align}
where $W_{\mu\nu}$ are the matrix-valued $\suW$ field strength tensors, $B_{\mu\nu}$ is the $\uY$ hypercharge field strength tensor, and we have also now included the non-Higgs part of the SM Lagrangian, which we denote $\Lag_{\textsc{sm},\, H=0}$ (by which notation we mean the usual SM Lagrangian with the \emph{elementary} Higgs doublet $H$ set equal to zero); $(J_A^0)^\mu \equiv \big[  \chi^\dagger \bar{\sigma}^\mu \chi - \allowbreak \chi^c \sigma^\mu (\chi^c)^\dagger \big]$ is the $U(1)_{\textrm{A}}$ axial current; and the covariant derivative is still given by \eqref{covarDeriv}.

As a final step before we pass to the Chiral Lagrangian, we rotate to the mass-eigenstate basis for the SM quarks via the usual CKM manipulations \cite{Cabibbo:1963ao,Kobayashi:1973xyz}. 
Following the exposition of \citeR{Schwartz:2014qft}, and letting $X\in\{U,\, D\}$ for the remainder in this paragraph (with $X$ always to be read consistently as either $U$ or $D$ in every formula), it is always possible to write $Y_{\textsc{x}} \equiv P_{\textsc{x}} Q_{\textsc{x}} y_{\textsc{x}} Q_{\textsc{x}}^\dagger$ where $P_{\textsc{x}}$ and $Q_{\textsc{x}}$ are unitary matrices in SM-generation space and $y_{\textsc{x}}$ are diagonal matrices whose entries are the real positive square roots of the eigenvalues of the Hermitian matrix $Y_{\textsc{x}}Y_{\textsc{x}}^\dagger$.
We define $V_{\textsc{ckm}} \equiv Q_{\textsc{u}}^\dagger Q_{\textsc{d}}$.
The necessary chiral quark rotation to bring the fields to the mass-eigenstate basis is then given by $X^c \rightarrow X^c Q_{\textsc{x}}^\dagger P_{\textsc{x}}^\dagger$ and $X \rightarrow Q_{\textsc{x}} X$, which shifts $\theta^0_{\textsc{qcd}} \rightarrow \theta^0_{\textsc{qcd}} - \arg\det Y_{\textsc{u}}Y_{\textsc{d}} \equiv \theta_{\textsc{qcd}}$.
The final result on the Lagrangian is
\begin{align}
{\cal L} & = {\cal L}_{\textsc{sm},\, H=0,\, \textsc{no quarks}} 
		+ i U^\dagger \bar{\sigma}^\mu \hat{D}_\mu U 
		+ i D^\dagger \bar{\sigma}^\mu \hat{D}_\mu D 
		+ i (D^c) \sigma^\mu \hat{D}_\mu (D^c)^\dagger 
		+ i (U^c) \sigma^\mu \hat{D}_\mu (U^c)^\dagger \nl
		+ \frac{g_2}{2} W_\mu^3 \lb( U^\dagger\bar{\sigma}^\mu U - D^\dagger \bar{\sigma}^\mu D \rb) 
		+ \frac{g_2}{\sqrt{2}} \lb( W_\mu^+ U^\dagger \bar{\sigma}^\mu V_{\textsc{ckm}} D + W_\mu^- D^\dagger \bar{\sigma}^\mu V_{\textsc{ckm}}^\dagger U \rb) \nl
		+  \frac{1}{2} (\partial_\mu \phi)^2 - V_\phi(\phi) + \frac{(g_s)^2}{16\pi^2} \lb[ \frac{\phi}{f} - \theta_{\textsc{qcd}} \rb] \Tr{ G^{\mu\nu} \tilde{G}_{\mu\nu} }  \nl
		+ i \chi^\dagger \bar{\sigma}^\mu  D_\mu \chi 
		+ i \chi^c  \sigma^\mu D_\mu (\chi^c)^\dagger  
		- \Tr{ M ( \chi \chi^c) } e^{i \phi/3F} + \text{h.c.} \nl
		+ \frac{ 1 }{\Lambda_y^2} \lb( D^c y_D D \rb)  \Tr{ P^2  (\chi \chi^c)  } e^{i \phi/3F} + \text{h.c.} 
		- \frac{ 1 }{\Lambda_y^2} \lb( U^c y_U U \rb) \Tr{ P_2 ( \chi  \chi^c ) } e^{i \phi/3F}  + \text{h.c.} \nl
		+ \frac{ 1 }{\Lambda_y^2} \lb( D^c y_D V_{\textsc{ckm}}^\dagger U \rb)  \Tr{ P^1 (\chi \chi^c)  } e^{i \phi/3F} + \text{h.c.} \nl
		+ \frac{ 1 }{\Lambda_y^2} \lb( U^c y_U V_{\textsc{ckm}} D \rb)  \Tr{ P_1 ( \chi  \chi^c ) } e^{i \phi/3F}   + \text{h.c.} \nl
		- \frac{\partial_\mu\phi }{6F} (J_A^0)^\mu
		-  \frac{\phi}{F} \lb[ \dfrac{(g_2)^2}{16\pi^2}  \frac{N}{3}  \Tr{ W_{\mu\nu} {\widetilde{W}}^{\mu\nu}}
										+ \dfrac{(g_1)^2}{16\pi^2} \frac{N}{3} \frac{1}{2}  B_{\mu\nu} {\widetilde{B}}^{\mu\nu}
									 \rb],
		  \label{eq:final_UV_L}
\end{align}
where $\hat{D}_\mu \equiv \partial_\mu - i g_s G_\mu - i g_1 B_\mu Y$ (we have explicitly extracted the $\suW$ gauge couplings to the quarks) with $G_{\mu}$ the matrix-valued $\suC$ gauge fields, and $B_\mu$ and $Y$ the $\uY$ hypercharge field and hypercharge operator respectively; $W_\mu^\pm \equiv \lb( W_\mu^1 \mp i W_\mu^2 \rb)/\sqrt{2}$; and the notation `\textsc{no quarks}' on the SM part of the Lagrangian indicates that we have explicitly extracted and displayed all the quark-dependent terms.

\section{Chiral Lagrangian}
\label{sect:chiral_lag}
To proceed with the analysis of our model, \eqref{final_UV_L}, we must now pass to the theory of the bound states of $\chi$ and $\chi^c$ fermions after the $\suTC$ group confines.
We therefore construct the Chiral Lagrangian (see, e.g., \citeR{Georgi:1984wem}) based on the global $\suLR \times U(1)_{\textrm{V}}$ (spurionic) TC-flavor symmetry exhibited by \eqref{final_UV_L}.
On confinement, $\suLR \times U(1)_{\textrm{V}} \rightarrow SU(3)_{\textrm{V}} \times U(1)_{\textrm{V}}$ owing to the spontaneous emergence of a chiral condensate $\langle \chi \chi^c \rangle \sim -\LambdaTC F_\pi^2\mathds{1}_3$ (na\"ive dimensional analysis [NDA] estimate \cite{Manohar:1983md,Georgi:1992dw}). 
The Chiral Lagrangian is therefore the theory of the eight `technipions' $\piTC^a$---the pseudo-Nambu--Goldstone bosons of the spontaneously broken global $SU(3)_{\textrm{A}}$ symmetry.
We assume that the excitation associated with the anomalous $U(1)_{\textrm{A}}$ symmetry, $\eta_{\textsc{tc}}'$, is massive enough to have been integrated out of the theory.

\subsection{Matrix-Valued Technipion Field \texorpdfstring{$\Umat$}{U}}
\label{sect:Umatrix}
The fundamental object in the construction of the Chiral Lagrangian is the matrix-valued field of the technipions, $\Umat$, which is assumed to transform in a $(\bm{3},\bm{\bar{3}})$ of the $\suLR$ TC-flavor group:%
\footnote{%
	See \appref{SU3exponentiation} for the general, closed-form expression for the matrix $\Umat$ in terms of the $\piTC^a$ fields.
	} %
\begin{align}
\Umat &\equiv \exp\lb[ \frac{2i}{F_\pi} \Pi \rb] &\text{where} &&
\Pi &\equiv \piTC^a T^a, \label{eq:Udefn}
\end{align}
with $T^a$ the $SU(3)$ generators, and where $F_\pi$ is the dimensionful compositeness scale.
We define $\eta_{\textsc{tc}} \equiv \piTC^8$, $i H^+ \equiv \lb( \piTC^4 - i \piTC^5 \rb)/\sqrt{2}$ and $i H^0 \equiv \lb( \piTC^6 - i \piTC^7 \rb)/\sqrt{2}$;
then $\Pi$ is explicitly given by 
\renewcommand{\arraystretch}{1.5}
\begin{align}
\Pi	&= \frac{1}{2} 	
		\lb( \begin{array}{c|c} 
			\frac{1}{\sqrt{3}} \eta_{\textsc{tc}} \mathds{1}_2 + \Tilde{\Pi} & i \sqrt{2} H \\ \hline
			-i\sqrt{2} H^\dagger & -\frac{2}{\sqrt{3}} \eta_{\textsc{tc}}
		\end{array} \rb), \label{eq:Pidefn}
\end{align}
\renewcommand{\arraystretch}{1.0}
where
\begin{align}
\tilde{\Pi}	&=  
		 \lb( \begin{array}{c|c} 
			\piTC^3  & \piTC^1 - i \piTC^2 \\ \hline
			\piTC^1+i\piTC^2 & - \piTC^3
		\end{array} \rb) & 
\text{and}&&
H &\equiv \begin{pmatrix} H^+ \\ H^0 \end{pmatrix}. \label{eq:Pi-Hdefn}
\end{align}
For reasons to become clear shortly, we will rewrite \eqref{Pi-Hdefn} as
\begin{align}
\tilde{\Pi}	&=  V_\xi^\dagger 
		 \lb( \begin{array}{c|c} 
			\piTC^0  & \sqrt{2} \piTC^+ \\ \hline
			\sqrt{2} \piTC^- & - \piTC^0
		\end{array} \rb) V_\xi & 
\text{and}&&
H &\equiv \frac{1}{\sqrt{2}} V_\xi^\dagger \begin{pmatrix} 0 \\ h \end{pmatrix}, \label{eq:Pi-Hdefn2}
\end{align}
where $V_\xi \equiv \exp\lb[ i \xi^{\tilde{a}}(x) \tau^{\tilde{a}} \rb]$.
Under $SU(3)_{\textrm{V}}$ transformations, $\Umat $ transforms as $\Umat \rightarrow V_3 \Umat  V_3^\dagger$, which implies that $\Pi \rightarrow V_3 \Pi V_3^\dagger$; in particular, since an $SU(2)$ subgroup of $SU(3)_{\textrm{V}}$ is gauged as $\suW$, an $\suW$ gauge transformation $V$ acts with
\begin{align}
V_3 &\equiv \begin{pmatrix} V & 0 \\ 0 & 1 \end{pmatrix} & \text{where} && V &\equiv \exp\lb[ i \alpha^{\tilde{a}}(x) \tau^{\tilde{a}} \rb];
\end{align}
alternatively,
\begin{align}
\tilde{\Pi} &\rightarrow V \tilde{\Pi} V^\dagger,  	& 
\quad H &\rightarrow V H,					&
\text{and} &							&
\eta_{\textsc{tc}} &\rightarrow \eta_{\textsc{tc}}.
\end{align}
We will work in a gauge where the $\xi^{\tilde{a}}$ in \eqref{Pi-Hdefn2} are gauged away [i.e., a local gauge transformation with $\alpha^{\tilde{a}} = \xi^{\tilde{a}}$ is made to \eqref{Pi-Hdefn2}].
That is, \eqref[s]{Udefn}, (\ref{eq:Pidefn}) and (\ref{eq:Pi-Hdefn2}) define the matrix-valued field $\Umat $ in terms of its five physical degrees of freedom $\piTC^0,\piTC^\pm,\eta_{\textsc{tc}}$, and $h$ provided we simply replace $V_\xi^{(\dagger)} \rightarrow \mathds{1}_2$ in \eqref{Pi-Hdefn2}.
After this gauge choice is made, it will turn out to be most convenient to also make the following $SO(2)$ rotation in field space
\begin{align}
\begin{pmatrix} \kappaTC \\ \omegaTC \end{pmatrix} &\equiv  \begin{pmatrix} \cos\vartheta & \sin\vartheta \\ - \sin\vartheta & \cos\vartheta \end{pmatrix} \begin{pmatrix} \piTC^0 \\ \eta_{\textsc{tc}} \end{pmatrix} &\text{with}&& \vartheta = \frac{\pi}{6} ,
\end{align}
and work instead with the five degrees of freedom $\omegaTC,\kappaTC,\piTC^\pm$, and $h$.

\subsection{Chiral Lagrangian}
\label{sect:chiral}

A careful analysis of the spurionic symmetries of \eqref{final_UV_L}, and the anomalies in the various axial currents, yields the Chiral Lagrangian corresponding to \eqref{final_UV_L}, at leading order in spurions and momenta: 
\begin{align}
{\cal L} & = {\cal L}_{\textsc{sm},\, H=0} 
+  \frac{1}{2} (\partial_\mu \phi)^2 - V_\phi(\phi) + \frac{(g_s)^2}{16\pi^2} \lb[ \frac{\phi}{f} - \theta_{\textsc{qcd}} \rb] \Tr{ G^{\mu\nu} \tilde{G}_{\mu\nu} }  \nl
+  \frac{F_\pi^2}{4} \Tr{ (D_\mu \Umat )^\dagger (D^\mu \Umat  ) } + c_m \LambdaTC F_\pi^2  \Tr{M \Umat } e^{i \phi /3F}  + \text{h.c.}  \nl 
- \frac{F_\pi}{\sqrt{2}} \lb( D^c y_D D \rb)  \Tr{ P^2  \Umat   } e^{i \phi /3F} + \text{h.c.}
+ \frac{F_\pi}{\sqrt{2}} \lb( U^c y_U U \rb) \Tr{ P_2 \Umat  } e^{i \phi /3F}   + \text{h.c.} \nl
- \frac{F_\pi}{\sqrt{2}} \lb( D^c y_D V_{\textsc{ckm}}^\dagger U \rb)  \Tr{ P^1 \Umat   } e^{i \phi /3F} + \text{h.c.}\nl
- \frac{F_\pi}{\sqrt{2}} \lb( U^c y_U V_{\textsc{ckm}} D \rb)  \Tr{ P_1 \Umat  } e^{i \phi /3F}   + \text{h.c.} \nl
- \frac{N}{8\pi^2} \lb[\begin{array}{l}
								 \dfrac{(g_2)^2}{2} \Tr{ \Pi_\phi \begin{pmatrix} \mathds{1}_2 & 0 \\ 0 & 0 \end{pmatrix} }  \Tr{ W_{\mu\nu} {\widetilde{W}}^{\mu\nu}} 
								 + (g_1)^2 \Tr{ \Pi_\phi  Y^2}  B_{\mu\nu} {\widetilde{B}}^{\mu\nu} \\[2ex]
								 + 2g_1g_2 \Tr{\Pi_\phi  Y T^{\tilde{b}} } B_{\mu\nu} {\widetilde{W}}^{\tilde{b}\, \mu\nu} 
								\end{array} \rb] , \label{eq:chiral_L}
\end{align}
where $\Pi_\phi \equiv - \frac{i}{2} \ln \Umat +  ( \phi / 6F ) \mathds{1}_3$; $D_\mu \Umat \equiv \partial_\mu \Umat - i \lb[ v_\mu , \Umat \rb]$ with $v_\mu \equiv g_1 B_\mu Y + g_2 W_\mu^{\tilde{a}} T^{\tilde{a}}$; $\LambdaTC \approx 4\pi F_\pi / \sqrt{N}$ is the cutoff scale for this effective description; $c_m$ is a perturbatively incalculable $\mathcal{O}(1)$ constant; we have absorbed the quark kinetic and gauge-coupling terms which were explicitly displayed in \eqref{final_UV_L} back into ${\cal L}_{\textsc{sm},\, H=0}$; and we have demanded canonically normalized Yukawa terms, which, up to a perturbatively incalculable constant, suggests $\Lambda_y^2 \sim \sqrt{2} F_\pi \LambdaTC \approx (\sqrt{2N}/4\pi) \cdot \LambdaTC^2$.
While this last relation would seem to indicate all flavor structures are generated at the same scale, we emphasize that this is not necessarily the case---different structures could be generated with a hierarchy of scales and/or Wilson coefficients, depending on the mechanism underlying flavor. 

\subsection{Yukawas}
\label{sect:yukawas}
It is straightforward to show that 
\begin{align}
\Lag &\supset \frac{F_\pi}{\sqrt{2}} \lb( U^c y_U U \rb) \Tr{ P_2 \Umat  } e^{i \phi /3F}  + \text{h.c.} \\
&\supset -  \frac{1}{\sqrt{2}} \lb( U^c y_U U \rb) h  \exp\lb[ i \frac{\phi}{3F} - i \frac{\kappaTC}{\sqrt{3} F_\pi}  \rb]\sinc\lb( \frac{\barpiTC}{F_\pi} \rb) + \text{h.c.} , \label{eq:upYukawa}
\end{align}
and
\begin{align}
\Lag &\supset - \frac{F_\pi}{\sqrt{2}} \lb( D^c y_D D \rb)  \Tr{ P^2  \Umat   } e^{i \phi /3F} + \text{h.c.} \\
&\supset -  \frac{1}{\sqrt{2}} \lb( D^c y_D D \rb) h  \exp\lb[ i \frac{\phi}{3F} - i \frac{\kappaTC}{\sqrt{3} F_\pi}  \rb]\sinc\lb( \frac{\barpiTC}{F_\pi} \rb) + \text{h.c.},\label{eq:downYukawa}
\end{align}
where 
\begin{align}
 \barpiTC &\equiv \sqrt{ h^2 + \omegaTC^2 } & \text{and} && \sinc \mkern2mu x\equiv \frac{\sin x}{x}\ , \label{eq:barpi-sinc}
\end{align}
and where these results are correct to all orders in $F_\pi$ for the uncharged fields, but we have ignored any (interaction) terms involving the electromagnetically charged (EM-charged) states $\sim \piTC^+ \piTC^-$. 
The other types of couplings of the $\Umat$ field to the SM quarks are (minimally) three-point interactions (after the $h$ gets a vev) involving the EM-charged technipions $\piTC^\pm$:
\begin{align}
\Lag &\supset - \frac{F_\pi}{\sqrt{2}} \lb( D^c y_D V_{\textsc{ckm}}^\dagger U \rb) \Tr{ P^1 \Umat } e^{i \phi /3F}  +\text{h.c.}  \\
 &\supset  - \frac{i}{2}  \lb( D^c y_D V_{\textsc{ckm}}^\dagger U \rb) \frac{ h \piTC^- }{F_\pi }  \exp\lb[ i \frac{\phi}{3F} - i \frac{\kappaTC}{\sqrt{3} F_\pi}  \rb] \lb[ 1 + \mathcal{O}(F_\pi^{-1}) \rb] +\text{h.c.} , \label{eq:P1upper}\\[1ex]
\Lag &\supset - \frac{F_\pi}{\sqrt{2}} \lb( U^c y_U V_{\textsc{ckm}} D \rb) \Tr{ P_1 \Umat } e^{i \phi /3F} +\text{h.c.} \\
& \supset  + \frac{i}{2}   \lb( U^c y_U V_{\textsc{ckm}} D \rb) \frac{ h \piTC^+ }{F_\pi } \exp\lb[ i \frac{\phi}{3F} - i \frac{\kappaTC}{\sqrt{3} F_\pi}  \rb] \lb[ 1 + \mathcal{O}(F_\pi^{-1}) \rb]  +\text{h.c.} \label{eq:P1lower} ,
 \end{align}
where we have kept only those terms with no additional charged fields.

Eqs.~(\ref{eq:upYukawa}) and (\ref{eq:downYukawa}) include, \emph{inter alia}, the Yukawas that give rise to the SM quark masses; however, they also contain unwanted phases unless $\langle  \phi/(3F) - \kappaTC/ (\sqrt{3} F_\pi) \rangle = 0$. 
We therefore perform a further chiral field redefinition on (every generation of) the SM quarks: 
\begin{align}
\{\, D ,\, D^c ,\, U ,\, U^c \, \} &\rightarrow \{\, D ,\, D^c ,\, U ,\, U^c \, \} \cdot \exp\lb[  - \frac{i}{2} \lb(  \frac{\phi}{3F} - \frac{\kappaTC}{\sqrt{3} F_\pi} \rb) \rb] .
\end{align}
This has a number of effects: (a) it removes the exponential factors in \eqref[s]{upYukawa} and (\ref{eq:downYukawa}), so that the Lagrangian contains the straightforward Yukawa terms 
\begin{align}
\Lag &\supset -  \frac{1}{\sqrt{2}} \lb( U^c y_U U + D^c y_D D \rb) h \sinc\lb( \frac{\barpiTC}{F_\pi} \rb) + \text{h.c.} ; \label{eq:Yukawas}
\end{align}
(b) it removes the explicit exponential factors in \eqref[s]{P1upper} and (\ref{eq:P1lower});
(c) the QCD $\theta$-angle shifts: $\theta_{\textsc{qcd}} \rightarrow \theta_{\textsc{qcd}} +  (6\kappaTC)/(\sqrt{3} F_\pi)- i (2\phi)/F$; and (d) an additional term is added to the Lagrangian: 
\begin{align}
\Lag \supset - \frac{1}{2} \lb[ U^c \sigma^\mu (U^c)^\dagger - U^\dagger \bar{\sigma}^\mu U + \lb( U^{(c)} \leftrightarrow D^{(c)}\rb) \rb] \partial_\mu \lb[ \frac{\phi}{3F} - \frac{\kappaTC}{\sqrt{3} F_\pi} \rb]. \label{eq:additional_term_chiral}
\end{align}

\subsection{Expanding the Chiral Lagrangian}
\label{sect:expanding}
In order to analyze \eqref{chiral_L} [as modified per the discussion in \sectref{yukawas}], it is necessary to write it out in terms of the physical degrees of freedom of $\Umat$: $\omegaTC$, $\kappaTC$, $\piTC^\pm$, and $h$. 
By making use of \eqref{Ufull} [see \appref{SU3exponentiation}], it is in principle possible to do this exactly in closed form. 
However, our primary interest here will be in those terms which have a bearing on the spectrum of the theory, and the effective potential for the neutral scalar fields.
As such, we will not need all the higher-order interaction terms in their full generality.

In addition to the Yukawa terms we already discussed in \sectref{yukawas}, the terms of interest to us are:
	(a) the kinetic and potential terms for the relaxion, and its coupling to QCD; 
	(b) the kinetic terms for the states $\omegaTC$, $\kappaTC$, and $h$;
	(c) the kinetic terms for the states $\piTC^\pm$, their kinetic mixing terms with $W_\mu^\pm$ owing to our gauge choice discussed in \sectref{Umatrix}, and the mass terms for the $W^\pm$ and $Z$ bosons;
	(d) the mass terms for the EM-charged technipions; and
	(e) the terms which give rise to the full tree-level potential for the EM-neutral scalars $\omegaTC$, $\kappaTC$, and $h$.
We discuss each of these terms in turn.

\indent\textbf{(a)} After the chiral rotation in \sectref{yukawas}, the relaxion-dependent terms in the first line of \eqref{chiral_L} are 
		\begin{align}
		{\cal L} &\supset  \frac{1}{2} (\partial_\mu \phi)^2 - V_\phi(\phi) + \frac{(g_s)^2}{16\pi^2} \lb[ \frac{\phi}{f} \lb( 1 + \frac{2f}{F} \rb) - \frac{2\sqrt{3}}{F_\pi} \kappaTC - \theta_{\textsc{qcd}} \rb] \Tr{ G^{\mu\nu} \tilde{G}_{\mu\nu} }. \label{eq:terms_a}
		\end{align}
		The term proportional to $f/F\ll1 $ can dropped.\\[1ex]
\indent\textbf{(b)} The kinetic terms for $\omegaTC$, $\kappaTC$, and $h$ are contained in the term \linebreak$\tfrac{F_\pi^2}{4} \Tr{ (D_\mu \Umat )^\dagger (D^\mu \Umat  ) } $ in \eqref{chiral_L}. 
		We are only interested in the two-derivative terms which are potentially quadratic in the fields after the EM-neutral scalars possibly obtain vevs; we can thus always neglect any terms which contain un-differentiated  $\piTC^\pm$ fields, but we need to keep terms to all orders in the un-differentiated $\omegaTC,\ \kappaTC$, and $h$ fields.
		It is straightforward to obtain these terms by sending $\piTC^\pm \rightarrow 0$ in the definition for $\Umat$ \emph{before} directly exponentiating and inserting the result into the relevant terms in \eqref{chiral_L}.
		The terms which arise are
		\begin{align}
		\Lag &\supset \frac{1}{2} (\partial_\mu \kappaTC)^2 
					+ \frac{1}{2} (\partial_\mu h)^2 \lb[ \frac{h^2}{\barpiTC^2}+\frac{\omegaTC^2}{\barpiTC^2} \sinc^{2}\lb( \frac{\barpiTC}{F_\pi} \rb)  \rb] \nl
					+ \frac{1}{2} (\partial_\mu \omegaTC)^2 \lb[ \frac{\omegaTC^2}{\barpiTC^2}+\frac{h^2}{\barpiTC^2} \sinc^{2}\lb( \frac{\barpiTC}{F_\pi} \rb)  \rb] 
					+ (\partial_\mu h)(\partial^\mu \omegaTC) \frac{h\omegaTC}{\barpiTC^2} \lb[1 - \sinc^{2}\lb( \frac{\barpiTC}{F_\pi} \rb)\rb].\label{eq:terms_b}
		\end{align}
\indent\textbf{(c)} Given our gauge choice, the $\piTC^\pm$ fields can possibly kinetically mix with the $W_\mu^\pm$. 
			The relevant kinetic terms for $\piTC^\pm$, the kinetic mixing terms, and the gauge boson mass terms are also contained in the term $\tfrac{F_\pi^2}{4} \Tr{ (D_\mu \Umat )^\dagger (D^\mu \Umat  ) } $ in \eqref{chiral_L}. 
		Similar arguments to those made at (b) apply about which terms need to be kept to find all possible contributions to the two-derivative, one-derivative--one-gauge-boson, or two-gauge-boson terms which are possibly quadratic in the fields after the EM-neutral scalars possibly obtain vevs.
		To extract these terms, we make use of the exact closed-form expression \eqref{Ufull}, expanded out to quadratic order in the charged technipion fields.
		The relevant terms are 
		\begin{align}
		\Lag &\supset \partial_\mu \piTC^+ \partial^\mu \piTC^- \lb( \begin{array}{l}
		2F_\pi^2 \dfrac{ h^2 + \lb( \omegaTC + \sqrt{3}\kappaTC \rb)^2}{(\barpiTC^2-3\kappaTC^2)^2} \lb[ 1 - \cos\lb( \tfrac{\sqrt{3}\kappaTC}{F_\pi} \rb) \cos\lb( \tfrac{\barpiTC}{F_\pi} \rb) \rb] \\[3ex]
		 - 2F_\pi \dfrac{\barpiTC^2 ( \omegaTC + 2\sqrt{3}\kappaTC) + 3\kappaTC^2 \omegaTC }{(\barpiTC^2-3\kappaTC^2)^2}\sin\lb( \tfrac{\sqrt{3}\kappaTC}{F_\pi} \rb) \sinc\lb( \tfrac{\barpiTC}{F_\pi} \rb)
																	\end{array} \rb)\nl
		- i g_2  \lb(  W_\mu^+ \partial^\mu \piTC^- -  W_\mu^- \partial^\mu \piTC^+ \rb) \lb(\begin{array}{l}
					 			 \dfrac{F_\pi^2 (\sqrt{3}\kappaTC+\omegaTC)}{\barpiTC^2-3\kappaTC^2} \lb[ 1 - \cos\lb( \tfrac{\sqrt{3}\kappaTC}{F_\pi} \rb) \cos \lb( \tfrac{\barpiTC}{F_\pi} \rb) \rb] \\[2ex]
								 - \dfrac{F_\pi (\barpiTC^2+\sqrt{3}\kappaTC\omegaTC)}{\barpiTC^2-3\kappaTC^2} \sin\lb( \tfrac{\sqrt{3}\kappaTC}{F_\pi} \rb) \sinc \lb( \tfrac{\barpiTC}{F_\pi} \rb) 
																			 \end{array}\rb)  \nl
		+ \frac{g_2^2 F_\pi^2}{2}  \lb[	1 - \cos\lb(\tfrac{\barpiTC}{F_\pi} \rb) \cos\lb(\tfrac{\sqrt{3}\kappaTC}{F_\pi} \rb) 
						- \dfrac{ \omegaTC }{F_\pi} \sinc\lb( \tfrac{\barpiTC}{F_\pi} \rb) \sin\lb(\tfrac{\sqrt{3}\kappaTC }{F_\pi}  \rb) 
						\rb] \, W_\mu^- W^{\mu\, +} 	\nl
		+ \frac{g_1^2+g_2^2 }{8} h^2 \sinc^2\lb( \tfrac{\barpiTC}{F_\pi} \rb) Z^2.\label{eq:terms_c}
		\end{align}
\noindent\textbf{(d)} The mass terms for the EM-charged technipions are obtained from the terms \linebreak$ c_m \LambdaTC F_\pi^2  \Tr{M \Umat } e^{i \phi /3F}  + \text{h.c.} $ in \eqref{chiral_L}. 
				The relevant terms in the expansion are those proportional to $\piTC^+\piTC^-$, and are again obtained using the exact closed-form expression \eqref{Ufull}, expanded out to quadratic order in the charged technipion fields:
				\begin{align}
				\Lag &\supset c_m \LambdaTC \piTC^+ \piTC^- \nonumber \\ &\times \lb[\begin{array}{l}
									- 4F_\pi^2 (m_L-m_N) \dfrac{h^2}{(\barpiTC^2-3\kappaTC^2)^2} 
										 \lb[  \cos\lb( \tfrac{2\kappaTC}{\sqrt{3}F_\pi} + \tfrac{\phi}{3F} \rb) 
										 	- \cos\lb( \tfrac{\barpiTC}{F_\pi} \rb) \cos\lb( \tfrac{\kappaTC}{\sqrt{3}F_\pi} - \tfrac{\phi}{3F} \rb)
										\rb] \\[2ex]
									+ 6\sqrt{3} F_\pi (m_L-m_N) \dfrac{ \kappaTC h^2 ( \barpiTC^2 - \kappaTC^2 ) }{\barpiTC^2 (\barpiTC^2 - 3\kappaTC^2)^2} 
											\sinc\lb( \tfrac{\barpiTC}{F_\pi} \rb) \sin\lb( \tfrac{\kappaTC}{\sqrt{3}F_\pi} - \tfrac{\phi}{3F} \rb) \\[2ex]
									+ 4 F_\pi m_L \dfrac{\sqrt{3}\kappaTC+\omegaTC}{\barpiTC^2-3\kappaTC^2}  
											\sin\lb( \tfrac{2\kappaTC}{\sqrt{3}F_\pi} + \tfrac{\phi}{3F} \rb) \\[2ex]
									+ 2 F_\pi \dfrac{ \sqrt{3} (m_L+m_N) \kappaTC h^2 + 2m_L \omegaTC(\barpiTC^2 + \sqrt{3}\kappaTC\omegaTC)}{\barpiTC^2(\barpiTC^2-3\kappaTC^2)}   \\[2ex]
											\qquad\qquad \times \cos\lb( \tfrac{\barpiTC}{F_\pi} \rb) \sin\lb( \tfrac{\kappaTC}{\sqrt{3}F_\pi} - \tfrac{\phi}{3F} \rb) \\[2ex]
									- 2 \dfrac{ (m_L+m_N) h^2 + 2m_L \omegaTC(\sqrt{3}\kappaTC+\omegaTC)}{(\barpiTC^2-3\kappaTC^2)}
											\sinc\lb( \tfrac{\barpiTC}{F_\pi} \rb) \cos\lb( \tfrac{\kappaTC}{\sqrt{3}F_\pi} - \tfrac{\phi}{3F} \rb)
									\end{array}  \rb]  \label{eq:terms_d} \\[-2ex]
				&\stackrel{\substack{\text{equal}\\\text{masses}\vspace{0.1cm}}}{\longrightarrow} 4c_m m \LambdaTC  \piTC^+ \piTC^- \nl \quad \times
						 \lb[\begin{array}{l}	
						 		F_\pi \dfrac{\sqrt{3}\kappaTC+\omegaTC}{\barpiTC^2-3\kappa^2}
										\lb[   
											\cos\lb( \dfrac{\barpiTC}{F_\pi} \rb) \sin\lb( \dfrac{\kappaTC}{\sqrt{3}F_\pi} - \dfrac{\phi}{3F} \rb) 
											+ \sin\lb( \dfrac{2\kappaTC}{\sqrt{3}F_\pi} + \dfrac{\phi}{3F} \rb) 
										\rb] \\[2ex]
								- \dfrac{  \barpiTC^2 + \sqrt{3}\kappaTC\omegaTC}{\barpiTC^2-3\kappaTC^2}
									\sinc\lb( \dfrac{\barpiTC}{F_\pi} \rb) \cos\lb( \dfrac{\kappaTC}{\sqrt{3}F_\pi} - \dfrac{\phi}{3F} \rb)
							\end{array}  \rb],\label{eq:terms_d_equal}
				\end{align}%
				where we have also displayed the equal-mass $m_L=m_N\equiv m$ limit of this result, as it will be needed later.\\[1ex]
\noindent\textbf{(e)} 
Finally, the terms that give the (tree-level) contribution to the scalar potential for the EM-neutral technipions are also contained in the terms $c_m \LambdaTC F_\pi^2  \Tr{M \Umat } e^{i \phi /3F}  + \text{h.c.} $ in \eqref{chiral_L}.
				The relevant terms are those with no EM-charged technipions [i.e., we again send $\piTC^\pm\rightarrow0$ in the expression for $\Umat$, before directly exponentiating and inserting the result into the relevant terms in \eqref{chiral_L}]:
				\begin{align}
				\Lag \supset  &+ 2F_\pi^2\LambdaTC c_m (m_L+m_N) \cos\lb( \tfrac{ \barpiTC }{F_\pi} \rb) \cos\lb( \tfrac{\kappaTC}{\sqrt{3}F_\pi} - \tfrac{\phi}{3F} \rb) \nonumber \\
							&+ 2F_\pi^2\LambdaTC c_m m_L \cos\lb( \tfrac{2\kappaTC}{\sqrt{3}F_\pi} + \tfrac{\phi}{3F}  \rb) \nonumber\\
							&+  2F_\pi^2\LambdaTC c_m (m_L-m_N) \frac{\omegaTC}{F_\pi} \sinc\lb( \tfrac{ \barpiTC }{F_\pi} \rb) 
									\sin\lb( \tfrac{\kappaTC}{\sqrt{3}F_\pi} - \tfrac{\phi}{3F}  \rb). \label{eq:terms_e}
				\end{align}

\section{Effective Potential and Electroweak Symmetry Breaking}
\label{sect:effective_potential}
Eqs.~(\ref{eq:Yukawas})--(\ref{eq:terms_e}) contain all the terms from the Chiral Lagrangian \eqref{chiral_L} which will be relevant for our further analysis. 
The immediate next step is to understand the effective potential for the EM-neutral scalars in more detail, in order to understand which of the EM-neutral scalars $\kappaTC$, $\omegaTC$, and $h$ obtain vevs as the relaxion field $\phi$ slow-rolls.%
\footnote{%
	Since the relaxion field $\phi$ is assumed to be slow-rolling down its potential until it stalls, we will always assume that the fields $\kappaTC$, $\omegaTC$, and $h$ take vevs such that the instantaneous minimum---with $\phi$ held fixed---of the effective potential is obtained.
	} %

\subsection{Effective Potential}
\label{sect:Veff}

After QCD quark confinement, and reading off the relevant terms from \eqref[s]{terms_a} and (\ref{eq:terms_e}), we have the following tree-level contributions to the effective potential (recalling that $\Lag \supset -V$):
\begin{align}
V_{\text{eff., tree}} &= V_\phi(\phi) + V_{\textsc{qcd}}(h,\kappaTC,\omegaTC,\phi) \nl
				- 2F_\pi^2\LambdaTC c_m (m_L+m_N) \cos\lb( \frac{ \barpiTC }{F_\pi} \rb) \cos\lb( \frac{\kappaTC}{\sqrt{3}F_\pi} - \frac{\phi}{3F} \rb) \nl
				- 2F_\pi^2\LambdaTC c_m m_L \cos\lb( \frac{2\kappaTC}{\sqrt{3}F_\pi} + \frac{\phi}{3F}  \rb) \nl
				-  2F_\pi^2\LambdaTC c_m (m_L-m_N) \frac{\omegaTC}{F_\pi} \sinc\lb( \frac{ \barpiTC }{F_\pi} \rb) 
						\sin\lb( \frac{\kappaTC}{\sqrt{3}F_\pi} - \frac{\phi}{3F}  \rb). \label{eq:Vefftree}
\end{align}

\subsubsection{QCD Contribution}
\label{sect:VQCD}
Although the detailed form of the QCD contribution to \eqref{Vefftree} depends on the exact details of QCD confinement and the SM quark masses, the only properties that will be relevant are that the potential is (a) periodic, and (b) proportional to the light SM quark masses, which owing to the Yukawa couplings \eqref{Yukawas} implies proportionality to $| h \sinc(\barpiTC/F_\pi) |$.
We will thus take the approximate form, based on \eqref[s]{Yukawas} and (\ref{eq:terms_a}), for the QCD contribution:
\begin{align}
V_{\textsc{qcd}}(h,\kappaTC,\omegaTC,\phi) \approx - \LambdaLow^3 \lb|\mkern1.5mu h \sinc\lb( \frac{\barpiTC}{F_\pi} \rb) \rb| \cos\lb[ \frac{\phi}{f} - \frac{2\sqrt{3}\kappaTC}{F_\pi} - \theta_{\textsc{qcd}} \rb]. \label{eq:VQCD}
\end{align}
In this normalization, $\LambdaLow^3 \, h \sinc\lb( \frac{\barpiTC}{F_\pi} \rb)  \sim m_\pi^2 f_\pi^2$.
For $h \sinc\lb( \frac{\barpiTC}{F_\pi} \rb) = v_{\textsc{sm}} \approx 246$\,GeV, we then have $\LambdaLow \approx 8.5$\,MeV.
Note also that the effective QCD $\theta$-angle is given by
\begin{align}
\theta_{\textsc{qcd}}^{\text{eff.}} &= \theta_{\textsc{qcd}} - \frac{\phi}{f} + \frac{2\sqrt{3}\kappaTC}{F_\pi} .
\end{align}
\subsubsection{Radiative Corrections}
\label{sect:radiative}
Important one-loop radiative corrections to the potential for the composite pNGB states arise from top quark loops owing to the large top Yukawa, $y_t \approx 1$.%
\footnote{%
	As mentioned in footnote \ref{ftnt:gauge_loops}, we ignore the subdominant gauge loops, as they do not qualitatively alter the dynamics of our model.
	} %
The impact of the top loops is in principle finite and calculable (e.g., on the lattice), but cannot be computed in the Chiral Lagrangian framework because the corrections are quadratically divergent, and are thus sensitive to physics at the cutoff scale $\LambdaTC$ of the low-energy effective description (which we cannot perturbatively match to the [known] UV completion as the latter is strongly coupled at the matching scale).
Nevertheless, we can estimate their size using NDA \cite{Georgi:1992dw}, and add to the effective potential a contribution
\begin{align}
V_{\text{eff.}} \supset - c_t N_c \frac{y_t^2}{16\pi^2} \LambdaTC^2 h^2 \sinc^2\lb( \frac{\barpiTC}{F_\pi} \rb), \label{eq:Vt}
\end{align}
where $c_t$ is an $\mathcal{O}(1)$ constant which is incalculable in perturbation theory, and $N_c=3$ is the number of QCD quark colors; note the dependence on $y_t^2 h^2 \sinc^2( \barpiTC/ F_\pi)$, which is proportional to $m_t^2$ per \eqref{Yukawas}. 
The sign here is crucially important, but is also not calculable within the Chiral Lagrangian framework for reasons similar to those advanced above about the size of the top loop correction; we nevertheless assume that the sign is negative, as is obtained from a na\"ive perturbative loop computation (see, e.g., \citeR{Galloway:2010bp} for discussion of this point).

\subsubsection{Equal-Mass Limit}
\label{sect:equal_mass}
For the remainder of the body of this paper we will work in the equal-mass limit $m_L = m_N$; this case is most amenable to straightforward analysis, and yields all the desired properties.
In \appref{unequal_masses}, we revisit the more complicated case of unequal masses, $m_L\neq m_N$.
The important conclusion from the analysis in \appref{unequal_masses} is that the equal-mass limit is not in any way special from the point of view of its physical properties: much the same qualitative picture of the EWSB dynamics is obtained for $m_L\neq m_N$ as for $m_L=m_N$, and we thus do not lose any qualitative features by making the simplifying equal-mass assumption.

Combining \eqref[s]{Vefftree}--(\ref{eq:Vt}), and setting $m_L = m_N\equiv m$, we obtain the following contributions to the one-loop effective potential:
\begin{align}
V_{\text{eff.}}&\supset V_\phi(\phi) - \LambdaLow^3 \lb| h \sinc\lb( \frac{\barpiTC}{F_\pi} \rb) \rb| \cos\lb[ \frac{\phi}{f} - \frac{2\sqrt{3}\kappaTC}{F_\pi} - \theta_{\textsc{qcd}} \rb] \nl
				- 2F_\pi^2\LambdaTC c_m m \lb[ 2 \cos\lb( \frac{ \barpiTC }{F_\pi} \rb) \cos\lb( \frac{\kappaTC}{\sqrt{3}F_\pi} - \frac{\phi}{3F} \rb)
										+ \cos\lb( \frac{2\kappaTC}{\sqrt{3}F_\pi} + \frac{\phi}{3F}  \rb) \rb] \nl
				- 2F_\pi^2 \LambdaTC c_m m\, \epsilon_t\, \frac{h^2}{F_\pi^2} \sinc^2\lb( \frac{\barpiTC}{F_\pi} \rb), \label{eq:Vfullequal}
\end{align}
where we have defined 
\begin{align}
\epsilon_t & \equiv \frac{c_t}{c_m} N_c\, \frac{y_t^2}{32\pi^2} \frac{\LambdaTC}{m} > 0. \label{eq:eps_defn}
\end{align}

\subsection{Electroweak-Symmetric Phase}
\label{sect:EWsym}
In the electroweak-symmetric (EW-symmetric) phase of the theory, it is straightforward to show that
\begin{align}
\hvev 		&=0, &
\kappaTCvev 	&=0, &
\omegaTCvev	&=0, \label{eq:EWsymvevs}
\end{align}
and, assuming slow-roll of the relaxion field $\phi$,
\begin{align}
\partial_t \phi \propto - \partial_\phi V\bigg|_{\hvev=\kappaTCvev=\omegaTCvev=0} &= - \partial_\phi V_\phi - 2 \frac{\LambdaTC F_\pi^2c_m m}{F} \sin\lb( \frac{\phi}{3F} \rb). \label{eq:phidot_EWsym}
\end{align}
We will return to a discussion of the rolling of the relaxion in \sectref{relaxion_potential}; for now let us focus on the other properties of this phase, for a fixed value of the relaxion field $\phi$.
It is straightforward to see from \eqref[s]{terms_b} and (\ref{eq:terms_c}) that $\kappaTC,\ \omegaTC,\ \piTC^\pm$, and $h$ have canonical kinetic terms, that there is no $\piTC^\pm$--$W_\mu^\pm$ kinetic mixing, and that the $W^\pm$ and $Z$ bosons are massless (as is of course required for this phase).
Moreover, ignoring in the squared-mass matrix%
\footnote{%
	Defined for the EM-neutral scalars as $M^2_{XY} \equiv \partial_X \partial_Y V_{\text{eff.}}|_{\hvev=\kappaTCvev=\omegaTCvev=0}$ with $X,Y\in\{h,\, \omegaTC,\, \kappaTC,\, \phi\}$; for the EM-charged scalars, one can simply read off the mass from \eqref{terms_d_equal}.
	} %
off-diagonal entries proportional to $F_\pi / F$ (which ratio is exponentially small) that mix $\phi$ with $\omegaTC$ and $\kappaTC$, that matrix is diagonal and the squared-masses of the scalars are 
\begin{align}
m^2_{\kappaTC} = m^2_{\omegaTC} = m^2_{\piTC^{\pm}} &= 4 c_m m \LambdaTC  \cos\lb( \frac{\phi}{3F} \rb),\\
\text{and} \qquad m^2_{h} &= 4 c_m m \LambdaTC \bigg[ \cos\lb( \frac{\phi}{3F} \rb) - \epsilon_t \bigg];
\label{eq:unbroken-spectrum}
\end{align}
since $\phi$ is still rolling in this phase, $m^2_{\phi} \equiv \partial_\phi^2 V|_{\hvev=\kappaTCvev=\omegaTCvev=0}$ has no physical interpretation. 

In exact parallel with the `cartoon' model of \sectref{cartoon}, we see from \eqref{unbroken-spectrum} that the EW-symmetric phase is thus stable so long as $\cos( \phi / 3F ) > \cos( \phicrit / 3F ) > 0$, where 
\begin{align}
\cos\lb( \frac{\phicrit}{3F} \rb) &\equiv \epsilon_t . \label{eq:phicritequal}
\end{align}

We emphasize the perhaps obvious point that the solution \eqref{EWsymvevs} exists independent of the values of any of the other parameters in the theory; this will be important to bear in mind when we discuss the evolution of the potential with changing $\phi$ in \sectref{EWSBsolutions}.

\subsection{Broken Phase}
\label{sect:EWbroken}
In the broken phase of the theory, we find that both $h$ and $\kappaTC$ obtain vevs, which are determined by the following relations:
\begin{align}
\cos\lb( \frac{ \hvev }{ F_\pi } \rb)	&= \frac{1}{\epsilon_t} \cos\lb[ \frac{ \phi }{3F} - \frac{1}{2} \arctan\lb( \frac{2\sin\lb(\frac{\phi}{3F} \rb) \lb[ \cos\lb(\frac{\phi}{3F} \rb) -  \epsilon_t  \rb]}{ \cos\lb(\frac{2\phi}{3F} \rb) + 2 \epsilon_t \cos\lb(\frac{\phi}{3F} \rb) } \rb) \rb], \label{eq:broken_cos_h}\\
\tan \lb( \frac{2\kappaTCvev}{\sqrt{3}F_\pi} \rb) 	&= \frac{2\sin\lb(\frac{\phi}{3F} \rb) \lb[ \cos\lb(\frac{\phi}{3F} \rb) -  \epsilon_t  \rb]}{ \cos\lb(\frac{2\phi}{3F} \rb) + 2 \epsilon_t \cos\lb(\frac{\phi}{3F} \rb) }\, , \label{eq:broken_tan_kappa}\\
\omegaTCvev	&=0. \label{eq:broken_omega}
\end{align}
Per \eqref{terms_b}, since $\omegaTCvev=0$, $h$ and $\kappaTC$ have canonical kinetic terms, and there is no $h$--$\mkern1mu\omegaTC$ kinetic mixing.
The field $\omegaTC$ does not however have a canonical kinetic term: $\Lag \supset \frac{1}{2} (\partial_\mu \omegaTC )^2 \sinc^2(\hvev/F_\pi)$; we will return to this point below.

\subsubsection{A Deeper Investigation of the EWSB Minimum}
\label{sect:EWSBsolutions}
Prior to any further examination of the properties of this phase (masses, etc.), it is worthwhile to examine the results \eqref[s]{broken_cos_h} and (\ref{eq:broken_tan_kappa}) in more detail in the vicinity of $\phi = \phicrit$, as this clarifies the physical situation tremendously.
Suppose that $\phi = \phicrit + 3F \cdot\delta$, where $|\delta| \ll 1$. 
Expanding \eqref[s]{broken_cos_h} and (\ref{eq:broken_tan_kappa}) in powers of $\delta$, we find
\begin{align}
\cos\lb( \frac{\hvev}{F_\pi} \rb) &= 1 - \frac{3\epsilon_t \delta \sqrt{1-\epsilon_t^2}}{4\epsilon_t^2-1} + \mathcal{O}(\delta^2) \label{eq:cos_h_delta_exp}\\
\tan\lb( \frac{2 \kappaTCvev}{\sqrt{3}F_\pi} \rb) &= - 2\delta \frac{1-\epsilon_t^2}{4\epsilon_t^2-1} + \mathcal{O}(\delta^2) .\label{eq:tan_kappa_delta_exp}
\end{align}
Clearly, \eqref{cos_h_delta_exp} has a real solution for $\hvev$ only if
\begin{align}
\frac{\delta}{4\epsilon_t^2-1} &\geq 0 &
\text{and}&&
&0<\epsilon_t\leq 1,
\end{align}
where we used that $\epsilon_t>0$ [\eqref{eps_defn}].
There are two regimes that satisfy these constraints: 
(a) $\delta > 0$ and $1/2< \epsilon_t\leq1$, and 
(b) $\delta < 0$ and $0<\epsilon_t<1/2$.
In either case (a) or (b), \eqref[s]{cos_h_delta_exp} and (\ref{eq:tan_kappa_delta_exp}) have two approximate solutions in the vicinity of $\hvev=\kappaTCvev=0$ (cf.~\sectref{cartoon}):
\begin{align}
\frac{\hvev}{F_\pi} &= \pm \lb[ 6 \delta \frac{ \epsilon_t \sqrt{1-\epsilon_t^2} } {4\epsilon_t^2 - 1 } + \mathcal{O}(\delta^2) \rb]^{\frac{1}{2}} &
\text{and}&&
\frac{\kappaTCvev}{F_\pi} &= -\sqrt{3} \delta \frac{1-\epsilon_t^2}{4\epsilon_t^2-1} + \mathcal{O}(\delta^2).\label{eq:exp_soln}
\end{align}
Note also that $\partial_h^2 V_{\text{eff.}}|_{\hvev=\kappaTCvev=\omegaTCvev=0} = - 4c_m m\delta\LambdaTC\sqrt{1-\epsilon_t^2} + \mathcal{O}(\delta^2)$.

Suppose then that $0<\epsilon_t<1/2$. 
As we noted above, the solution at $\hvev=\kappaTCvev=0$ also exists for any value of the parameters.
If additionally $\delta<0$, then both solutions \eqref{exp_soln} exist [case (b)], for a total of three solutions for $\hvev$ and $\kappaTCvev$ in the vicinity of $\hvev=\kappaTCvev=0$ (and $\omegaTCvev=0$).
These solutions merge as $\delta \rightarrow 0$ from below.
Once $\delta > 0$, the solutions \eqref{exp_soln} no longer exist, leaving only the solution $\hvev=\kappaTCvev=0$ in the vicinity of $\hvev=\kappaTCvev=0$.
Moreover, since $\partial_h^2 V_{\text{eff.}}|_{\hvev=\kappaTCvev=\omegaTCvev=0}$ is positive for $\delta <0$ and negative for $\delta >0$, the solution $\hvev=\kappaTCvev=0$ is stable for $\delta<0$ and unstable for $\delta >0$.
Further analysis shows that the solutions \eqref{exp_soln} are unstable if $0<\epsilon_t<1/2$.
Therefore, we find that for $0<\epsilon_t<1/2$, the model exhibits a so-called \emph{subcritical pitchfork bifurcation} at $\delta =0$ (see, e.g., \citeR{Strogatz:2014ndc}).
As $\delta$ approaches zero from below, a stable solution exists at the origin in field space; two additional solutions---both unstable---exist nearby in field space.
As $\delta$ gets nearer zero, the two unstable solutions approach the stable one, and they merge at $\delta =0$ (i.e., $\phi = \phicrit$).
Once $\delta >0$, there are no stable solutions left in the vicinity of the original stable solution.
Indeed, in this case, once $\delta >0$, the nearest minimum of the potential occurs for $|\hvev/F_\pi| \approx \pi$ and $|\kappaTCvev/F_\pi| \sim \mathcal{O}(1)$.
This is clearly the incorrect behavior for a reasonable EWSB transition.

Consider then the other case, $1/2<\epsilon_t\leq1$. 
For $\delta < 0$, the solutions \eqref{exp_soln} do not exist, and the only solution in the vicinity of $\hvev=\kappaTCvev=0$ is $\hvev=\kappaTCvev=0$ itself.
If $\delta>0$, then both solutions \eqref{exp_soln} exist [case (a)], for a total of three solutions for $\hvev$ and $\kappaTCvev$ in the vicinity of $\hvev=\kappaTCvev=0$ (and $\omegaTCvev=0$).
These solutions separate from each other as $\delta$ grows more positive.
Moreover, the solution $\hvev=\kappaTCvev=0$ is stable for $\delta<0$ and unstable for $\delta >0$.
Further analysis shows that the solutions \eqref{exp_soln} are stable if $1/2<\epsilon_t\leq 1$.
Therefore, we find that for $1/2<\epsilon_t\leq 1$, the model exhibits a so-called \emph{supercritical pitchfork bifurcation} at $\delta =0$.
For $\delta<0$, a stable solution exists at the origin in field space, and no other solutions exist nearby.
Once $\delta>0$, two new stable solutions appear in the vicinity of the origin in field space, and the solution at the origin becomes unstable.
The system will relax to one or the other of these new stable solutions, which slowly separate from $\hvev=\kappaTCvev=0$ as $\delta$ becomes increasingly positive.
This is the behavior we need, and closely mirrors the behavior of the `cartoon' model of \sectref{cartoon}; see \figref{cartoon_plots}.

An alternative analysis is also instructive. 
Consider $V_{\text{eff.}}$ evaluated at $\omegaTCvev=0$ and with $\kappaTCvev$ fixed at the solution \eqref{broken_tan_kappa}.
Expanding $V_{\text{eff.}}$ in powers of $h$ for fixed $\delta$, we find
\begin{align}
\frac{V_\text{eff.}\big|_{\substack{\kappaTCvev \text{ from \eqref{broken_tan_kappa}}\\\omegaTCvev=0}}}{ c_m \LambdaTC F_\pi^2 m} &\approx \lb( -6 \epsilon_t + 6 \delta \sqrt{1-\epsilon_t^2} \rb) - 2 \delta \sqrt{1-\epsilon_t^2} \lb( \frac{h^2}{F_\pi^2}  \rb)+ \frac{ 4\epsilon_t^2 - 1 }{ 6 \epsilon_t } \lb( \frac{h^4}{F_\pi^4} \rb) + \cdots.
\end{align}
Therefore, for $1/2<\epsilon_t\leq1$, the quartic coupling is positive, with the $h$ squared-mass parameter positive for $\delta<0$ and negative for $\delta>0$, exactly as required to obtain a slow separation of the EWSB minimum from the EW-symmetric minimum as $\delta$ increases through zero.
For $0<\epsilon_t<1/2$, the $h$ squared-mass parameter is still positive for $\delta<0$ and negative for $\delta>0$; however, the quartic coupling is negative in both cases.
Thus, the moment the $h$ squared-mass parameter runs negative as $\delta$ increases through zero, the $h$ field rolls off to a large field value.%
\footnote{%
	Strictly speaking, the expansion for fourth-order in $h$ does not allow one to make this latter conclusion because, e.g., the sixth-order term could stabilize a nearby minimum.
	Our conclusion here is nevertheless correct, and is based on evaluation of the full (unexpanded) potential.
	} %

A parameter space restriction is thus required [cf.~\eqref{CartoonMrestr}]:
\begin{align}
1/2 &< \epsilon_t \leq 1 & \Leftrightarrow && \dfrac{c_tN_cy_t^2}{32\pi^2c_m} &\leq \dfrac{m}{\LambdaTC} < \dfrac{c_t N_c y_t^2}{16\pi^2c_m}. \label{eq:eps_constraint}
\end{align}
The technifermion masses may thus be no larger than a loop factor smaller than $\LambdaTC$; such a choice is, however, technically natural.
Note also that this implies a very mild restriction on the value of $\phicrit$: $0\leq (\phicrit / F) < \pi$.

\subsubsection{Properties of the Broken Phase}
\label{sect:EWSBproperties}
Again, we will return to a discussion of the rolling of the relaxion in \sectref{relaxion_potential}; for now let us focus again on the other properties of this phase for a fixed value of the relaxion field $\phi$.

Canonically normalizing the $\omegaTC$ field by sending $\omegaTC \rightarrow \omegaTC / \sinc(\hvev/F_\pi)$, and evaluating the broken-phase scalar squared-mass matrix
\begin{align}
M_{XY}^2 &\equiv \lb. \frac{\partial^2 V_{\text{eff.}}}{\partial X \partial Y} \rb|_{\substack{  \hvev \text{ from \eqref{broken_cos_h}} \\ \kappaTCvev \text{ from \eqref{broken_tan_kappa}} \\  \omegaTCvev=0 \\ \phi=\phicrit+3F\delta } }  &\text{with}&&
X,Y \in \{  h,\, \kappaTC,\, \omegaTC,\, \phi \},
\end{align}
ignoring all terms suppressed by one of more powers of the (exponentially small) ratio $F_\pi/F$, and ignoring (small) QCD corrections everywhere except in the squared-mass of the relaxion field, we find that the squared-masses of the scalars%
\footnote{%
		The $h$ and $\kappaTC$ fields mix; the physical Higgs is mostly $h$; the `physical $\kappaTC$' is the mostly $\kappaTC$ state.
		Note that if we did not ignore the terms $\sim F_\pi/F$ in the mass matrix, the $\phi$ would also mix with the $h$ and $\kappaTC$, with a mixing angle $\sim F_\pi/F \ll 1$.
		This mixing of the CP-even and CP-odd scalars is allowed since the $\phi$ vev is non-zero, which breaks CP.
	} %
are
\begin{align}
m_{\text{phys.~Higgs}}^2 &= \frac{2}{3}c_m m\LambdaTC \nl \times \lb[\begin{array}{l}
															2 \cos\lb( \frac{2\kappaTCvev}{\sqrt{3}F_\pi} + \frac{\phi}{3F} \rb) + 4 \cos\lb( \frac{\hvev}{F_\pi} \rb) \cos\lb( \frac{\kappaTCvev}{\sqrt{3}F_\pi} - \frac{\phi}{3F} \rb) - 3 \epsilon_t \cos\lb( \frac{2\hvev}{F_\pi} \rb) \\[2ex]
															- \lb(\begin{array}{l}
																	 \lb[\begin{array}{l} 
																	 	2 \cos\lb( \frac{2\kappaTCvev}{\sqrt{3}F_\pi} + \frac{\phi}{3F} \rb) - 2 \cos\lb( \frac{\hvev}{F_\pi} \rb) \cos\lb( \frac{\kappaTCvev}{\sqrt{3}F_\pi} - \frac{\phi}{3F} \rb) \\[1ex]
																		+ \, 3\, \epsilon_t \cos\lb( \frac{2\hvev}{F_\pi} \rb) 
																		\end{array} \rb]^2 \\[3ex]
																	 + 12 \sin^2\lb( \frac{\hvev}{F_\pi} \rb) \sin^2\lb( \frac{\kappaTCvev}{\sqrt{3}F_\pi} - \frac{\phi}{3F} \rb)
																 \end{array} \rb)^{\frac{1}{2}}	 
															\end{array}\rb] \\
					&= \lb( 8c_m m\LambdaTC \sqrt{1-\epsilon_t^2} \rb) \delta + \mathcal{O}(\delta^2) , \\[3ex]
m_{\text{phys.}~\kappaTC }^2 &= \frac{2}{3}c_m m\LambdaTC \nl \times \lb[\begin{array}{l}
															2 \cos\lb( \frac{2\kappaTCvev}{\sqrt{3}F_\pi} + \frac{\phi}{3F} \rb) + 4 \cos\lb( \frac{\hvev}{F_\pi} \rb) \cos\lb( \frac{\kappaTCvev}{\sqrt{3}F_\pi} - \frac{\phi}{3F} \rb) - 3 \epsilon_t \cos\lb( \frac{2\hvev}{F_\pi} \rb) \\[2ex]
															+ \lb(\begin{array}{l}
																	 \lb[\begin{array}{l} 
																	 	2 \cos\lb( \frac{2\kappaTCvev}{\sqrt{3}F_\pi} + \frac{\phi}{3F} \rb) - 2 \cos\lb( \frac{\hvev}{F_\pi} \rb) \cos\lb( \frac{\kappaTCvev}{\sqrt{3}F_\pi} - \frac{\phi}{3F} \rb) \\[1ex]
																		+ \, 3\, \epsilon_t \cos\lb( \frac{2\hvev}{F_\pi} \rb) 
																		\end{array} \rb]^2 \\[3ex]
																	 + 12 \sin^2\lb( \frac{\hvev}{F_\pi} \rb) \sin^2\lb( \frac{\kappaTCvev}{\sqrt{3}F_\pi} - \frac{\phi}{3F} \rb)
																 \end{array} \rb)^{\frac{1}{2}}	 
															\end{array}\rb] \\
						&= 4c_m m\LambdaTC \epsilon_t  \lb[ 1 -  4 \delta  \frac{(2\epsilon_t^2-1)  \sqrt{1-\epsilon_t^2}}{\epsilon_t(4\epsilon_t^2-1)} + \mathcal{O}(\delta^2) \rb], \\[3ex]
m_{ \omegaTC }^2 &= 4c_m m\LambdaTC \frac{ \cos\lb( \frac{\kappaTCvev}{\sqrt{3} F_\pi} - \frac{\phi}{3F} \rb) + \epsilon_t \lb[ \sinc\lb(\frac{\hvev}{F_\pi}\rb) - \cos\lb(\frac{\hvev}{F_\pi}\rb) \rb]}{\sinc\lb(\frac{\hvev}{F_\pi} \rb)}\\
			&= 4c_m m\LambdaTC \epsilon_t \qquad (\text{exactly}),\label{eq:mwsq2}
\end{align}
and
\begin{align}
m_{\phi}^2	&\simeq \lb[ \frac{\partial^2V_\phi(\phi)}{\partial\phi^2}  + \frac{ \LambdaLow^3 F_\pi^2}{f^2} \sin\lb( \frac{\hvev}{F_\pi} \rb) \cos\lb( \frac{\phi}{f}  - \frac{2\sqrt{3}\kappaTCvev}{F_\pi} - \theta_{\textsc{qcd}} \rb) \rb]_{\substack{  \hvev \text{ from \eqref{broken_cos_h}} \\ \kappaTCvev \text{ from \eqref{broken_tan_kappa}} \\ \phi=\phicrit+3F\delta } },
\end{align}
where in the first two results we have used \eqref[s]{broken_cos_h} and (\ref{eq:broken_tan_kappa}), and $\phi = \phicrit + 3F\cdot\mkern1.5mu\delta$, have expanded in powers of $\delta$, and have kept only leading terms, as we expect that the QCD barriers will stall the relaxion in the vicinity of $\phicrit$ (i.e., at $0<\delta \ll 1$); see \sectref{relaxion_potential}.
The expression for $m_{ \omegaTC }^2$ at \eqref{mwsq2} is exact owing to the relation $ \cos\lb( \frac{\kappaTCvev}{\sqrt{3} F_\pi} - \frac{\phi}{3F} \rb) =\epsilon_t \cos\lb(\frac{\hvev}{F_\pi}\rb)$ in the broken phase, which can easily be verified using \eqref[s]{broken_cos_h} and (\ref{eq:broken_tan_kappa}).
Note also that $m_{\phi}^2$ is only interpretable as the present-day squared-mass of the relaxion field once it has stopped rolling after the post-inflation slope-drop discussed in \sectref{cartoonRelaxionV} has occurred; see \sectref{relaxion_potential}. 

It is also straightforward to read off the $W$ and $Z$-boson squared-masses from \eqref{terms_c} (note that we had to keep the $\mathcal{O}(\delta^2)$ terms in the expansions \eqref{exp_soln} in order to obtain the $\mathcal{O}(\delta^2)$ terms here correctly):
\begin{align}
m_W^2 &= \frac{g_2^2}{2} F_\pi^2 \lb[ 1 - \cos\lb( \frac{\hvev}{F_\pi} \rb) \cos\lb( \frac{\sqrt{3}\kappaTCvev}{F_\pi} \rb) \rb] \\
		&= \frac{3}{2} g_2^2 F_\pi^2\, \delta \lb[ \frac{\epsilon_t\sqrt{1-\epsilon_t^2}}{4\epsilon_t^2-1} + \frac{3}{2} \delta \frac{2-3\epsilon_t^2+2\epsilon_t^4}{(4\epsilon_t^4-1)^2} + \mathcal{O}(\delta^2) \rb],\\
m_Z^2 &= \frac{g_1^2+g_2^2}{4} F_\pi^2 \sin^2\lb( \frac{\hvev}{F_\pi} \rb) \\
		&= \frac{3}{2} (g_1^2+g_2^2) F_\pi^2\, \delta \lb[ \frac{\epsilon_t\sqrt{1-\epsilon_t^2}}{4\epsilon_t^2-1} + \frac{3}{2} \delta \frac{1-2\epsilon_t^2+2\epsilon_t^4}{(4\epsilon_t^4-1)^2} + \mathcal{O}(\delta^2) \rb],
\end{align}
which imply a tree-level contribution to the $T$ parameter of 
\begin{align}
\alpha_e T &\equiv \frac{1}{m_W^2} \lb[ \Pi_{W^+W^-}(0) - c_{\textsc{w}}^2\, \Pi_{ZZ}(0) \rb] \\
&= 1 - \frac{g_2^2}{g_1^2+g_2^2} \frac{m_Z^2}{m_W^2}\\
&= 1 - \frac{1}{2} \frac{\sin^2\lb( \frac{\hvev}{F_\pi} \rb)}{1 - \cos\lb( \frac{\hvev}{F_\pi} \rb) \cos\lb( \frac{\sqrt{3}\kappaTCvev}{F_\pi} \rb)}  \\
&= \frac{3}{2} \delta \frac{\sqrt{1-\epsilon_t^2}}{\epsilon_t(4\epsilon_t^2-1)} + \mathcal{O}(\delta^2) ,
\end{align}
where $\alpha_e$ is measured at the $Z$-pole, and $c_{\textsc{w}}\equiv g_2 / \sqrt{g_1^2+g_2^2}$ is the cosine of the weak mixing angle.
Note that $\alpha_e T \propto \xi$ per \eqref[s]{xi} and (\ref{eq:exp_soln}) (cf.~eq.~(16) of \citeR{Giudice:2007fh}).

Owing to (a) the fact that $\piTC^\pm$ do not have canonical kinetic terms, and (b) the kinetic mixing of the $\piTC^\pm$ with the $W_\mu^\pm$, two manipulations are required before the $\piTC^\pm$ masses can be read off from \eqref[s]{terms_c} and (\ref{eq:terms_d_equal}): (1) we send $W^\pm_\mu \rightarrow W^\pm_\mu \pm i\alpha \partial_\mu \piTC^\pm$, with $\alpha$ chosen to eliminate the kinetic mixing term in \eqref{terms_c}:
\begin{align}
\alpha &= \frac{2}{g_2} \frac{ \hvev \sin\lb( \frac{\hvev}{F_\pi} \rb) \sin\lb( \frac{\sqrt{3}\kappaTCvev}{F_\pi} \rb) -\sqrt{3}\kappaTCvev \lb[ 1 -  \cos\lb( \frac{\hvev}{F_\pi} \rb) \cos\lb( \frac{\sqrt{3}\kappaTCvev}{F_\pi} \rb) \rb]}{ \Big( \hvev^2 - 3\kappaTCvev^2 \Big) \lb[ 1 -  \cos\lb( \frac{\hvev}{F_\pi} \rb) \cos\lb( \frac{\sqrt{3}\kappaTCvev}{F_\pi} \rb) \rb] } ,
\end{align}
which does not impact the $W$ mass but does modify the $\piTC^\pm$ kinetic term (and of course induces couplings to the $\piTC^\pm$ for all fields that couple to the $W_\mu^\pm$); and (2) we rescale the $\piTC^\pm$ fields to achieve a canonical kinetic term:
\begin{align}
\piTC^\pm \rightarrow \piTC^\pm \times \frac{ \hvev^2 - 3\kappaTCvev^2 }{\sqrt{2} F_\pi \hvev } \frac{ \sqrt{  1 -  \cos\lb( \frac{\hvev}{F_\pi} \rb) \cos\lb( \frac{\sqrt{3}\kappaTCvev}{F_\pi} \rb) } }{ \cos\lb( \frac{\sqrt{3}\kappaTCvev}{F_\pi} \rb)  - \cos\lb( \frac{\hvev}{F_\pi} \rb) }.
\end{align}
The squared-mass of the $\piTC^\pm$ can then be read off from the rescaled \eqref{terms_d_equal} as
\begin{align}
m_{\piTC^\pm}^2 &= 2c_m m \LambdaTC  \frac{ \hvev^2-3\kappaTCvev^2}{\hvev^2} \frac{ 1 -  \cos\lb( \frac{\hvev}{F_\pi} \rb) \cos\lb( \frac{\sqrt{3}\kappaTCvev}{F_\pi} \rb) }{ \lb[ \cos\lb( \frac{\hvev}{F_\pi} \rb) - \cos\lb( \frac{\sqrt{3}\kappaTCvev}{F_\pi} \rb) \rb]^2	}
\nl  \times \lb[ \begin{array}{l} 
			 \dfrac{\hvev}{F_\pi} \sin\lb( \dfrac{\hvev}{F_\pi} \rb) \cos\lb( \dfrac{\kappaTCvev}{\sqrt{3}F_\pi} - \dfrac{\phi}{3F} \rb)	\\[2ex]
			 -  \dfrac{\sqrt{3}\kappaTCvev}{F_\pi} \lb[   \cos\lb( \dfrac{\hvev}{F_\pi} \rb) \sin\lb( \dfrac{\kappaTCvev}{\sqrt{3}F_\pi} - \dfrac{\phi}{3F} \rb) + \sin\lb( \dfrac{2\kappaTCvev}{\sqrt{3}F_\pi} + \dfrac{\phi}{3F} \rb)  \rb]  
			 \end{array} \rb]   \\[2ex]
		&= 2c_m m \LambdaTC \epsilon_t \frac{ \hvev^2-3\kappaTCvev^2}{F_\pi^2} \frac{ 1 -  \cos\lb( \frac{\hvev}{F_\pi} \rb) \cos\lb( \frac{\sqrt{3}\kappaTCvev}{F_\pi} \rb) }{ \lb[ \cos\lb( \frac{\hvev}{F_\pi} \rb) - \cos\lb( \frac{\sqrt{3}\kappaTCvev}{F_\pi} \rb) \rb]^2	} \sinc\lb( \dfrac{\hvev}{F_\pi} \rb) \cos\lb( \dfrac{\hvev}{F_\pi} \rb) \label{eq:simplifyintermediate} \\ 
&= 4c_m m\LambdaTC \epsilon_t  \lb[ 1 -  \frac{1}{2} \delta  \frac{(13\epsilon_t^2-6)  \sqrt{1-\epsilon_t^2}}{\epsilon_t(4\epsilon_t^2-1)} + \mathcal{O}(\delta^2) \rb],\label{eq:mchargedexpdelta}
\end{align}
where at \eqref{simplifyintermediate} we have used the exact relations $ \cos\lb( \tfrac{\kappaTCvev}{\sqrt{3}F_\pi} - \tfrac{\phi}{3F} \rb)=\epsilon_t\cos\lb( \tfrac{\hvev}{F_\pi} \rb)$ and $\cos\lb( \tfrac{\hvev}{F_\pi} \rb) \sin\lb( \tfrac{\kappaTCvev}{\sqrt{3}F_\pi} - \tfrac{\phi}{3F} \rb) + \sin\lb( \tfrac{2\kappaTCvev}{\sqrt{3}F_\pi} + \tfrac{\phi}{3F} \rb) =0$, which follow from \eqref[s]{broken_cos_h} and (\ref{eq:broken_tan_kappa}). 
At \eqref{mchargedexpdelta}, we have used \eqref[s]{broken_cos_h} and (\ref{eq:broken_tan_kappa}), and have expanded in powers of $\delta$.

\subsection{Summary}
\label{sect:EWsummary}
In summary, we have found that for the parameter range $1/2<\epsilon_t\leq 1$, the theory has a stable EW-symmetric vacuum solution $\hvev=\kappaTCvev=\omegaTCvev=0$ while $\cos(\phi/3F) > \cos(\phicrit/3F) \equiv \epsilon_t$.
This vacuum solution destabilizes if $\phi$ is larger than $\phicrit$, and if $\phi=\phicrit+3F\delta$ (with $0<\delta \ll 1$), we find a Higgs-vev $\hvev \propto \sqrt{\delta} F_\pi$, a $\kappaTC$-vev $\kappaTCvev \propto \delta F_\pi$, a light physical Higgs mass ($m_h^2 \propto m \LambdaTC  \delta$), light $W$ and $Z$ masses ($m_{W,Z}^2 \propto F_\pi^2 \delta$), and four heavy states ($M^2 \propto m \LambdaTC$)---two of these states are EM-neutral and two are charged.
A tree-level $T$ parameter $\alpha_e T \propto \delta$ is generated.

\section{Relaxion Potential}
\label{sect:relaxion_potential}
In order to exploit the observations summarized in \sectref{EWsummary}, we desire to have the relaxion field initially in the range $\phi \in [0,\phicrit)$, and have the field slow-roll out to larger values of $\phi$ over time: $\partial_t \phi >0$.
In exactly the same fashion as discussed in \sectref{cartoon}, this will trigger dynamical EWSB as $\phi$ crosses $\phicrit$, giving rise to increasingly large QCD barriers to the rolling, which will stall the relaxion shortly after it crosses $\phicrit$, while $0<\delta \ll 1$, per the mechanism of \citeR{Graham:2015cka}.

To this end, examine again \eqref{phidot_EWsym}, which gives the gradient of the potential with respect to $\phi$ in the EW-symmetric phase.
As the second term in \eqref{phidot_EWsym} is negative on $\phi \in [0,\phicrit)$, the first term must be made positive to obtain the correct rolling direction.

Following \citeR{Graham:2015cka} and our discussion in \sectref{cartoonRelaxionV}, we add a linear term for the $\phi$, which explicitly breaks the residual discrete $\phi$ shift symmetry. 
As in \sectref{cartoon}, we write this term as follows (the additional factor of 2 here compared to $V_\phi(\phi)$ in \sectref{cartoonRelaxionV} is merely a convenient rescaling of $\gamma$):
\begin{align}
V_\phi(\phi) &= - \, \gamma \,   \frac{2\LambdaTC F_\pi^2c_m m }{F}\, \phi , \label{eq:linear_potential}
\end{align}
where the free numerical prefactor $\gamma = \gamma(\sigma)$ is again assumed to take a value $\gamma_i \sim 10^{10}$ during inflation (see \sectref{cartoonRelaxionV}). 
This implies that, during inflation, [cf.~\eqref{dVphiCartoon}]
\begin{align}
\partial_t \phi \propto - \partial_\phi V\big|_{\text{EW-symmetric}} &= 2\frac{\LambdaTC F_\pi^2c_m m}{F} \lb[ \gamma_i - \sin\lb( \frac{\phi}{3F} \rb) \rb]. \label{eq:phidot_EWsym_linear}
\end{align}
Taking $\gamma_i \gg 1$ certainly guarantees that $\partial_t\phi >0$ in the EW-symmetric phase.

In the broken phase, we require the relaxion to stop rolling once the QCD barriers become sufficiently large. 
The stopping condition $\partial_\phi V_{\text{eff.}}= 0$ is
\begin{align}
&\frac{\LambdaLow^3}{f} h \sinc\lb( \frac{\barpiTC}{F_\pi} \rb) \sin\lb[ \frac{\phi}{f} - \frac{2\sqrt{3}\kappaTCvev}{F_\pi} - \theta_{\textsc{qcd}} \rb] \nl
\approx  \frac{2\LambdaTC F_\pi^2 c_m m}{F} \lb[ \gamma + \frac{2}{3} \cos\lb( \frac{\barpiTC}{F_\pi} \rb) \sin\lb( \frac{\kappaTCvev}{\sqrt{3}F_\pi} - \frac{\phi}{3F} \rb) - \frac{1}{3} \sin\lb( \frac{2\kappaTCvev}{\sqrt{3}F_\pi} + \frac{\phi}{3F} \rb) \rb].\label{eq:stopping_cond}
\end{align}
Here, during inflation $\gamma = \gamma_i$ for the initial stalling of the relaxion, but $\gamma \rightarrow 0$ when the post-inflation slope-drop occurs and $\phi$ settles to its new minimum.
Evaluating this at $\phi = \phicrit + 3F\delta$ in the broken phase as defined by \eqref[s]{broken_cos_h}--(\ref{eq:broken_omega}), and expanding in powers of $\delta$ everywhere \emph{except} for the $\phi/f$ term in the argument of the sine term on the LHS of \eqref{stopping_cond}, we find during inflation that the relaxion stalls when
\begin{align}
&\frac{2 \LambdaTC F_\pi^2 c_m m }{F} \lb[ \gamma_i - \sqrt{1-\epsilon_t^2} \rb] \nl
\approx \sqrt{\delta} \frac{\sqrt{6}F_\pi\LambdaLow^3}{f} \frac{\lb(1-\epsilon_t^2\rb)^{\frac{1}{4}} \sqrt{\epsilon_t}}{\sqrt{4\epsilon_t^2-1}} \sin\lb[ \frac{3F}{f} \Big(\mkern-2mu\arccos(\epsilon_t)  + \delta \Big) - \theta_{\textsc{qcd}} \rb] + \mathcal{O}(\delta) .\label{eq:stopping_cond2}
\end{align}
This cannot be solved for $\delta$ exactly in closed form, and no small-$\delta$ expansion of the argument of the sine term is possible owing to the large $\delta$-prefactor proportional to $F/f \gg 1$.
However, we can make progress by assuming that the sine factor on the RHS of \eqref{stopping_cond2} is equal to 1 to obtain the approximate solution for the relaxion stalling during inflation:
\begin{align}
\delta \approx \frac{2}{3} \frac{ f^2 F_\pi^2 c_m^2 m^2 \LambdaTC^2 }{ F^2 \LambdaLow^6 } \frac{ 4\epsilon_t^2-1}{\epsilon_t \sqrt{1-\epsilon_t^2}} \lb[ \gamma_i - \sqrt{1-\epsilon_t^2} \rb]^2 \approx \frac{2}{3} \frac{ f^2 F_\pi^2 c_m^2 m^2 \LambdaTC^2 }{ F^2 \LambdaLow^6 } \frac{ 4\epsilon_t^2-1}{\epsilon_t \sqrt{1-\epsilon_t^2}} \mkern2mu \gamma_i^2, \label{eq:delta_approx}
\end{align}
where in the latter approximate equality we have used $\gamma_i \gg \sqrt{1-\epsilon_t^2} \sim \mathcal{O}(1)$.
Self-consistency of the expansion effectively demands that, up to $\mathcal{O}(1)$ factors,\linebreak $F \gg \gamma_i f  F_\pi c_m m \LambdaTC /  \LambdaLow^3$.
Since we have that $\LambdaLow^3 \hvev \sim \LambdaQCD^4$ [cf.~\eqref[s]{cartoonVQCD} and (\ref{eq:VQCD})], so that $\LambdaLow^{-3} \sim \sqrt{\xi} F_\pi \LambdaQCD^{-4}$, we require here that $F \gg \sqrt{\xi} \gamma_i f  F_\pi^2 c_m m \LambdaTC /  \LambdaQCD^4$; this should be compared to \eqref{slopeMatchingCartoon}, which indicates that in the cartoon model we had $F \sim \gamma_i f  F_\pi^2 c_m m \LambdaTC /  \Lambda^4$: the parametrics examined for the cartoon model thus correctly imply $\xi \ll 1$.

By the intermediate value theorem, the actual solution of \eqref{stopping_cond2} must occur for a value of $\delta$ no more than $\Delta\delta = \frac{2\pi}{3} \frac{f}{F} \ll 1$ greater than the approximate solution \eqref{delta_approx}.
Although it is not obvious parametrically, the shift $\Delta\delta$ turns out to be numerically small compared to $\delta$.
Eq.~(\ref{eq:delta_approx}) is thus a sufficient approximate solution to be used everywhere \emph{except} when an expression depends on a (co)sine factor with an argument containing a contribution proportional to $\phi/f$.
The only other place that this occurs is in the relaxion squared-mass [note that terms proportional to $ (F_\pi/F)^2 \ll1$ have been ignored here]:
\begin{align}
m_\phi^2 & \approx  \frac{\LambdaLow^3F_\pi}{f^2}  \sin\lb( \frac{\hvev}{F_\pi} \rb) \cos\lb[  \theta_{\textsc{qcd}} - \frac{\phi}{f} + \frac{2\sqrt{3}\kappaTCvev}{F_\pi} \rb]_{\substack{  \hvev \text{ from \eqref{broken_cos_h}} \\\kappaTCvev \text{ from \eqref{broken_tan_kappa}} }}. \label{eq:relaxion_mass1}
\end{align}
However, for precisely the reason that this expression in sensitive to shifts of size $\Delta \delta \sim f/F$ (i.e., $\Delta \phi \sim f$), the relaxion mass will change after the post-inflation slope-drop as the relaxion rolls a distance of order $|\Delta \phi| \sim f$ to its new settling point (see \sectref{cartoonRelaxionV}); on the other hand, all the other estimates we have made will not be significantly impacted by this small change in $\phi$ due to the slope drop.
Since the argument of the cosine factor in \eqref{relaxion_mass1} is just $\theta^{\text{eff.}}_{\textsc{qcd}}$ [which from comparing \eqref{stopping_cond} with $\gamma = \gamma_i$ and with $\gamma \rightarrow 0$, is easily seen to be a factor of $\gamma_i\sim10^{10}$ smaller than its $\mathcal{O}(1)$ value at initial stalling, cf.~\eqref{thetaQCDestCartoon}], it follows that the post-slope-drop relaxion mass can be estimated by setting the cosine factor to 1, and using \eqref[s]{exp_soln} and (\ref{eq:delta_approx}) in the $\sin(\hvev/F_\pi)$ term:
\begin{align}
m_\phi^2 & \approx  \frac{2\LambdaTC F_\pi^2 c_m m}{ fF}\mkern2mu \gamma_i \approx \frac{\LambdaLow^3\hvev}{f^2}  \approx \frac{m_\pi^2 f_\pi^2}{f^2}  \qquad\qquad (\text{post slope-drop}), \label{eq:phimass} 
\end{align}
where we have used the stopping relation estimated from \eqref{stopping_cond} in the second step above, and the estimates $\LambdaLow^3\hvev \sim \Lambda^4 \sim m_\pi^2 f_\pi^2$ in the third step [see the discussions just below \eqref[s]{slopeMatchingCartoongamma1} and (\ref{eq:delta_approx})].
Note that the expression appearing on the RHS of the first approximate equality in \eqref{phimass} is $F/f \gg 1$ larger than the $\mathcal{O}(F_\pi^2/F^2)$ terms we neglected in \eqref{relaxion_mass1}.
We thus see that the relaxion mass is expected to obey the standard scaling relation of a generic QCD axion. 

\section{Summary and Numerical Results}
\label{sect:summary_numerical}
In this section, we present a summary of our analytical results, as well as selected numerical results.

In order to present the analytical results in the cleanest fashion possible, we first exchange all appearances of $m$ for $\epsilon_t$ using \eqref{eps_defn}, and we replace $\LambdaTC \rightarrow 4\pi F_\pi / \sqrt{N}$.
Next, keeping the terms at $\mathcal{O}(\delta^2)$ in the expansion for $\hvev^2/F_\pi^2$ which were not shown explicitly at \eqref{exp_soln}, we invert that expansion to obtain $\delta$ as a power series in $\hvev^2/F_\pi^2$ correct to $\mathcal{O}(\hvev^4/F_\pi^4)$, and insert this inverted expansion into the various results from \sectref[s]{effective_potential} and \ref{sect:relaxion_potential} that had previously been expanded in powers of $\delta$.
In this fashion, all results other than $\hvev^2/F_\pi^2$ can be expressed as a power series in $\hvev^2/F_\pi^2$, while $\hvev^2/F_\pi^2$ is expressed as a power series in $\delta$, which is estimated by \eqref{delta_approx}.
In the broken phase, this procedure leaves us with the following results:
\begin{align}
\delta &\approx \lb( c_t y_t^2 \frac{N_c}{N} \rb)^2 \frac{ f^2 F_\pi^6  }{ F^2 \LambdaLow^6 } \frac{ \gamma_i^2 }{\epsilon_t^2 }  \lb[\frac{4\epsilon_t^2-1}{6\epsilon_t\sqrt{1-\epsilon_t^2}}  \rb] ,\\
\frac{\hvev^2}{F_\pi^2} &=  \frac{6\epsilon_t\sqrt{1-\epsilon_t^2}}{4\epsilon_t^2-1} \delta + \frac{3(2\epsilon_t^4-2\epsilon_t^2+3)}{(4\epsilon_t^2-1)^2} \delta^2 + \cdots, \label{eq:hvevexpansion} \\[1ex]
\frac{\kappaTCvev}{F_\pi} &= - \frac{\sqrt{1-\epsilon_t^2}}{2\sqrt{3}\mkern2mu \epsilon_t}\, \frac{\hvev^2}{F_\pi^2} + \cdots ,\\[1ex]
\frac{m}{F_\pi} &\equiv \frac{1}{8\pi} \lb( \frac{c_t}{c_m} y_t^2 \frac{N_c}{\sqrt{N}} \rb) \frac{1}{\epsilon_t}\, ,\\[1ex]
\frac{m_t}{F_\pi} &= \frac{y_t}{\sqrt{2}} \frac{\hvev}{F_\pi} \lb[ 1 - \frac{1}{6} \frac{\hvev^2}{F_\pi^2} + \cdots \rb], \label{eq:mtop}\\[1ex]
\frac{m_{\text{phys.~Higgs}}^2 }{F_\pi^2} &= \frac{2}{3} \lb( c_ty_t^2 \frac{N_c}{N} \rb) \frac{4\epsilon_t^2-1}{\epsilon_t^2}\frac{ \hvev^2 }{F_\pi^2}\lb[ 1 - \frac{2}{3\epsilon_t^2} \frac{\hvev^2}{F_\pi^2} + \cdots \rb], \label{eq:mphyshiggsepssoln} \\
\frac{m_{\text{phys.~}\kappaTC}^2}{F_\pi^2} \approx \frac{ m_{\piTC^\pm}^2 }{F_\pi^2}  \approx \frac{ m_{\omegaTC}^2}{F_\pi^2}&= 2 \lb(c_t y_t^2 \frac{N_c}{N} \rb),  \\[1ex]
\frac{m_W^2}{F_\pi^2} &= \frac{g_2^2}{4} \frac{\hvev^2}{F_\pi^2}
 \lb[ 1 -  \frac{1}{3}\lb( 1- \frac{3}{4\epsilon_t^2} \rb) \frac{\hvev^2}{F_\pi^2}+ \cdots \rb], \label{eq:W_mass} \\[1ex]
 \frac{m_Z^2}{F_\pi^2} &= \frac{g_1^2+g_2^2}{4} \frac{\hvev^2}{F_\pi^2}
  \lb[ 1 -  \frac{1}{3}   \frac{\hvev^2}{F_\pi^2}+ \cdots \rb], \label{eq:Z_mass} \\
\widehat{T} \equiv \alpha_e T &=   \frac{1}{4\epsilon_t^2}\frac{\hvev^2}{F_\pi^2}+ \cdots , \\
G_F = \frac{g_2^2}{4\sqrt{2} m_W^2} &= \frac{1}{\sqrt{2} \hvev^2} \lb[ 1 + \frac{1}{3} \lb( 1 - \frac{3}{4\epsilon_t^2} \rb) \frac{\hvev^2}{F_\pi^2} + \cdots \rb], \label{eq:GFexp} \\[1ex]
m_\phi^2 &\approx \frac{\hvev \LambdaLow^3}{f^2}  \approx \frac{m_\pi^2 f_\pi^2}{f^2} \qquad \text{(post slope-drop)}.
\end{align}

\begin{figure}[t!]
\centering
\includegraphics[width=\textwidth]{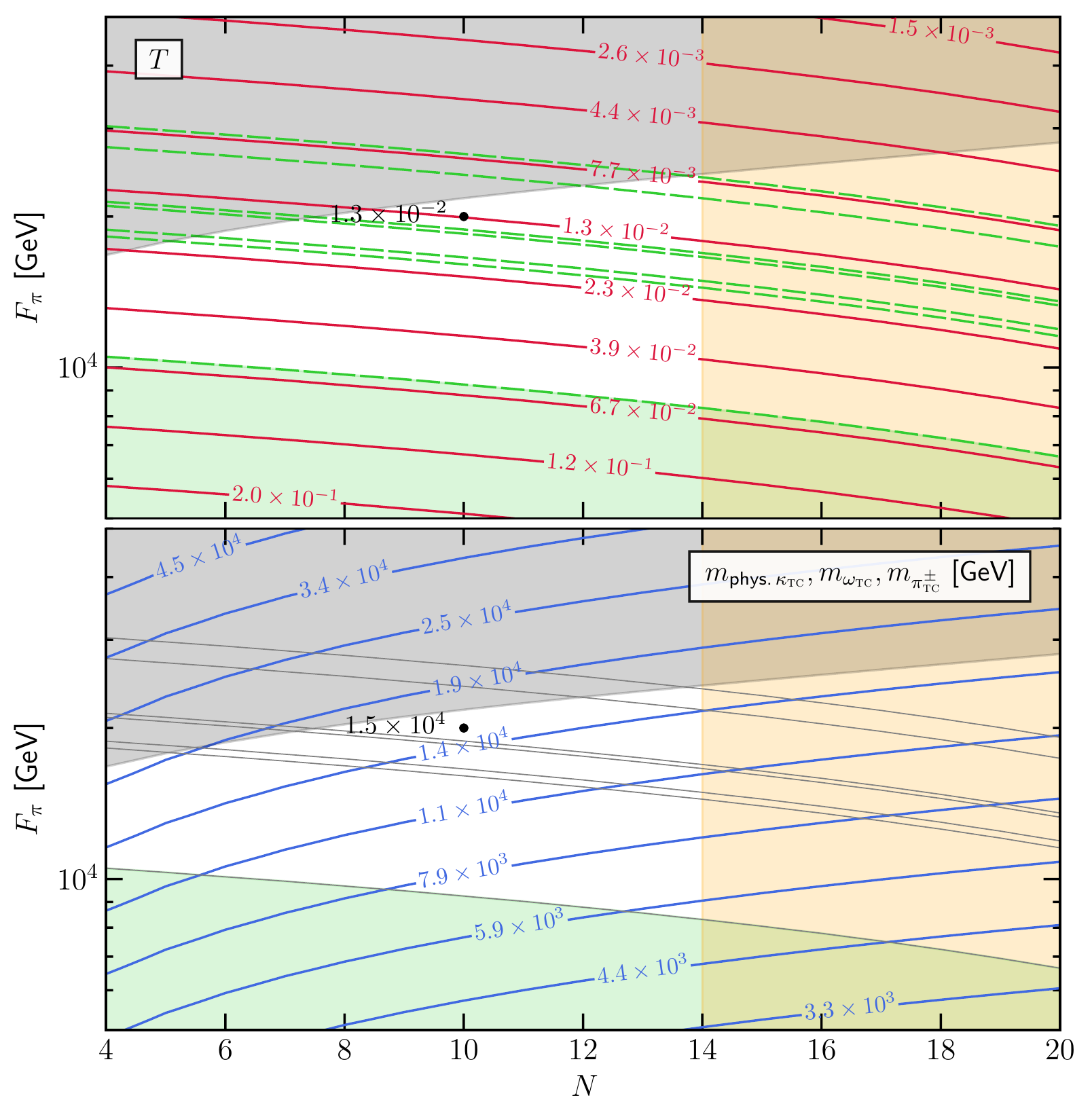}
\caption{\label{fig:numerics_pheno}%
	Numerical results for $T$ and $m_{\text{phys. }\kappaTC}\approx m_\omega\approx m_{\piTC^\pm}$, as a function of $N$ and $F_\pi$, with $c_m=c_t=1$, $N_c=3$, $\Lambda_{\text{low}} = 8.5\,$MeV, $f=10^{11}$\,GeV, $\gamma_i = 10^{10}$, and $\alpha_e(m_Z) \approx 1/127.950(17)$ \cite{Olive:2016xmw} fixed, and with $y_t,\ \hvev,$ and $\epsilon_t$ adjusted to obtain $G_F= 1.1663787(6)\eten{-5}$\,GeV${}^{-2}$ \cite{Olive:2016xmw}, $m_t = 173.21(51)(71)$\,GeV \cite{Olive:2016xmw}, and $m_{\text{phys. Higgs}} = 125.09(24)$\,GeV \cite{Olive:2016xmw}.
	The dashed (green) lines in the left panel are the one-parameter 95\%-confidence exclusion regions on the $T$ parameter (taken at fixed $S=U=0$ and computed using the results in Table 1 of \citeR{Fedderke:2015txa}\up{a}), for a variety of current and future colliders: from bottom to top, these lines represent the `Current', `CEPC ``Baseline''',  `ILC', `FCC-ee-Z', `CEPC ``Improved $\Gamma_Z,\, \sin^2\theta$''',  `CEPC ``Improved $\Gamma_Z,\, \sin^2\theta,\, m_t$''', and `FCC-ee-t' projections of \citeR{Fan:2014vta}.
	These same contours are indicated by light (grey) lines on the right panel.
	The green shaded region on each panel is excluded by the current $T$-parameter constraints.
	The grey shaded region in each panel is where the consistency relation \eqref{full_model_consistency} is not satisfied.
	The orange shaded region indicates $N>14$, where Landau poles in $g_{1}$ or $g_{2}$ may appear below the Planck scale (see \sectref{CH-sector}).
	Also shown on each panel (black dots) is the parameter point $N=10$ and $F_\pi = 20$\,TeV, along with the value of the respective quantity at that point; this parameter point is singled out for discussion in the text.\vspace{0.15cm}\newline%
	\up{a} {\footnotesize To be explicit, the one-parameter 95\%-confidence upper bound on the $T$ parameter (taken at $S=U=0$), and assuming that the best fit point is $(S,T)=(0,0)$, is given by $1.96\, \sigma_{\textsc{t}} \sqrt{1-\rho_{\textsc{ST}}^2}$, where $\sigma_{\textsc{t}}$ and $\rho_{\textsc{ST}}$ are given in Table 1 of \citeR{Fedderke:2015txa}, which supplies a convenient numerical parametrization of the results of \citeR{Fan:2014vta}.}
	}
\end{figure}

\begin{figure}[p]
\centering
\includegraphics[width=\textheight,angle=90]{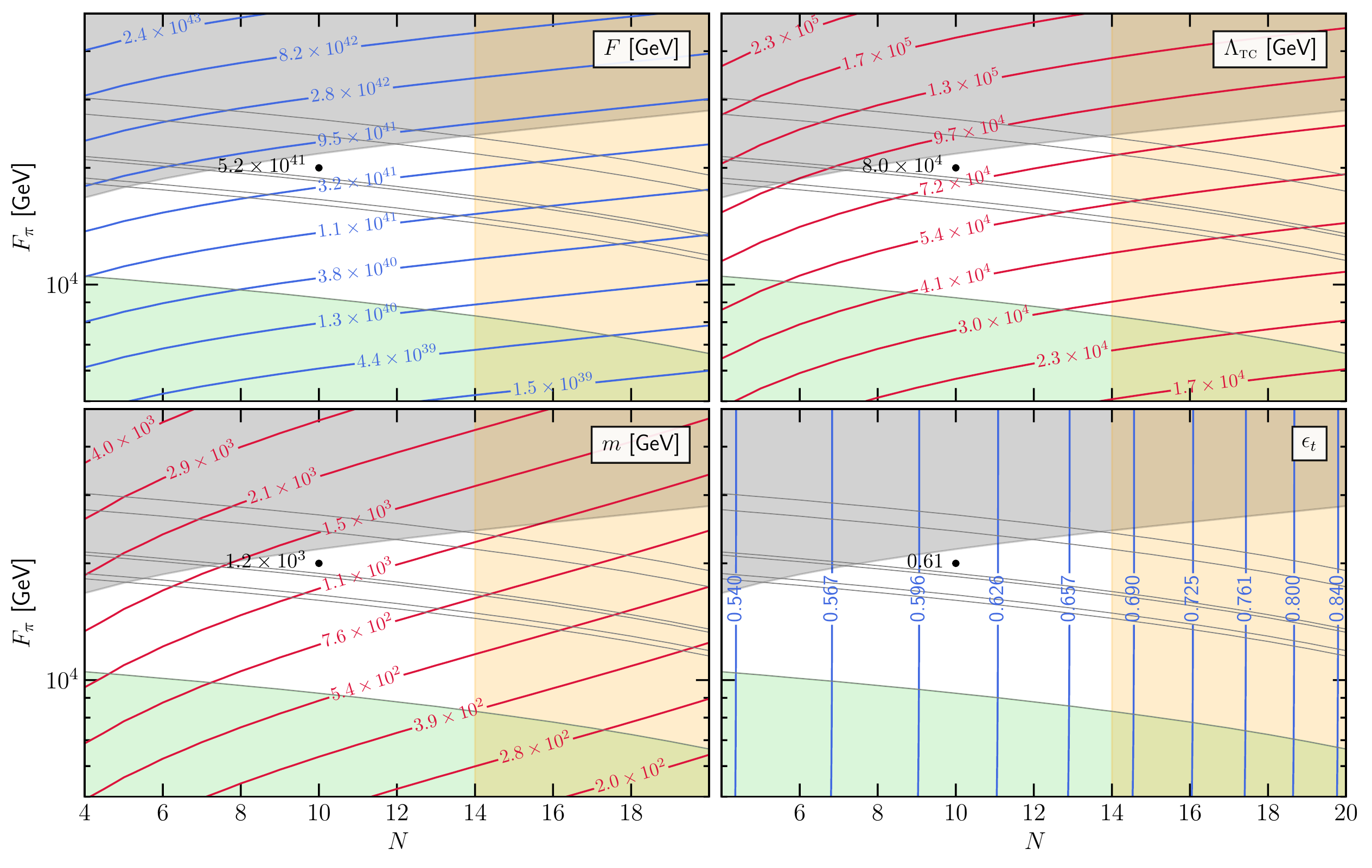}
\caption{\label{fig:numerics_model}%
	As for \figref{numerics_pheno}, but showing the phenomenologically less interesting results for $F$, $\LambdaTC$, $m$, and $\epsilon_t$, as a function of $N$ and $F_\pi$.
		}
\end{figure}
\clearpage

To investigate these results numerically, we fix $c_m=c_t=1$, $N_c=3$, $\LambdaLow = 8.5\,$MeV, $f=10^{11}$\,GeV, and $\gamma_i = 10^{10}$.
Using as input the values $G_F= 1.1663787(6)\eten{-5}$\,GeV${}^{-2}$ \cite{Olive:2016xmw}, $m_t = 173.21(51)(71)$\,GeV \cite{Olive:2016xmw}, and $m_{\text{phys. Higgs}} = 125.09(24)$\,GeV \cite{Olive:2016xmw}, we scan over $N$ and $F_\pi$, solving for $y_t$, $\hvev$, and $\epsilon_t$ using \eqref[s]{mtop}, (\ref{eq:mphyshiggsepssoln}), and (\ref{eq:GFexp}). 
We also fix $\alpha_e(m_Z) \approx 1/127.950(17)$ \cite{Olive:2016xmw} to be able to compute $T$ using the values of $\epsilon_t$ and $\hvev$ thus obtained.%
\footnote{%
            We do not require the values of $g_1$ and $g_2$ to discuss the relevant phenomenology.
            They could be obtained by, e.g., additionally fixing $m_Z=91.1876(21)$\,GeV \cite{Olive:2016xmw}, or by performing a global EW fit with the additional tree-level shifts indicated above accounted for; however, as is well-known even in the SM, it would be necessary to include the one-loop SM corrections in order to obtain accurate values here. 
	} %
The results of this numerical analysis are shown in \figref[s]{numerics_pheno} and \ref{fig:numerics_model}; the former shows the phenomenologically interesting results for $T$ and $m_{\text{phys.~}\kappaTC} \approx m_{\piTC^\pm}  \approx m_{\omegaTC}$, while the latter shows the values of $F$, $m$, $\LambdaTC$, and $\epsilon_t$ that are required at each point in parameter space.
Also shown in \figref{numerics_pheno} are the current and projected 95\%-confidence one-parameter upper limits on the $T$ parameter (taken at fixed $S=U=0$) for a variety of proposed collider configurations for the ILC, CEPC, and FCC-ee (these limits are taken from the presentation in \citeR{Fedderke:2015txa} of the limits examined in \citeR{Fan:2014vta}, and assume that the best-fit point for the global electroweak fit is at $(S,T)=(0,0)$; cf.~\citeR{Baak:2014ora}).

A benchmark parameter point of interest is $N=10$ and $F_\pi = 20$\,TeV, indicated by the black dots on \figref[s]{numerics_pheno} and \ref{fig:numerics_model}.
At this point, and with $c_m,\ c_t,\ N_c,\ \LambdaLow,\ f,$ and $\gamma_i$ fixed as above, we have $\LambdaTC = 80$\,TeV, $F = 5.2\eten{41}$\,GeV, $T = 1.3\eten{-2}$, $m_{\text{phys. }\kappaTC} \approx m_{\omegaTC} \approx m_{\piTC^\pm} = 15$\,TeV, $m = 1.2$\,TeV, $\epsilon_t = 0.61$, $y_t = 0.99$, and $\hvev = 246$\,GeV. 
Additionally, we find $\theta_{\textsc{qcd}}^{\text{eff.}} \approx \frac{\pi}{2}$ at the initial stalling point.
Post slope-drop, we have $m_\phi \approx 120$\,$\mu$eV and $|\theta_{\textsc{qcd}}^{\text{eff.}}|=7.9\eten{-11}$, which is small enough to evade the constraint from the neutron electric dipole moment (EDM) limits, $|d_n| \sim 3\eten{-16} |\theta_{\textsc{qcd}}^{\text{eff.}}| < 3\eten{-26}$\,e\,cm \cite{Crewther:1979pi,Olive:2016xmw}.

In \figref[s]{numerics_pheno} and \ref{fig:numerics_model}, we also show the limits of the parameter region in which the self-consistency conditions discussed in \sectref{cartoonSelfConsistency} are satisfied.
The essential content of the conditions for the cartoon model are captured by \eqref{LambdaTC_limit} which, taking\linebreak $\LambdaTC = 4\pi F_\pi / \sqrt{N}$, expresses a constraint on the upper limit of $F_\pi$ of about 20\,TeV. 
For the full model, essentially the same parametric estimate is obtained, which the exception of a weak $\epsilon_t$-dependence:
\begin{align}
F_\pi &\lesssim \lb( \frac{\sqrt[3]{3} N }{ 2\gamma_i c_t y_t^2N_c} \rb)^{\frac{1}{4}} (2\epsilon_t)^{\frac{1}{4}} \lb( \frac{\LambdaQCD^4 \MPlred^3}{f} \rb)^{\frac{1}{6}} \approx (20\,\text{TeV}) \times \lb( \frac{N}{10} \rb)^{\frac{1}{4}} \times \lb( \frac{\epsilon_t}{0.61} \rb)^{\frac{1}{4}}  \label{eq:full_model_consistency}.
\end{align}
For $N=10$, this translates to a lower bound $\xi \gtrsim 1.2\eten{-4}$.

The massive bound states here are on the order of 10\,TeV, and are charged only under the electroweak gauge group; as such they would be extremely hard to see at the proposed SPPC and FCC-hh high-energy hadron colliders.
The $T$ parameter is however a highly relevant probe: although current experiments are not sensitive to values of $T$ as small as those obtained at the benchmark point $N=10$ and $F_\pi = 20$\,TeV, with improved $Z$-pole measurements and a top-threshold scan, both CEPC and FCC-ee would be able to exclude at 95\% confidence not only this benchmark point, but almost the entirety of the model parameter space in which the relaxion consistency conditions are satisfied, and in which Landau poles in the couplings $g_{1,2}$ are not expected below the Planck scale.

\section{Other Considerations}
\label{sect:other_considerations}
The cosmological relaxation mechanism generates a technically natural weak scale when the cutoff of the SM effective field theory is at much higher scales.
In our context, we have seen that we are able to push the cutoff $\LambdaTC$ to scales of order 100\,TeV.
Despite the high scale of new physics, it is possible to imagine low energy probes of scenario, including precision electroweak, flavor, and CP tests, as well as signatures of heavy technibaryon composite dark matter.
Similar considerations would follow from a tuned composite Higgs model (see, e.g., \citeR[s]{Vecchi:2013iza,Barnard:2014tla}); searches for the relaxion are therefore crucial to test the scenario, although connecting the low- and high-energy dynamics may be challenging.
In this section we will make a few remarks concerning these issues. 

\subsection{Flavor}
\label{sect:flavor_short}
In this work we have taken a bottom-up approach to flavor, writing only the minimal effective operators that generate the Yukawa couplings; none of our detailed model conclusions depend sensitively on the exact UV mechanism leading to these couplings.
The relaxion mechanism allows us to push $\LambdaTC$ to higher scales than in typical quasi-natural composite Higgs models, and this implies that the scale of flavor dynamics may also be at a higher scale.
This generally eases the severe constraints from anomalous flavor changing neutral currents (FCNCs) and potentially allows for simpler UV models of flavor (i.e., without necessarily requiring large anomalous dimensions of the composite operators, walking dynamics, etc.~\cite{Holdom:1984sk,Yamawaki:1985zg,Appelquist:1986an,Akiba:1985rr,Appelquist:1987fc,Appelquist:1986tr,Luty:2004ye,Rattazzi:2008pe}).
Nevertheless, it is not possible in our scenario to push the flavor scale arbitrarily high due to the self-consistency conditions in the inflation sector.
In particular, the upper limit on the cutoff in our model ($\sim 10^2$\,TeV) is too low to be automatically safe from current flavor constraints on composite Higgs models; we thus expect additional UV structure will be required in the flavor sector.
Thus it may still be possible to have experimental signatures of flavor and CP violation within the reach of current and future experiments.

\subsection{Relaxion Phenomenology}
\label{sect:relaxion_pheno}
A chief prediction of this scenario is the existence of a QCD (rel-)axion, which can be tested through a variety of techniques, depending on its underlying couplings to the SM; see, e.g., \citeR{Jaeckel:2015txa} for a review of axion phenomenology.
The classic probes of the relaxion--photon coupling include helioscopes, light-shining-through-walls experiments, and observations of a variety of astrophysical systems in which relaxions may be produced.
Furthermore, since the effective QCD $\theta$-angle is expected to be small but non-zero, it may be possible to probe the relaxion--gluon coupling with improved measurements of the static neutron EDM.
The relaxion may also form some or all of the dark matter, and there are numerous proposals expected to make significant inroads in the axion dark matter parameter space; see, e.g., \citeR[s]{Asztalos:2003px,Budker:2013hfa,Arvanitaki:2014dfa,Rybka:2014cya,Graham:2015ifn,Kahn:2016aff,Barbieri:2016vwg,Arvanitaki:2016fyj,Chung:2016ysi,TheMADMAXWorkingGroup:2016hpc,Stadnik:2013raa,Stadnik:2014tta,Stadnik:2015kia}.

\subsection{Custodial Models}
\label{sect:custodial_models}
The coset studied in this work does not admit a custodial symmetry, leading generically to a large $T$ parameter if the compositeness scale is below about 10\,TeV (see \figref{numerics_pheno}).
As we have shown, the relaxion mechanism can allow a high compositeness scale, but this means it will be challenging to directly search for the additional composite pNGBs and resonances at colliders.%
\footnote{%
	See also, e.g., \citeR[s]{Croon:2015wba,Banerjee:2017qod} for some non-minimal composite Higgs scenarios in which the resonances are made heavier compared to the compositeness scale, also allowing evasion of direct search bounds with alleviated tuning.
	} %
On the other hand, one can certainly consider cosets that are custodially symmetric.
For such models, it is conceivable that the compositeness scale is as low as a few TeV, and that the relaxion addresses only a mild little hierarchy.
This would potentially open the window for observation of new heavy composite particles at the LHC and future hadron colliders. 

\subsection{Composite Dark Matter}
\label{sect:composite_DM}
 Another potentially interesting consequence of the general scenario is heavy composite dark matter in the form of a technibaryon (see, e.g., \citeR[s]{Nussinov:1985xr,Chivukula:1989qb,Barr:1990ca,Bagnasco:1993st,Gudnason:2006yj,Antipin:2015xia,Kribs:2016cew}).
The technibaryon in our scenario is expected to have a mass $m_{B} \sim N \LambdaTC  \sim 100\mkern1.5mu$--$1000$\,TeV, which is, interestingly, in the correct range for a thermal relic cosmology.
In general composite Higgs models, one must take care to ensure that the lightest technibaryon is a neutral state if a dark-matter interpretation is desired.
Furthermore, if the lightest state carries hypercharge, the scenario may face strong constraints from direct detection experiments due to the tree-level $Z$-boson exchange.
Possible phenomenological implications range from scattering in direct detection experiments, to indirect signals from decaying dark matter.  
  
\subsection{Inflation sector}
\label{sect:inflation_sector}
The success of our scenario relies on an exponentially long period of low-scale inflation.
As realized in the original relaxion paper~\cite{Graham:2015cka}, the construction of an explicit model of inflation with these features that is consistent with cosmological observations from Planck \cite{Ade:2015xua} and other experiments presents a challenging task, and may very well bring with it new naturalness questions that would need to be addressed.
Moreover, the viability of the slope-drop mechanism cannot be evaluated without an explicit construction.
While these issues concerning the inflation sector go beyond the scope of our work, they clearly represent a critical open problem and we encourage further model building efforts addressing this sector.
See, e.g., \citeR[s]{Hardy:2015laa,Patil:2015oxa,DiChiara:2015euo,Hook:2016mqo,Higaki:2016cqb,Choi:2016kke,You:2017kah,Evans:2017bjs} for further work on this issue.

\section{Conclusion}
\label{sect:conclusion}
In this paper, we have examined how the cosmological relaxation mechanism of \citeR{Graham:2015cka} can be utilized to dynamically generate the little hierarchy in a composite Higgs model based on underlying strong $\suTC$ technicolor dynamics with three Dirac fermion flavors, leading to the global chiral symmetry breaking $\suLR \times U(1)_{\textrm{V}} \rightarrow SU(3)_{\textrm{V}}\times U(1)_{\textrm{V}}$. 
The relaxion was given anomaly-like couplings to both the technicolor and QCD gauge groups, with the former giving rise to the requisite coupling of the relaxion to the pNGB Higgs in the low-energy Chiral Lagrangian description, and the latter giving rise to the back-reaction on the relaxion slope required to stall the relaxion rolling after QCD chiral symmetry breaking and confinement \cite{Graham:2015cka}.
The hierarchy of axion decay constants $f \ll F$ required by our model was engineered using a clockwork mechanism \cite{Kaplan:2015fuy,Choi:2014rja,Choi:2015fiu}.
We found that an additional potential term for the relaxion is required in this model to obtain the correct slow-roll direction for the relaxion field during the requisite exponentially long period of low-scale inflation.
With the additional term in the potential, the technifermion masses---a source of explicit global chiral symmetry breaking---are scanned as the relaxion field rolls, which in turn results in the scanning of the term in the Higgs potential which opposes EWSB.
Eventually, this scanning results in the importance of the dominant top-loop radiative corrections to the Higgs potential increasing relative to the contributions arising from the technifermion masses, leading to dynamical EWSB provided the technifermion masses are chosen to be roughly a loop factor below the cutoff scale of the composite theory.
We utilized the post-inflation slope-drop mechanism of \citeR{Graham:2015cka} to obtain an acceptably small QCD $\theta$-angle in this framework.

We conclude that little hierarchies on the order of $\xi \equiv \langle h \rangle^2 / F_\pi^2 \sim \mathcal{O}(10^{-4})$ can be generated by our model (i.e., $F_\pi \sim 20$\,TeV, with $\LambdaTC \sim 80$\,TeV for $N=10$) while remaining within the region of parameter space in which the relaxion model is self-consistent, and without running afoul of the QCD $\theta$-angle constraint.
Phenomenological signatures of this (custodial violating) model include an electroweak $T$ parameter large enough that high-precision measurements at proposed $e^+e^-$ Higgs factories could explore essentially the entire viable parameter space for the model at the 95\%-confidence exclusion level; a set of electroweak-charged states with masses on the order of 10\,TeV, which would be challenging to observe at next-generation hadron colliders owing to large backgrounds and low rates; observables related to the existence of a QCD-like axion; and---depending on how this part of the model is implemented in detail---possibly also additional strongly charged states associated with the clockwork mechanism.
Additionally, within the general scenario of composite Higgs models with cutoffs on the order of 100\,TeV (whether tuned, or arising from relaxation as we have considered here), there may be interesting signatures associated with flavor physics, or technibaryon dark matter.

\acknowledgments
We would like to thank Andrea Tesi for many useful discussions, and collaboration at an early stage of this project. 
The work of B.B.\ is supported in part by the U.S.\ Department of Energy under grant No.~DE-SC0015634, and in part by PITT PACC. 
The work of M.A.F.\ is supported in part by the Kavli Institute for Cosmological Physics at the University of Chicago through grant NSF PHY-1125897 and an endowment from the Kavli Foundation and its founder Fred Kavli.
The work of L.-T.W.\ is supported by the U.S. Department of Energy under grant No.~DE-SC0013642.  
We would also like to thank the Aspen Center for Physics for hospitality, where part of the work was completed.
The Aspen Center for Physics is supported by the NSF under Grant No.~PHYS-1066293.

\appendix

\section{\texorpdfstring{$\bm{SU(3)}$}{SU(3)} Exponentiation}
\label{app:SU3exponentiation}
The generic form of an exponentiated $SU(3)$ matrix $\Umat$ in terms of the pion fields can be expressed in closed form \cite{Curtright:2015iba}.
If we define
\begin{align}
H &\equiv 2 \frac{ \piTC ^a T^a }{ \sqrt{ \piTC^a \piTC^a } }\,, &
\theta &\equiv \frac{\sqrt{\piTC^a\piTC^a} }{F_\pi}\,, &
\text{and} &&
\varphi &\equiv \frac{1}{3} \lb[ \arccos\lb( \frac{3\sqrt{3}}{2} \det H \rb) - \frac{\pi}{2} \rb],
\end{align}
then
\begin{align}
\Umat &\equiv \exp\lb[ \frac{2i}{F_\pi} \piTC^a  T^a \rb] = \exp[i\theta H ] \\
&= \sum_{k=0}^2 
			\lb[ 	H^2 + \frac{2}{\sqrt{3}} H \sin\lb( \varphi + \tfrac{2\pi k}{3} \rb)  
				- \frac{1}{3} \mathds{1}_3 \lb( 1 + 2 \cos\lb[ 2 \lb( \varphi + \tfrac{2\pi k}{3} \rb) \rb] \rb)
			\rb] \nl \qquad \; \times \frac{ \exp\lb[ \frac{2}{\sqrt{3}} i \theta \sin\lb( \varphi + \frac{2\pi k }{3} \rb) \rb] }{ 1 - 2 \cos\lb[ 2 \lb( \varphi + \frac{2\pi k}{3} \rb) \rb] }. \label{eq:Ufull}
\end{align}
This expression is of limited direct utility owing to its complexity; it is however of great utility in providing a closed-form expression for $\Umat$ that can be expanded out to find certain relevant terms, as outlined in, e.g., \sectref{effective_potential}.

\section{Unequal Masses, \texorpdfstring{$\bm{m_L\neq m_N}$}{mL =/= mN}}
\label{app:unequal_masses}
In \sectref{equal_mass}, we specialized to the case $m_L=m_N\equiv m$, as this simplified the presentation of the analysis of the effective potential in the main body of the paper.
The aim of this appendix is to demonstrate that qualitatively the same EWSB picture is obtained for the case $m_L \neq m_N$.

In particular, we will demonstrate that over a non-negligible region of parameter space, the most important characteristics of the equal-mass case carry over to the unequal-mass case: (a) an initially stable EW-symmetric solution is destabilized as $\phi$ rolls through a critical value, (b) the destabilization is due to the squared-eigenmass corresponding to the $h$-field changing sign when evaluated at the EW-symmetric solution, while the other squared-eigenmasses remain positive there, and (c) $\phi = \phicrit$ is a supercritical pitchfork bifurcation point for the system.

As in \sectref{equal_mass}, the one-loop effective potential is again obtained by combining \eqref[s]{Vefftree}--(\ref{eq:Vt}), except we now define
\begin{align}
\epsilon_t \equiv \dfrac{3 c_t N_c |y_t|^2 \LambdaTC}{32\pi^2 c_m ( 2m_L + m_N )},
\end{align}
which reduces to the definition \eqref{eps_defn} in the $m_L=m_N\equiv m$ limit;
the top-loop contribution to the potential is then written as [cf.~\eqref[s]{Vt} and (\ref{eq:Vfullequal})]
\begin{align}
V \supset - \frac{2}{3}F_\pi^2 \LambdaTC c_m (2m_L+m_N)\, \epsilon_t\, \frac{h^2}{F_\pi^2} \sinc^2\lb( \frac{\barpiTC}{F_\pi} \rb).
\end{align}
We also define
\begin{align}
m_N \equiv z\mkern1mu m_L.
\end{align}

In this appendix, we will ignore the $V_\phi(\phi)$ and $V_{\textsc{qcd}}$ contributions to the potential, and present an analysis of the EWSB dynamics analogous to that in \sectref[s]{EWsym} and \ref{sect:EWbroken}, working in the limit $\phi = \phicrit + 3F\delta$ where $|\delta| \ll 1$; the value of $\phicrit$ for the unequal mass case will be defined at \eqref{cosphicritunequal} below.

Evaluating the minimization conditions $\partial_X V = 0$ for $X\in\{ h,\, \kappaTC ,\, \omegaTC \}$, we find that the EW-symmetric phase is given by 
\begin{align}
\hvev&=0, &
\text{and}&&
\kappaTCvev &= \frac{1}{\sqrt{3}} \omegaTCvev \qquad [\text{i.e., } \langle \pi^0_{\textsc{tc}} \rangle=0,\ \langle \eta_{\textsc{tc}} \rangle \neq 0],
\end{align}
with $\omegaTCvev$ implicitly defined as a function of $\phi$ by (for $z\neq1$)
\begin{align}
\sin\lb(\frac{\phi}{3F} \rb) &= \pm \frac{ \sin\lb(\frac{2\omegaTCvev}{3F_\pi}\rb) + z \sin\lb(\frac{4\omegaTCvev}{3F_\pi}\rb) }{\sqrt{ 1 + z^2 - 2z \cos\lb( \frac{2\omegaTCvev}{F_\pi} \rb) } } & \text{for}&& \cos\lb(\frac{2\omegaTCvev}{3F_\pi} \rb) \lessgtr z \cos\lb(\frac{4\omegaTCvev}{3F_\pi} \rb). \label{eq:omegasol}
\end{align}
There are two regions of parameter space to consider.

\subsection{Case 1}
The first case is $\cos\lb(2\omegaTCvev/3F_\pi \rb) < z \cos\lb(4\omegaTCvev/3F_\pi \rb) $; this can only obtained in the vicinity of $\phicrit$ if $z>1$.

We will phrase the physical squared-eigenmasses in terms of $\omegaTCvev$ instead of $\phi$ because inverting \eqref{omegasol} in closed form is not straightforward.
In the EW-symmetric phase, the three squared-eigenmasses of the canonically normalized%
\footnote{%
	At $\hvev=0$ and $\omegaTCvev \neq 0$, the $h$ field has a non-canonical kinetic term $\Lag \supset \frac{1}{2} (\partial_\mu h)^2 \sinc^2( \omegaTCvev / F_\pi )$; cf.~\eqref{terms_b}.
	We must thus rescale $h \rightarrow h / \sinc( \omegaTCvev / F_\pi )$ to canonically normalize.
	The result for $m_h^2$ at \eqref{m1case1} is shown after this rescaling has been effected.
	} %
$h$--$\kappaTC$--$\omegaTC$ system are, as a function of $\omegaTCvev$,
\begin{align}
m_h^2 &= \frac{2}{3} c_m \LambdaTC m_L\lb[ \frac{3(z^2-1)}{\sqrt{z^2+1-2z\cos\lb(\tfrac{2\omegaTCvev}{F_\pi}\rb)}} - 2(z+2) \epsilon_t \rb],
 \label{eq:m1case1} \\ 
m_2^2 &= 4 c_m \LambdaTC m_L \frac{z\cos\lb(\tfrac{2\omegaTCvev}{F_\pi}\rb)-1}{\sqrt{z^2+1-2z\cos\lb(\tfrac{2\omegaTCvev}{F_\pi}\rb)}}\,, \\
m_3^2 &= \frac{4}{3} c_m \LambdaTC m_L \frac{(2z+1)(z-1)+2z\sin^2\lb(\tfrac{\omegaTCvev}{F_\pi}\rb)}{\sqrt{z^2+1-2z\cos\lb(\tfrac{2\omegaTCvev}{F_\pi}\rb)}}\, .
\end{align}
It is easy to see that $m_3^2$ is always positive on $z>1$; $m_h^2$ will change sign at $\phi=\phicrit$ [condition (a)] if
\begin{align}
\frac{3(z-1)}{2(z+2)}&<\epsilon_t<\frac{3(z+1)}{2(z+2)},
\end{align}
while demanding that $m_2^2$ is still positive at this point [condition (b)] implies that 
\begin{align}
\epsilon_t > \frac{3\sqrt{z^2-1}}{2(z+2)}.
\end{align}
Here, $\phicrit$ is given by
\begin{align}
\cos\lb(\frac{\phicrit}{3F}\rb) &= \lb\{ 1 -  \frac{4\epsilon_t^2(z+2)^2}{9(z^2-1)^2} \lb[ \begin{array}{l}
	 \sin\lb\{ \dfrac{1}{3} \arccos\lb[ \dfrac{1}{2z} \lb( 1+z^2-\dfrac{9(z^2-1)^2}{4\epsilon_t^2(z+2)^2} \rb) \rb] \rb\} \\[4ex]
	 + z \sin\lb\{ \dfrac{2}{3} \arccos\lb[ \dfrac{1}{2z} \lb( 1+z^2-\dfrac{9(z^2-1)^2}{4\epsilon_t^2(z+2)^2} \rb)  \rb] \rb\}  \end{array}\rb]^2 \rb\}^{\frac{1}{2}}\label{eq:cosphicritunequal}\\[1ex]
	 &\stackrel{z\simeq1}{=}  \epsilon_t - \frac{z-1}{6\epsilon_t} +\cdots . 
\end{align}

It remains to satisfy condition (c): that the destabilization point at $\phi = \phicrit$ is a supercritical pitchfork bifurcation.
As in the equal-mass case, this can be done by demonstrating that, in addition to $\hvev=0$, two other solutions for $\hvev$ exist for $\delta >0$, and that these separate from $\hvev=0$ as $\delta$ increases in size.
Because the solution with $\hvev=0$ is stable for $\phi < \phicrit$ and unstable for $\phi >\phicrit$, the existence of two such additional extrema in the potential in the vicinity for $\hvev=0$ for $\phi > \phicrit$ (i.e., $\delta >0$) suffices to demonstrate the supercriticality of the pitchfork bifurcation. 

We need only study the vevs in the perturbative limit $0<\delta \ll 1$.
We will expand in formal power series
\begin{align}
\omegaTCvev &= \omegaTCvev^{\text{EW-sym}}(\delta) + \sum_{j= 0} \delta^{2j+1} \omegaTCvev^{(2j+1)}, \label{eq:case1omegaexp} \\
\kappaTCvev &= \kappaTCvev^{\text{EW-sym}}(\delta) + \sum_{j= 0} \delta^{2j+1} \kappaTCvev^{(2j+1)}, \label{eq:case1kappaexp} \\
\hvev &= \sqrt{\delta} \sum_{j=0} \delta^j \hvev^{(j)}, \label{eq:case1hexp}
\end{align}
where $\omegaTCvev^{\text{EW-sym}}(\delta)$ and $\kappaTCvev^{\text{EW-sym}}(\delta)$ are the vevs for $\omegaTC$ and $\kappaTC$ in the\linebreak EW-symmetric phase, so that the coefficients $\hvev^{(j)}$, $\kappaTCvev^{(2j+1)}$, and $\omegaTCvev^{(2j+1)}$\linebreak parametrize the deviations from the EW-symmetric solution.
In the EW-symmetric phase we know that we have $\kappaTCvev^{\text{EW-sym}}(\delta) = (1/\sqrt{3}) \omegaTCvev^{\text{EW-sym}}(\delta)$.
We expand
\begin{align}
\omegaTCvev^{\text{EW-sym}}(\delta) &\equiv \omegaTC^{\text{crit.}} + \delta \cdot \omegaTCvev^{\text{EW-sym slope}}+ \cdots, \label{eq:omegaEWsymcase1exp}
\end{align}
where $\omegaTC^{\text{crit.}}$ is the solution to $m_h^2=0$ [see \eqref{m1case1}]:
\begin{align}
\omegaTC^{\text{crit.}} &= \frac{F_\pi}{2} \arccos\lb\{ \frac{1}{2z}\lb[ 1+z^2-\lb( \frac{3(z^2-1)}{2\epsilon_t(z+2)}\rb)^2 \rb] \rb\}, \label{eq:omegacritcase1}
\end{align}
and $\omegaTCvev^{\text{EW-sym slope}}$ is obtained by inserting the expansion \eqref{omegaEWsymcase1exp} and the definition $\phi = \phicrit+3F\delta$ into the EW-symmetric solution \eqref{omegasol}, expanding in powers of $\delta$ and equating coefficients:
\begin{align}
\omegaTCvev^{\text{EW-sym slope}} &=  \frac{9F_\pi(z^2-1)}{3(z^2-1)+4(z+2)^2\epsilon_t^2} . \label{eq:omegacritslopecase1}
\end{align}

To find the values of the other expansion coefficients in \eqref[s]{case1omegaexp}--(\ref{eq:case1hexp}), we substitute all relevant expansions into the extremization conditions $\partial_X V =0$ for $X\in\{h,\, \kappaTC,\allowbreak \omegaTC\}$ and expand in powers of $\delta$ to find a formal power series whose individual coefficients we equate to zero to obtain a hierarchical system of equations which we can solve recursively to find $\hvev^{(j)}$, $\kappaTCvev^{(2j+1)}$, and $\omegaTCvev^{(2j+1)}$; the lowest-order solutions $\hvev^{(0)}$, $\kappaTCvev^{(1)}$, and $\omegaTCvev^{(1)}$ suffice to determine the nature of the bifurcation.
The resulting expressions correctly reduce to the results \eqref{exp_soln} in the $z\rightarrow 1$ equal-mass limit, but are too lengthy to display here explicitly.
Nevertheless, analysis of these expressions allows us to conclude that the two additional solutions for $\hvev^{(0)}$ exist for $0<\delta \ll 1$, and that the pitchfork bifurcation is thus supercritical, if the following conditions are met [we additionally impose $\epsilon_t>0$ by definition]:
\begin{align}
\Bigg[ 0 < &\ \epsilon_t < \frac{3\sqrt{z^2-1}}{2(z+2)} &
&\textsc{or} &
\frac{3z}{2(z+2)} &<  \epsilon_t < \frac{3(z+1)}{2(z+2)} \Bigg] &
\textsc{and}&&
\frac{3(z-1)}{2(z+2)} &< \epsilon_t < \frac{3(z+1)}{2(z+2)} .
\end{align}
The most stringent constraint for the region $z>1$ is thus 
\begin{align}
 \frac{3z}{2(z+2)} &< \epsilon_t < \frac{3(z+1)}{2(z+2)}\ .
\end{align}

\subsection{Case 2}
The second case is $\cos\lb(2\omegaTCvev/3F_\pi \rb) > z \cos\lb(4\omegaTCvev/3F_\pi \rb) $.
The relevant part of this region of parameter space occurs for $z<1$ when $\phi \approx \phicrit$; although there is also some part of this region of parameter space at $z>1$ when $\phi \approx \phicrit$, it turns out not to be interesting for our purposes, and we will not discuss it.

We follow a similar analysis as for the previous case.
In the EW-symmetric phase, the three squared-eigenmasses of the canonically normalized $h$--$\kappaTC$--$\omegaTC$ system are, as a function of $\omegaTCvev$,
\begin{align}
m_h^2 &= \frac{2}{3} c_m \LambdaTC m_L   \lb[ \frac{3(1-z^2)}{\sqrt{z^2+1-2z\cos\lb(\tfrac{2\omegaTCvev}{F_\pi}\rb)}} - 2(z+2) \epsilon_t \rb],
 \label{eq:m1case2}\\
m_2^2 &=   \frac{4}{3} c_m \LambdaTC m_L \frac{1-2z^2+z\cos\lb(\tfrac{2\omegaTCvev}{F_\pi}\rb)}{\sqrt{z^2+1-2z\cos\lb(\tfrac{2\omegaTCvev}{F_\pi}\rb)}}\, , 
\end{align}
and
\begin{align}
m_3^2 &= 4 c_m \LambdaTC m_L \frac{(1-z)+2z\sin^2\lb(\tfrac{\omegaTCvev}{F_\pi}\rb)}{\sqrt{z^2+1-2z\cos\lb(\tfrac{2\omegaTCvev}{F_\pi}\rb)}}\, .
\end{align}
It is again easy to see that $m_3^2$ is always positive on $0<z<1$; $m_h^2$ will change sign at $\phi=\phicrit$ [condition (a)] if
\begin{align}
\frac{3(1-z)}{2(z+2)}&<\epsilon_t<\frac{3(z+1)}{2(z+2)} ,
\end{align}
while demanding that $m_2^2$ is still positive at this point [condition (b)] implies that 
\begin{align}
\epsilon_t > \frac{\sqrt{3}\sqrt{1-z^2}}{2(z+2)} ,
\end{align}
although this latter constraint is only applicable on $\frac{1}{2} < z < 1$, where a solution for $m_2^2=0$ exists.
The value of $\phicrit$ is given by \eqref{cosphicritunequal} without change.

We perform the same perturbative expansion and solution as for the previous case to find the requirements to satisfy condition (c), supercriticality of the pitchfork bifurcation.
The value of $\omegaTC^{\text{crit.}}$ changes sign relative to the solution in \eqref{omegacritcase1}, while $\omegaTCvev^{\text{EW-sym slope}}$ is still given by \eqref{omegacritslopecase1}.
The solutions for $\hvev^{(0)}$, $\kappaTCvev^{(1)}$, and $\omegaTCvev^{(1)}$ are again too lengthy to show here explicitly.

Once again, the relevant conclusion that can be drawn from this procedure is that the two additional solutions for $\hvev^{(0)}$ exist for $0<\delta\ll1$, and that the bifurcation is thus supercritical, if the following conditions are met [we additionally impose $\epsilon_t>0$ by definition]:
\begin{align}
\Bigg[ 0 < &\ \epsilon_t < \frac{3\sqrt{1-z^2}}{2(z+2)} &
&\textsc{or} &
\frac{3z}{2(z+2)} &<  \epsilon_t < \frac{3(z+1)}{2(z+2)} \Bigg] &
\textsc{and}&&
\frac{3(1-z)}{2(z+2)} &< \epsilon_t < \frac{3(z+1)}{2(z+2)} .
\end{align}
The most stringent constraint for the region $z<1$ is thus 
\begin{align}
\text{max} \lb\{ \frac{3(1-z)}{2(z+2)} , \frac{3z}{2(z+2)}\rb\} &< \epsilon_t < \frac{3(z+1)}{2(z+2)} .
\end{align}

\subsection{Summary}
\begin{figure}[t]
\centering
\includegraphics[width=\textwidth]{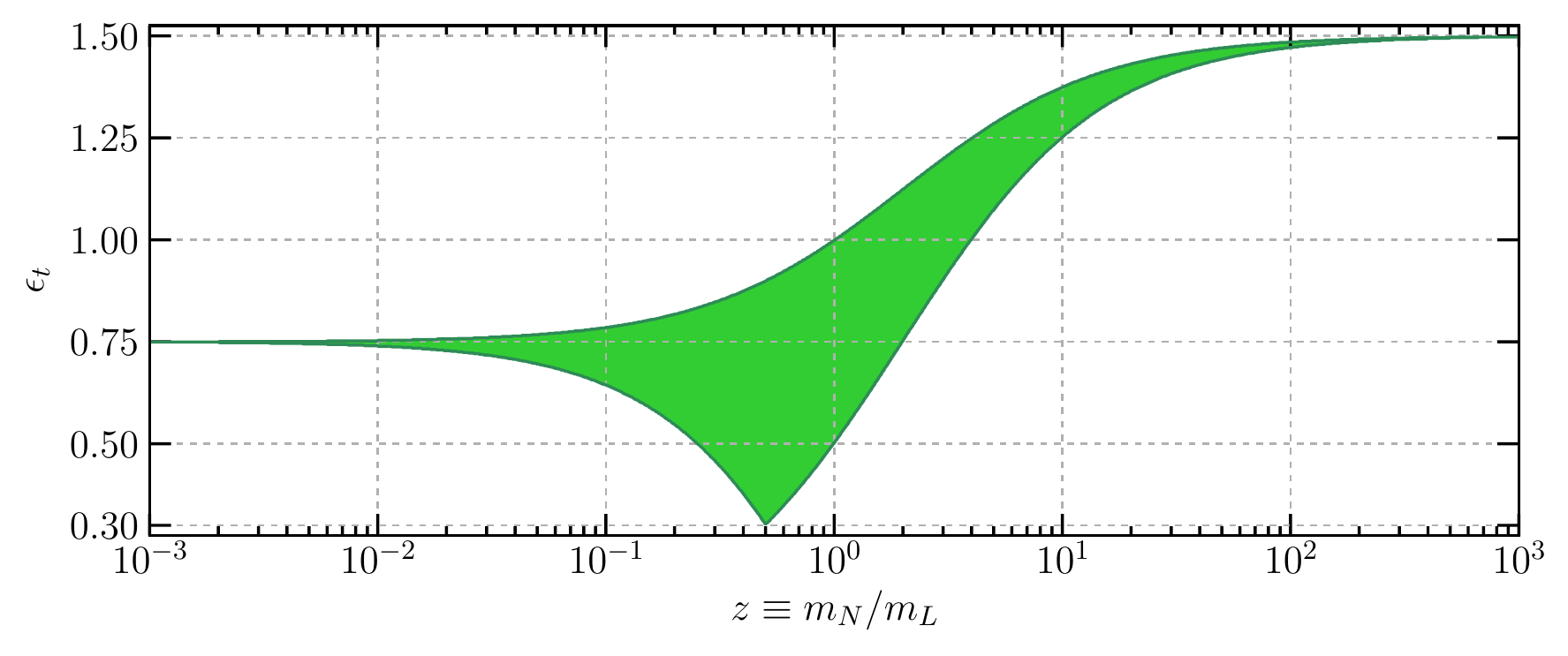}
\caption{\label{fig:unequal_mass_case_region}%
	The green shaded area on this $z$--$\epsilon_t$ plane is the region of parameter space satisfying \eqref{finalunequalconstr}, in which qualitatively similar EWSB dynamics as for the equal-mass case examined in the main body of the paper are obtained.
	}
\end{figure}

In summary, the unequal-mass case exhibits qualitatively similar behavior to the equal-mass case analyzed in the main text when the following conditions are met:
\begin{align}
\text{max} \lb\{ \frac{3|z-1|}{2(z+2)} , \frac{3z}{2(z+2)}\rb\} &< \epsilon_t < \frac{3(z+1)}{2(z+2)} . \label{eq:finalunequalconstr}
\end{align}
This region of parameter space is displayed in \figref{unequal_mass_case_region}.
Note that for $z=1$, the constraint correctly reduces to $1/2<\epsilon_t<1$; cf.~\eqref{eps_constraint}.

\phantomsection
\addcontentsline{toc}{section}{References}
\bibliographystyle{JHEP}
\bibliography{bfw}

\end{document}